\documentclass[prb,aps,english,onecolumn,notitlepage]{revtex4-2}
\usepackage{epsfig}
\usepackage{amsmath}
\usepackage{mathtools}
\usepackage{xcolor}

\DeclarePairedDelimiter\ket{\lvert}{\rangle}
\DeclarePairedDelimiterX\braket[2]{\langle}{\rangle}{#1 \delimsize\vert #2}
\usepackage{sidecap}
\usepackage[figuresleft]{rotating}
\usepackage{caption}
\usepackage{float}
\usepackage{appendix}
\usepackage{subfigure}
\usepackage{graphicx}
\usepackage{hyperref}
\usepackage{array,multirow}
\usepackage{multirow}

\begin{document}
\title{Surface induced odd-frequency spin-triplet superconductivity as a veritable signature of Majorana bound states}
\author{Subhajit Pal}
\author{Colin Benjamin} \email{colin.nano@gmail.com}
\affiliation{School of Physical Sciences, National Institute of Science Education \& Research, Jatni-752050, India.}
\affiliation{Homi Bhabha National Institute, Training School Complex, AnushaktiNagar, Mumbai, 400094, India.}
\begin{abstract}
We predict surface-induced odd-frequency (odd-$\nu$) spin-triplet superconducting pairing can be a veritable signature of Majorana bound states (MBS) in a Josephson nodal $p$-wave superconductor ($p_{x}$)-spin flipper (SF)-nodal $p$-wave superconductor ($p_{x}$) junction. Remarkably, in a $p_{x}$-SF-$p_{x}$ Josephson junction three distinct phases emerge: the topological phase featuring MBS, the topological phase without MBS, and the trivial phase devoid of MBS. Surface odd-$\nu$ spin-triplet pairing is induced only in the topological regime when MBS appears. In contrast, surface-induced even-frequency (even-$\nu$) spin-triplet pairing is finite regardless of the existence of MBS. {Importantly, we find the surface induced odd-$\nu$ spin-triplet pairing is immune to disorder in the topological phase featuring MBS, while in the trivial phase the surface induced even-$\nu$ spin-triplet pairing is affected by disorder}. Our study offers a potential means for distinguishing the topological phase featuring MBS from both the trivial phase as well as the topological phase devoid of MBS, primarily through the observation of induced surface odd-$\nu$ spin-triplet superconductivity.
\end{abstract}
\maketitle
\section{INTRODUCTION}
Odd-$\nu$ pairing is a novel superconducting state wherein the electrons in a Cooper pair have distinct time coordinates along with distinct position coordinates. Generally, two electrons in a Cooper pair are formed simultaneously, categorizing it as an even-$\nu$ pairing state\cite{bar}. Even-$\nu$ pairing state can be subdivided into two categories: even-$\nu$ spin-singlet (SS) and even-$\nu$ spin-triplet (ST) pairing states. Even-$\nu$ SS pairing is exemplified by $s$ and $d$-wave pairings, while $p$-wave pairing serves as an illustration of even-$\nu$ ST pairing\cite{sigr}. For odd-$\nu$ pairing state, Cooper pair wavefunction or pair amplitude is odd in relative time coordinate or frequency of the two Cooper pair electrons\cite{linder,tana,jca}. An odd-$\nu$ pairing state can be further categorized as odd-$\nu$ SS state and odd-$\nu$ ST state. ST states, in turn, can be classified into two subcategories: equal ST (EST) states, represented by $\ket{\uparrow\uparrow}$, $\ket{\downarrow\downarrow}$, and mixed ST (MST) states, represented by $\ket{\uparrow\downarrow}+\ket{\downarrow\uparrow}$. 

{In historical context, the concept of odd-$\nu$ ST pairing dates back to Berezinskii's 1974 proposal in $^{3}$He\cite{beri} and subsequent predictions in disordered systems\cite{trk,bel}. Balatsky and Abrahams later proposed the existence of odd-$\nu$ SS pairing in superconductors with broken time-reversal and parity symmetries\cite{balat}. Subsequent investigations extended to various systems, including a two-channel Kondo system\cite{eme}, the 1D $t-J-h$ model\cite{bon}, the 2D Hubbard model\cite{bulu}, and heavy fermion compounds\cite{cole}. The existence of bulk odd-$\nu$ pairing has been suggested theoretically through the Majorana scanning tunneling microscope\cite{oka}. Experimental indications have also been observed through phenomena such as the Kerr effect\cite{ers,kom} and the paramagnetic Meissner effect\cite{adi,ali,yta}. Moreover, the presence of bulk odd-$\nu$ pairing induced by magnetic impurities has been demonstrated in $s$-wave superconductors\cite{perr,kuzz}.} Earlier, it was assumed that odd-$\nu$ state was an intrinsic effect\cite{balat,abr} but later it was discovered that odd-$\nu$ state can be generated in superconducting junctions\cite{tanak,vol,tak,yoko,ann,linn,lindd2,pal,voll,berg3,kopu,amb1,cre,burr,lof,vol1,fom,buz,hwa,tamu,berg1,berg2,dutta} as well as under time-dependent fields\cite{tri1,tri2}.

Odd-$\nu$ pairing holds intrinsic significance due to its remarkable departure from conventional superconductivity. Odd-$\nu$ spin-polarized Cooper pairs exhibit resilience against both the Pauli limiting field and impurity scattering, in contrast to conventional even-$\nu$ Cooper pairs, which are only robust against impurity scattering\cite{linder}. The exceptional robustness of odd-$\nu$ pairing positions it as an attractive candidate for applications in superconducting spintronics\cite{LIND}.

In this work, we consider a $p_{x}$-SF-$p_{x}$ Josephson junction (JJ), with the nodal $p_{x}$ superconductor featuring even-$\nu$ EST pairing in its Cooper pair. Odd-$\nu$ EST pairing arises from the breaking of spatial parity\cite{tanak} at the $p_{x}$-$p_{x}$ interface. We see that spin-flip scattering induces odd-$\nu$ MST pairing in our setup. JJ's with $p_{x}$ superconductors are theoretically predicted to harbor Majorana fermions\cite{kit}. The Majorana fermion is a particle that is its own antiparticle and has attracted a lot of attention due to its potential application in topological quantum computation\cite{nay,sds,sch}. In the case of $p_{x}$-SF-$p_{x}$ JJ, there is a sign change in energy-bound states when Majorana fermions occur\cite{kyon}.

Our main motivation in this work is to distinguish MBS via the surface-induced odd-$\nu$ MST correlations in the presence of spin-flip scattering. Although surface odd-$\nu$ EST and surface odd-$\nu$ MST correlations are finite and large in the topological regimes when zero energy MBS occur, they vanish in their absence. Surface even-$\nu$ EST \& MST correlations remain finite in both the presence and absence of MBS, rendering them ineffective for MBS detection. Surface-induced odd-$\nu$ ST correlations are a signature of MBS in $p_{x}$ JJ's with a spin-flipper.

In a recent paper\cite{tsi}, it was seen that both odd-$\nu$ EST and even-$\nu$ EST correlations are induced in short $p_{x}$-N-$p_{x}$ JJ. In Ref.~\cite{tsi}, neither even-$\nu$ MST nor odd-$\nu$ MST correlations are present since spin-flip scattering does not exist. In contrast, in our work both odd-/even-$\nu$ EST and odd-/even-$\nu$ MST correlations are induced via the spin flipper. In a related work\cite{yas}, the authors discuss the relation between Majorana fermion and odd-$\nu$ Cooper pair in the case of a disordered superconductor-metal-superconductor JJ of nanowires when the nanowire is in the topologically non-trivial regime. In the absence of spin-flip scattering, it was seen that odd-$\nu$ EST pairing exists whenever MBS appears, while, odd-$\nu$ MST pairing does not arise. Further, the connection between MBS and odd-$\nu$ pairing has also been explored in Kitaev chain systems\cite{taka} and in spin-polarized nanowires coupled to Majorana zero modes\cite{Lee,Kuzm}. In all these Refs.~\cite{tsi}-\cite{Kuzm}, odd-$\nu$ MST pairing vanishes, while in our paper, due to the interplay of both MBS and spin-flip scattering, odd-$\nu$ MST pairing is finite, thereby enhancing their utility in MBS detection.

The remainder of the paper is structured as follows: in the subsequent section, we present our setup and discuss the theoretical background. In section III, we outline the method for computing energy-bound states and Josephson currents. In the same section, we also discuss the results for energy-bound states and Josephson current. Moving on to Section IV, we delve into the relationship between odd-$\nu$ ST superconductivity and MBS in a $p_{x}$-SF-$p_{x}$ JJ. {Next in section V, we discuss the impact of disorder on surface induced odd-$\nu$ spin-triplet pairing.} We analyze the results in {section VI}, offering a comparative summary of odd- and even-$\nu$ correlations induced by the presence of MBS. {In section VII, we present some tools to experimentally detect odd-$\nu$ spin-triplet pairing via the local density of states (LDOS), local magnetization density of states (LMDOS) and enhanced DC Josephson current and  in this section we also summarize our work}. For completeness, we {provide} the wavefunctions and boundary conditions of our setup, along with detailed calculations of Green's functions, analytical formulas for even- and odd-$\nu$ pairing amplitudes, {and the details for calculating the LDOS, LMDOS, and the total DC Josephson current in Appendices A-E.}
\section{MODEL}
We consider a JJ where a spin flipper (SF) is embedded between two nodal $p_{x}$ superconductors, as depicted in Fig.~1. The SF's Hamiltonian is described as follows\cite{AJP,Liu,Maru,FC,ysr}:
\begin{equation}
H_{\mbox{SF}}=-J_{0}\delta(x)\vec{s}\cdot\vec{\mathcal{S}}.
\label{flipper}
\end{equation}
We address this problem by solving a time-independent Schr\"{o}dinger equation, which has been adapted to incorporate a Bogoliubov-de Gennes (BdG) Hamiltonian. The BdG Hamiltonian for  $p_{x}$-SF-$p_{x}$ JJ, as shown in Fig.~1, is expressed as follows:
\begin{equation}
H_{BdG}^{\mbox{$p_{x}$-SF-$p_{x}$}}(x)=
\begin{pmatrix}
H_{P}\hat{I} & \Delta_{J}p  \hat{\sigma}_{x} \\
\Delta_{J}^{*}p \hat{\sigma}_{x} & -H_{P}\hat{I}
\end{pmatrix},
\label{hamm}
\end{equation}
with $H_{P}=-\frac{\hbar^2}{2m^{*}}\frac{\partial^2}{\partial x^2}-J_{0}\delta(x)\vec s\cdot\vec{\mathcal{S}}-\mu_{p_{x}}'$. $-\frac{\hbar^2}{2m^{*}}\frac{\partial^2}{\partial x^2}$ represents electron-like quasiparticle's kinetic energy operator with an effective mass denoted as $m^{*}$, $J_{0}$ denotes the exchange interaction between the spins of the electron-like quasiparticle ($\vec{s}$) and SF ($\vec{\mathcal{S}}$) and, the third term corresponds to the chemical potential of a $p_{x}$-superconductor. $\hat{I}$ represents identity matrix, {$p=-i\hbar\frac{\partial}{\partial x}$ is the momentum operator} and, $\hat{\sigma}$ being the Pauli matrices. Further, the gap parameter $\Delta_{J}$ has the following form $\Delta_{J}=\Delta_{p_{x}}'[e^{i\varphi_{L}}\theta(-x)+e^{i\varphi_{R}}\theta(x)]$ with $\Delta_{p_{x}}'$ is the pairing potential for $p_{x}$ superconductor and $\theta(x)$ represents the unit step function. $\varphi_{L}$ is the superconducting phase for left superconductor, while $\varphi_{R}$ is the superconducting phase for right superconductor. $\varphi=\varphi_{R}-\varphi_{L}$ is the phase difference between two superconductors. In this paper, the SF's spin magnetic moment $m'$ can have possible values, $m'=-\mathcal{S},-\mathcal{S}+1,...,\mathcal{S}$. For example, if $\mathcal{S}=1/2$, then $m'$ has two possible values, $m'=1/2,-1/2$. Similarly, if $\mathcal{S}=3/2$, then $m'=3/2,1/2,-1/2,-3/2$. We compute various measurable quantities like Josephson current or the pairing magnitude {or LDOS/LMDOS} for each of the $2\mathcal{S}+1$ possible values of $m'$ for SF's spin $\mathcal{S}$ and finally take an average overall $m'$ values. When a spin-up electron-like quasiparticle (ELQ) interacts with the SF with a spin of $\mathcal{S}=1/2$, it results in a product state $\frac{m'}{2}\big(\ket{\uparrow}_{ELQ}\otimes\ket{\uparrow}_{SF}\big)$ if the SF's spin is in the up state ($m'=1/2$). However, if the SF is in the spin-down state ($m' = -1/2$), an entangled state may emerge after scattering, represented as
\Big($\overbrace{\frac{f}{2}(\ket{\downarrow}_{ELQ}\otimes\ket{\uparrow}_{SF})}^\textrm{Mutual spin flip}+\overbrace{\frac{m'}{2}(\ket{\uparrow}_{ELQ}\otimes\ket{\downarrow}_{SF})}^\textrm{No flip}$\Big). In the subsequent scattering event at the SF, if a spin-down electron-like quasiparticle is incident, it encounters the SF in either the spin-up or spin-down state. If the SF is in the spin-up state, a spin flip occurs, once again leading to the formation of an entangled state. As a result, measurable quantities like Josephson current or the pairing magnitude are determined by averaging over these two processes: spin-flip and no-flip. This approach applies similarly to SF's spin states of $\mathcal{S}=3/2$ and beyond, with measurable quantities calculated by averaging over all possible values of $m'$. In this paper, we have used dimensionless parameters $J=\frac{2m^{*}J_{0}}{k_{\mu_{p_{x}}}}$ to quantify the exchange coupling strength\cite{AJP}, where $k_{\mu_{p_{x}}}=\sqrt{\frac{2m^{*}\mu_{p_{x}}'}{\hbar^2}}$ is the Fermi momentum.
\begin{figure}[h]
\centering{\includegraphics[width=0.5\textwidth]{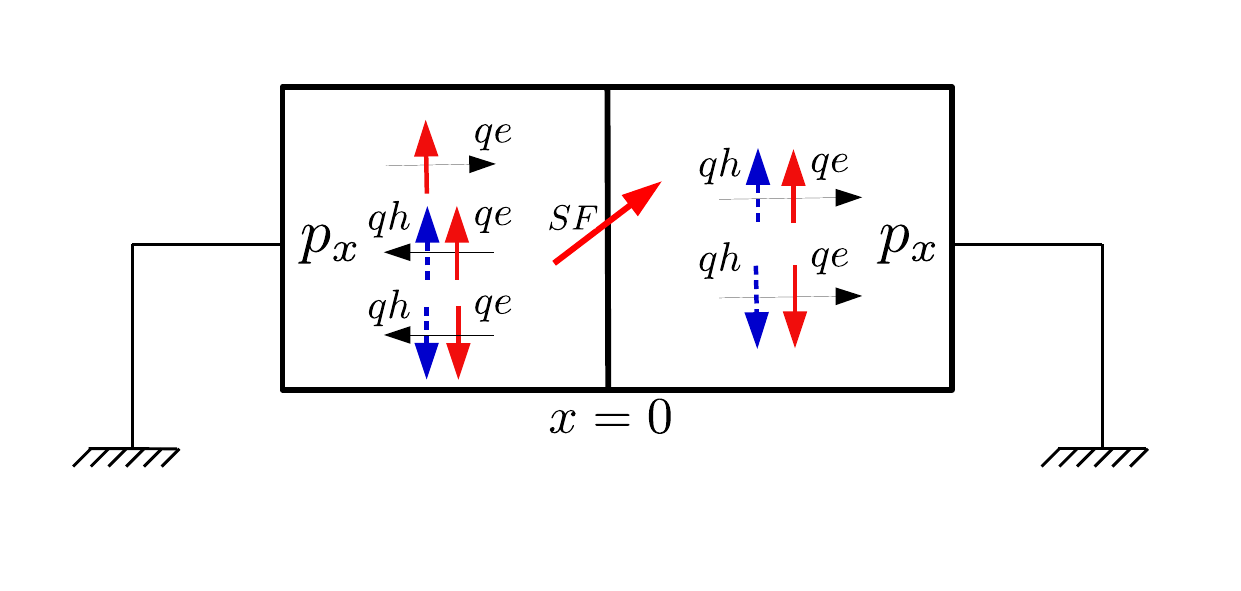}}
\vskip -0.2 in \caption{\small \sl Josephson junction composed of a SF at $x=0$ embedded between two nodal $p_{x}$ superconductors. Two $p_{x}$ superconductors are grounded. The scattering process of an incident spin-up electron-like quasiparticle {(qe)} is depicted and SF's spin being oriented along any arbitrary direction.}
\end{figure}

By diagonalizing the Hamiltonian \eqref{hamm}, we obtain the wavefunctions in distinct regions of the $p_{x}$-SF-$p_{x}$ JJ corresponding to different scattering processes. The detailed wavefunctions are provided in Appendix A. The wavevectors $q_{e,h}$ in $p_{x}$ superconductor can be found from
\begin{equation}
\label{eq5}
\nu^2=(q_{e,h}^2-\mu_{p_{x}}')^2+(\Delta_{p_{x}}' q_{e,h})^2,\,\,\,\,\,\,\,\, \mbox{(assuming $\hbar=2m^{*}=1$)}.
\end{equation}
The general solution of Eq.~\eqref{eq5} for $q_{e,h}$ is given as,
\begin{equation}
\label{eqq5}
q_{e,h}=\pm\sqrt{\frac{-\Delta_{p_{x}}'^2+2\mu_{p_{x}}'\pm\sqrt{4\nu^2+\Delta_{p_{x}}'^4-4\Delta_{p_{x}}'^2\mu_{p_{x}}'}}{2}}.
\end{equation}
In Eq.~\eqref{eqq5}, $q_{e}$ and $q_{h}$  have two values. The positive value of $q_{e}$ ($q_{h}$) represents the electron-like (hole-like) quasiparticle moving from left (right) to right (left), while the negative value of $q_{e}$ ($q_{h}$) represents the electron-like (hole-like) quasiparticle moving from right (left) to left (right).
Expressions for $q_{e,h}$ for different values of chemical potential $\mu_{p_{x}}'\leq\Delta_{p_{x}}'^2/4$, $\Delta_{p_{x}}'^2/4\leq\mu_{p_{x}}'\leq\Delta_{p_{x}}'^2/2$ and, $\mu_{p_{x}}'\geq\Delta_{p_{x}}'^2/2$ and energy $E$ (frequency $\nu$) are provided in Table I of Ref.~\cite{setiawan}. At $\mu_{p_{x}}'=0$, the energy spectrum for $p_{x}$ superconductor is gapless (see Fig.~1 of Ref.~\cite{setiawan}) which indicates the topological transition between the trivial ($\mu_{p_{x}}'<0$) and topological ($\mu_{p_{x}}'>0$) phases. The gap in the energy spectrum of nodal $p_{x}$ superconductor is given as
\begin{equation}
E_{c}=\begin{cases}|\mu_{p_{x}}'|,\,\,\mu_{p_{x}}'<0,\,\,\,\,\,\,\,\,\,\mbox{(trivial regime)}\\
\begin{rcases}
\mu_{p_{x}}',\,\, 0<\mu_{p_{x}}'<\Delta_{p_{x}}'^2/2,\\
\Delta_{p_{x}}'\sqrt{\mu_{p_{x}}'-\Delta_{p_{x}}'^2/4},\,\,\mu_{p_{x}}'>\Delta_{p_{x}}'^2/2.
\end{rcases}\,\,\,\,\,\,\,\,\,\mbox{(topological regime)}
\end{cases}
\end{equation}
In the subsequent sections, we will employ normalized pairing potential, denoted as $\Delta_{p_{x}}=\frac{2m^{*}\Delta_{p_{x}}'}{\hbar^2 k_{\mu_{p_{x}}}}$, and normalized chemical potential, denoted as $\mu_{p_{x}}=\frac{\hbar^2\mu_{p_{x}}'}{m^{*}\Delta_{p_{x}}^2}$.
\section{MAJORANA BOUND STATES AND JOSEPHSON CURRENT}
To compute energy bound states in $p_{x}$-SF-$p_{x}$ JJ we neglect the contribution from incoming quasiparticles in the wavefunctions\cite{enok,anun}, see Eq.~\eqref{waveJ} in Appendix A and substitute these wavefunctions into the boundary conditions, see Eq.~\eqref{bc1} in Appendix A. We will get $8$ equations, with
\begin{equation}
\label{det}
Py=0,
\end{equation}
and wherein $y=[r_{\uparrow\uparrow}^{e'e'},r_{\uparrow\downarrow}^{e'e'},r_{\uparrow\uparrow}^{e'h'},r_{\uparrow\downarrow}^{e'h'},\tilde{t}_{\uparrow\uparrow}^{e'e'},\tilde{t}_{\uparrow\downarrow}^{e'e'},\tilde{t}_{\uparrow\uparrow}^{e'h'},\tilde{t}_{\uparrow\downarrow}^{e'h'}]^{T}$ is a $8\times1$ column matrix and $P$ is a $8\times8$ matrix. For the nontrivial solution of Eq.~\eqref{det}, the det P should be zero and we obtain bound state energies $E_{l}(l=1,...,8)=\pm E_{n}(n=1,...,4)$. Since we consider a short JJ, the Josephson total current is equal to the Josephson bound current, which can be obtained from bound state energies\cite{golu},
\begin{equation}
I=-\frac{2e}{\hbar}\sum_{n=1}^{4}\tanh\Big(\frac{E_{n}}{2k_{B}T}\Big)\frac{dE_{n}}{d\varphi}.
\end{equation}

In Figs.~2(a), (b), bound state energies are plotted as a function of phase difference $\varphi$ in both topological (Fig.~2(a)) and trivial (Fig.~2(b)) regimes. From Fig.~2(a), we see that in the topological regime at $\varphi=\pm\pi$, energy-bound states change their sign due to their $4\pi$ periodicity, which indicates the presence of Majorana zero modes inside the junction\cite{kyon}. These $4\pi$ periodic energy bound states occur because of the coupling of two Majorana fermions at zero energy. However, in the trivial regime energy bound states do not change their sign, see Fig.~2(b), and they are $2\pi$ periodic, which indicates the absence of MBS. In Fig.~2(c), we plot Josephson current versus phase difference ($\varphi$) for both topological and trivial regimes. We notice that in the topological regime, Josephson current is also $4\pi$ periodic and becomes maximum at $\varphi=\pi$ when MBS occur. However, Josephson current is $2\pi$ periodic and vanishes at $\varphi=\pm\pi$ in the trivial regime when MBS do not occur.
\begin{figure}[h]
\centering{\includegraphics[width=0.99\textwidth]{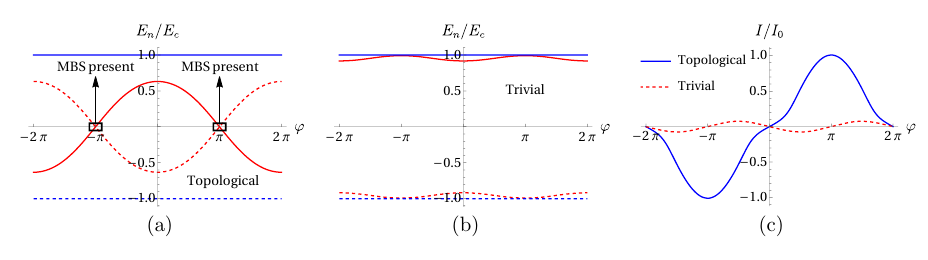}}
\vskip -0.2 in \caption{\small \sl Energy bound states as a function of phase difference $\varphi$ for (a) topological regime, (b) trivial regime. (c) Josephson current as a function of phase difference $\varphi$. Parameters: $\mathcal{S}=1/2$, $J=3$, $\mu_{p_{x}}=1$ (for topological regime), $\mu_{p_{x}}=-1$ (for trivial regime), $\Delta_{p_{x}}=\sqrt{2}$, $E_{c}=1$, $I_{0}=eE_{c}/\hbar$, $T=0$. In (a) and (b) we choose the particular case of $\mathcal{S}=-m'=1/2$.}
\end{figure}
\section{ODD-FREQUENCY PAIRING AND MAJORANA BOUND STATES}
The primary objective of this study is to investigate the potential correlation between the presence of MBS and the emergence of odd-$\nu$ pairing correlations. For this reason, we form retarded Green's function, denoted as $\mathcal{G}^{r}(x,\bar{x},\nu)$, for the setup depicted in Fig.~1, incorporating the interface scattering processes\cite{mcm}. We adopt the methodology established in Refs.~\cite{cay1} and \cite{cay2}, and the details of Green's functions calculation are mentioned in Appendix B. We focus on the anomalous component of $\mathcal{G}^{r}$, which gives the pairing amplitudes. Anomalous component of $\mathcal{G}^{r}(x,\bar{x},\nu)$ is expressed as,
\begin{equation}
\mathcal{G}^{r}_{eh}(x,\bar{x},\nu)=i\sum_{l=0}^{3}f_{l}^{r}\sigma_{l}\sigma_{2}.
\label{green}
\end{equation}
In Eq.~\eqref{green}, $\sigma_{0}$ represents the Identity matrix and, $\sigma_{l}$ ($l=1,2,3$) denote the Pauli matrices. Further, $f_{0}^{r}$ is spin-singlet ($\uparrow\downarrow-\downarrow\uparrow$) pairing amplitude, $f_{1,2}^{r}$ represent equal spin-triplet ($\downarrow\downarrow\pm\uparrow\uparrow$) pairing amplitudes and finally $f_{3}^{r}$ being the mixed spin-triplet ($\uparrow\downarrow+\downarrow\uparrow$) pairing amplitude. The EST pairing amplitudes $\uparrow\uparrow$ and $\downarrow\downarrow$ are calculated as $f_{\uparrow\uparrow}=if_{2}^{r}-f_{1}^{r}$ and $f_{\downarrow\downarrow}=if_{2}^{r}+f_{1}^{r}$, respectively. Even and odd-$\nu$ pairing amplitudes are obtained from,
\begin{equation}
\label{EVENODD}
f^{E}_{l}(x,\bar{x},\nu)=\frac{{f^{r}_{l}(x,\bar{x},\nu)+f^{a}_{l}(x,\bar{x},-\nu)}}{2},\,\,\,\mbox{and}\,\,\,
f^{O}_{l}(x,\bar{x},\nu)=\frac{{f^{r}_{l}(x,\bar{x},\nu)-f^{a}_{l}(x,\bar{x},-\nu)}}{2},
\end{equation}
$f_{l}^{a}$ are related to the advanced Green's function and determined from $\mathcal{G}^{a}(x,\bar{x},\nu)=[\mathcal{G}^{r}(\bar{x},x,\nu)]^{\dagger}$\cite{cay1}. The even and odd-$\nu$ EST pairing amplitudes are calculated as,
\begin{equation}
\label{eo1}
\mbox{Even-$\nu$ EST:}\,\,\,\,f_{\uparrow\uparrow}^{E}=if_{2}^{E}-f_{1}^{E}\,\,\,\,\mbox{and,}\,\,\,\, f_{\downarrow\downarrow}^{E}=if_{2}^{E}+f_{1}^{E};\,\,\,\,\mbox{Odd-$\nu$ EST:}\,\,\,\,
f_{\uparrow\uparrow}^{O}=if_{2}^{O}-f_{1}^{O}\,\,\,\,\mbox{and,}\,\,\,\, f_{\downarrow\downarrow}^{O}=if_{2}^{O}+f_{1}^{O}.
\end{equation}
The even and odd-$\nu$ MST pairing amplitudes are calculated as,
\begin{equation}
\label{eoo1}
\mbox{Even-$\nu$ MST:}\,\,\,\,f^{E}_{3}(x,\bar{x},\nu)=\frac{{f^{r}_{3}(x,\bar{x},\nu)+f^{a}_{3}(x,\bar{x},-\nu)}}{2};\,\,\,\,
\mbox{Odd-$\nu$ MST:}\,\,\,\,f^{O}_{3}(x,\bar{x},\nu)=\frac{{f^{r}_{3}(x,\bar{x},\nu)-f^{a}_{3}(x,\bar{x},-\nu)}}{2}.
\end{equation}
\subsection{Equal spin-triplet (EST) superconducting correlations}
We calculate even- and odd-$\nu$ pairing amplitudes using Eqs.~\eqref{green}-\eqref{eoo1}, see \cite{mathematica} for detailed calculations. Spin-singlet (SS) pairing is zero in the setup considered in Fig.~1, however, spin-triplet (ST) pairing is finite. ST pairing is of two kinds, equal spin-triplet (EST) pairing and mixed spin-triplet (MST) pairing. Since, in our work, $p_{x}$ superconductor has equal spin-triplet Cooper pair, thus EST pairing already exists in
this setup, however, MST pairing is induced due to the spin flipper. Here we show that even-$\nu$ EST and odd-$\nu$ EST pairings are induced in our setup due to the $p_{x}$ superconductor. Similar to Fig.~2, we look at two cases: (a) the trivial regime and, (b) the topological regime.
In the topological regime, MBS appear at $\nu\rightarrow0$ and $\varphi=\pm\pi$ as depicted in Fig.~2(a). To understand the nature of even-$\nu$ EST and odd-$\nu$ EST correlations when MBS appear in the topological regime, we compute even-$\nu$ EST and odd-$\nu$ EST correlations at the $\nu\rightarrow0$ limit in both trivial and topological regimes. However, the explicit expressions for even-$\nu$ EST and odd-$\nu$ EST correlations at finite $\nu$ are provided in Appendix C. In the \textbf{trivial} regime, for $\nu\rightarrow0$, the coherence
factors $\gamma_e=\gamma_h=I$, normal reflection amplitudes, $r_{\uparrow\uparrow}^{e'e'}= r_{\downarrow\downarrow}^{e'e'}= r_{\uparrow\uparrow}^{h'h'}= r_{\downarrow\downarrow}^{h'h'}=0$ and, Andreev reflection amplitudes, $r_{\uparrow\uparrow}^{e'h'}=r_{\uparrow\uparrow}^{h'e'}$ and,  $r_{\downarrow\downarrow}^{e'h'}=r_{\downarrow\downarrow}^{h'e'}$, see Appendix C. We should mention that the Andreev reflection amplitudes $r_{\uparrow\uparrow}^{e'h'}$, $r_{\uparrow\uparrow}^{h'e'}$, $r_{\downarrow\downarrow}^{e'h'}$, $r_{\downarrow\downarrow}^{h'e'}$ are purely real and its because of the perfect Andreev reflection seen at $\nu\rightarrow0$ limit. In $\nu\rightarrow0$ limit, both bulk and surface even-$\nu$ EST pairing are non-zero in the left superconductor and given as:
\begin{align}
\mbox{Bulk even-$\nu$ EST:}\,\,\label{even-equal-pp-zero}
f_{\uparrow\uparrow}^{E,B}(x,\bar{x},\nu\rightarrow0)&=\frac{\eta}{2(q_{h}-q_{e})}\Big(\big(e^{iq_{e}|x-\bar{x}|}-e^{-iq_{h}|x-\bar{x}|}\big)\mbox{sgn}(x-\bar{x})\Big)=f_{\downarrow\downarrow}^{E,B}(x,\bar{x},\nu\rightarrow0),\,\nonumber\\&\mbox{for}\,\, x<0,\\
\mbox{Surface even-$\nu$ EST:}\,\,
\label{even-equal-pp1-zero}
f_{\uparrow\uparrow}^{E,S}(x,\bar{x},\nu\rightarrow0)&=\frac{\eta}{4(q_{h}-q_{e})}\Big(\big(r_{\uparrow\uparrow}^{e'h'}+r_{\downarrow\downarrow}^{e'h'}\big)\big(e^{-i(q_{e}\bar{x}-q_{h}x)}-e^{-i(q_{e}x-q_{h}\bar{x})}\big)\Big)=f_{\downarrow\downarrow}^{E,S}(x,\bar{x},\nu\rightarrow0),\,\nonumber\\&\mbox{for}\,\, x<0,
\end{align}
\normalsize
where $\eta=\frac{2m^{*}}{\hbar^2}$, while bulk and surface components of odd-$\nu$ EST pairing vanishes. In the above equations, we have separated the pairing amplitudes into bulk (B) and surface (S) components, where bulk components do not depend on interface scattering amplitudes. In the \textbf{topological} regime, at $\nu\rightarrow0$, $\gamma_h=\gamma_e^{*}$, $q_e\gamma_e=(q_h\gamma_h)^{*}$, $q_h\gamma_e=(q_e\gamma_h)^{*}$, $r^{e'e'}_{\uparrow\uparrow}=r^{h'h' *}_{\uparrow\uparrow}$, $r^{e'e'}_{\downarrow\downarrow}=r^{h'h' *}_{\downarrow\downarrow}$ and there are two cases: MBS \textbf{absent} ($\varphi\neq\pi$), and MBS \textbf{present} ($\varphi=\pi$). When MBS are \textbf{absent} ($\varphi\neq\pi$), normal reflection amplitudes, $r_{\uparrow\uparrow}^{e'e'}= r_{\downarrow\downarrow}^{e'e'}= r_{\uparrow\uparrow}^{h'h'}= r_{\downarrow\downarrow}^{h'h'}=0$ and, Andreev reflection amplitudes, $r_{\uparrow\uparrow}^{e'h'}= r_{\downarrow\downarrow}^{e'h'}= r_{\uparrow\uparrow}^{h'e'}= r_{\downarrow\downarrow}^{h'e'}$ (all are imaginary). Again there is a perfect Andreev reflection at $\nu\rightarrow0$ limit. We find that similar to the trivial regime, in the \textbf{topological} regime when MBS are \textbf{absent}, both bulk and surface even-$\nu$ EST pairings are non-zero in the left superconductor
and given as:
\begin{align}
\label{even-equal-topo-pp-zero}
\mbox{Bulk even-$\nu$ EST:}\,\,f_{\uparrow\uparrow}^{E,B}(x,\bar{x},\nu\rightarrow0)&=
\frac{\eta}{8}\Bigg[\frac{1}{\text{Im}[q_e\gamma_e]}+\frac{|\gamma_e|^2}{\text{Im}[q_h\gamma_e]}\Bigg]\Bigg[\Big(e^{-iq_{h}|x-\bar{x}|}-e^{iq_{e}|x-\bar{x}|}\Big)\mbox{sgn}(x-\bar{x})\Bigg]\nonumber\\&=f_{\downarrow\downarrow}^{E,B}(x,\bar{x},\nu\rightarrow0),\,
\mbox{for}\,\, x<0,\\
\label{even-equal-topo-pp1-zero}
\mbox{Surface even-$\nu$ EST:}\,\,f_{\uparrow\uparrow}^{E,S}(x,\bar{x},\nu\rightarrow0)&=\frac{\eta}{8}r_{\uparrow\uparrow}^{e'h'}\Bigg(\Bigg[\frac{1}{\text{Im}[q_e\gamma_e]}-\frac{(\gamma_e^{*})^2}{\text{Im}[q_h\gamma_e]}\Bigg]e^{-i(q_{e}\bar{x}-q_{h}x)}-\Bigg[\frac{1}{\text{Im}[q_e\gamma_e]}-\frac{\gamma_e^2}{\text{Im}[q_h\gamma_e]}\Bigg]\nonumber\\&\times e^{-i(q_{e}x-q_{h}\bar{x})}\Bigg)=f_{\downarrow\downarrow}^{E,S}(x,\bar{x},\nu\rightarrow0),\,\,
\mbox{for}\,\, x<0,
\end{align}
while at $\nu\rightarrow0$ limit odd-$\nu$ EST pairing vanishes when MBS are absent. In the \textbf{topological} regime, when MBS are \textbf{present} ($\varphi=\pi$), normal reflection amplitudes satisfy $r^{e'e'}_{\uparrow\uparrow}=r^{h'h' *}_{\uparrow\uparrow}$, $r^{e'e'}_{\downarrow\downarrow}=r^{h'h' *}_{\downarrow\downarrow}$ and, Andreev reflection amplitudes satisfy $r^{e'h'}_{\uparrow\uparrow}=r^{h'e' *}_{\uparrow\uparrow}$, $r^{e'h'}_{\downarrow\downarrow}=r^{h'e' *}_{\downarrow\downarrow}$ with $\text{Im}[r^{e'h'}_{\uparrow\uparrow}]\gg1$, $\text{Im}[r^{e'h'}_{\downarrow\downarrow}]\gg1$. We find that similar to the trivial regime and topological regime when MBS are absent, in the topological regime when MBS are \textbf{present}, both bulk and surface even-$\nu$ EST pairing are non-zero in the left superconductor and given as:
\begin{align}
\label{even-equal-topo-pp-zero-mbs}
\mbox{Bulk even-$\nu$ EST:}\,\,f_{\uparrow\uparrow}^{E,B}(x,\bar{x},\nu\rightarrow0)&=
\frac{\eta}{8}\Bigg[\frac{1}{\text{Im}[q_e\gamma_e]}+\frac{|\gamma_e|^2}{\text{Im}[q_h\gamma_e]}\Bigg]\Bigg[\Big(e^{-iq_{h}|x-\bar{x}|}-e^{iq_{e}|x-\bar{x}|}\Big)\mbox{sgn}(x-\bar{x})\Bigg]\nonumber\\&=f_{\downarrow\downarrow}^{E,B}(x,\bar{x},\nu\rightarrow0),\,
\mbox{for}\,x<0,\\
\label{even-equal-topo-pp1-zero-mbs}
\mbox{Surface even-$\nu$ EST:}\,\,f_{\uparrow\uparrow}^{E,S}(x,\bar{x},\nu\rightarrow0)&=
\frac{\eta}{16}\Bigg(\big(r_{\uparrow\uparrow}^{h'e'}+r_{\downarrow\downarrow}^{h'e'}\big)\Bigg[\frac{1}{\text{Im}[q_e\gamma_e]}+\frac{\gamma_e^2}{\text{Im}[q_h\gamma_e]}\Bigg]e^{-i(q_{e}x-q_{h}\bar{x})}-\big(r_{\uparrow\uparrow}^{e'h'}+r_{\downarrow\downarrow}^{e'h'}\big)\nonumber\\&\times\Bigg[\frac{1}{\text{Im}[q_e\gamma_e]}+\frac{(\gamma_e^{*})^2}{\text{Im}[q_h\gamma_e]}\Bigg]e^{i(q_{h}x-q_{e}\bar{x})}\Bigg)=f_{\downarrow\downarrow}^{E,S}(x,\bar{x},\nu\rightarrow0),\,\, \mbox{for}\,\, x<0.
\end{align}
However, in contrast to the scenarios when MBS are absent, in the topological regime when MBS are \textbf{present}, surface odd-$\nu$ EST correlations are finite with vanishing bulk odd-$\nu$ EST correlations.
\begin{align}
\mbox{Surface odd-$\nu$ EST:}\,\,f_{\uparrow\uparrow}^{O,S}(x,\bar{x},\nu\rightarrow0)&=
\frac{\eta}{16}\Bigg(\big(r_{\uparrow\uparrow}^{e'e'}+r_{\downarrow\downarrow}^{e'e'}\big)\Bigg[\frac{1}{\text{Im}[q_e\gamma_e]}+\frac{|\gamma_e|^2}{\text{Im}[q_h\gamma_e]}\Bigg] e^{-iq_{e}(x+\bar{x})}-\big(r_{\uparrow\uparrow}^{h'h'}+r_{\downarrow\downarrow}^{h'h'}\big)\nonumber\\&\times\Bigg[\frac{1}{\text{Im}[q_e\gamma_e]}+\frac{|\gamma_e|^2}{\text{Im}[q_h\gamma_e]}\Bigg]e^{iq_{h}(x+\bar{x})}-\big(r_{\uparrow\uparrow}^{h'e'}+r_{\downarrow\downarrow}^{h'e'}\big)\Bigg[\frac{1}{\text{Im}[q_e\gamma_e]}-\frac{\gamma_e^2}{\text{Im}[q_h\gamma_e]}\Bigg]\nonumber\\&\times e^{-i(q_{e}x-q_{h}\bar{x})}+\big(r_{\uparrow\uparrow}^{e'h'}+r_{\downarrow\downarrow}^{e'h'}\big)\Bigg[\frac{1}{\text{Im}[q_e\gamma_e]}-\frac{(\gamma_e^{*})^2}{\text{Im}[q_h\gamma_e]}\Bigg]e^{i(q_{h}x-q_{e}\bar{x})}\Bigg)\nonumber\\&=f_{\downarrow\downarrow}^{O,S}(x,\bar{x},\nu\rightarrow0),\,\,
\mbox{for}\,\, x<0.
\label{odd-equal-topo-pp-zero-mbs}
\end{align}
This finite surface odd-$\nu$ EST correlation can act as a signature of MBS. Both bulk and surface contribution of odd-$\nu$ EST correlations vanish in the trivial regime and in the topological regime when MBS are absent at $\nu\rightarrow0$ limit, while only bulk contribution of odd-$\nu$ EST correlations vanish in the topological regime when MBS are present. Bulk and surface contributions of even-$\nu$ EST and odd-$\nu$ EST correlations at $\nu\rightarrow0$ within the left superconductor are presented in Fig.~3. We choose three cases: (a) trivial regime, (b) topological regime when MBS are absent, and (c) topological regime when MBS are present. From Fig.~3, it is evident that in the trivial regime, both bulk and surface contributions of even-$\nu$ EST correlations are finite and exhibit a decay without any oscillation with vanishing odd-$\nu$ EST correlations. However, in the topological regime when MBS are absent, both bulk and surface contributions of even-$\nu$ EST correlations are nonzero and exhibit an oscillatory decay with vanishing odd-$\nu$ EST correlations. In the topological regime, when MBS are present, surface-induced odd-$\nu$ EST correlations are enhanced with vanishing bulk EST correlations.
\begin{figure}[ht]
\centering{\includegraphics[width=0.99\textwidth]{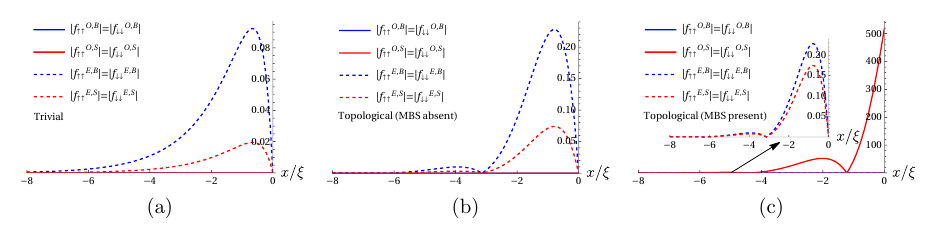}}
\vskip -0.2 in \caption{\small \sl The absolute values of the bulk and surface contributions to even-$\nu$ EST and odd-$\nu$ EST correlations within the left superconductor vs. position $x$ for (a) trivial regime, (b) topological regime when MBS are absent ($\varphi=\pi/5$) and, (c) topological regime when MBS are present ($\varphi=\pi$). Parameters: $\mathcal{S}=1/2$, $J=3$, $\Delta_{p_{x}}=\sqrt{2}$, $\mu_{p_{x}}=-1$ (for (a)), $\mu_{p_{x}}=1$ (for (b) and (c)), $\bar{x}=0$, $\varphi=\pi/5$ (for (a) and (b)), $\varphi=\pi$ (for (c)), $\nu\rightarrow0$. In Figs.~3(a-c), $x$ has been scaled by the superconducting coherence length $\xi$ for dimensionless representation.}
\end{figure}
However, bulk and surface contributions of even-$\nu$ EST correlations are suppressed. In the trivial regime, at $\nu\rightarrow0$, both $q_{e}$ and $q_{h}$ are purely imaginary in Eqs.~\eqref{even-equal-pp-zero}, \eqref{even-equal-pp1-zero}, therefore even-$\nu$ EST
and odd-$\nu$ EST correlations show a decay without any oscillation.
In the topological regime, at $\nu\rightarrow0$, both $q_{e}$ and $q_{h}$ have both real and imaginary components ($q_{h}=q_{e}^{*}$) in Eqs.~\eqref{even-equal-topo-pp-zero}-\eqref{odd-equal-topo-pp-zero-mbs}, therefore even-$\nu$ EST and odd-$\nu$ EST correlations exhibit a decay with oscillation inside the superconductor. Even-$\nu$ EST correlations are finite both in the presence as well as in the absence of MBS and, thus, do not help in detecting MBS. However, surface odd-$\nu$ EST correlations are finite in the presence of MBS, making them a crucial indicator in detecting MBS.
\subsection{Mixed spin-triplet (MST) superconducting correlations}
We now compute the induced even-$\nu$ MST and odd-$\nu$ MST correlations in our setup. In our work as mentioned before MST correlations are induced in the presence of spin flipper. Similar to Fig.~2, we examine two scenarios: (a) the trivial regime, and (b) the topological regime. To understand the behavior of even-$\nu$ MST and odd-$\nu$ MST correlations when MBS occur, we consider the $\nu\rightarrow0$ limit for both trivial and topological regimes. However, the analytical expressions for
even-$\nu$ MST and odd-$\nu$ MST correlations at finite $\nu$ are mentioned in Appendix C. In the \textbf{trivial} regime, at $\nu\rightarrow0$, the coherence factors $\gamma_e=\gamma_h=I$, normal reflection amplitudes, $r_{\uparrow\downarrow}^{e'e'}= r_{\downarrow\uparrow}^{e'e'}= r_{\uparrow\downarrow}^{h'h'}= r_{\downarrow\uparrow}^{h'h'}=0$, and, Andreev reflection amplitudes, $r_{\uparrow\downarrow}^{e'h'}=r_{\uparrow\downarrow}^{h'e'}$ and,  $r_{\downarrow\uparrow}^{e'h'}=r_{\downarrow\uparrow}^{h'e'}$. The Andreev reflection amplitudes are purely real. We find that bulk MST correlations vanish for both even-$\nu$ and odd-$\nu$, and surface MST correlations are non-zero for even-$\nu$ while for odd-$\nu$, they too disappear.
\begin{align}
\label{even-mixed-pp1-zero}
\mbox{Surface even-$\nu$ MST:}\,\,f_{3}^{E,S}(x,\bar{x},\nu\rightarrow0)&=\frac{\eta}{4(q_{h}-q_{e})}\Big((r_{\uparrow\downarrow}^{e'h'}+r_{\downarrow\uparrow}^{e'h'})\big(e^{-i(q_{e}\bar{x}-q_{h}x)}-e^{-i(q_{e}x-q_{h}\bar{x})}\big)\Big),\mbox{for}\,x<0.
\end{align}
In the \textbf{topological} regime, at $\nu\rightarrow0$, $\gamma_h=\gamma_e^{*}$, $q_e\gamma_e=(q_h\gamma_h)^{*}$, $q_h\gamma_e=(q_e\gamma_h)^{*}$, $r^{e'e'}_{\uparrow\downarrow}=r^{h'h' *}_{\uparrow\downarrow}$, $r^{e'e'}_{\downarrow\uparrow}=r^{h'h' *}_{\downarrow\uparrow}$ and when MBS are \textbf{absent} ($\varphi\neq\pi$), normal and Andreev reflection amplitudes $r_{\uparrow\downarrow}^{e'e'}= r_{\downarrow\uparrow}^{e'e'}= r_{\uparrow\downarrow}^{h'h'}= r_{\downarrow\uparrow}^{h'h'}= r_{\uparrow\downarrow}^{e'h'}= r_{\downarrow\uparrow}^{e'h'}= r_{\uparrow\downarrow}^{h'e'}= r_{\downarrow\uparrow}^{h'e'}=0$. We find that all correlations, both bulk and surface, for either odd-$\nu$ or even-$\nu$ vanish. Finally, in the \textbf{topological} regime, at $\nu\rightarrow0$, when MBS are \textbf{present} ($\varphi=\pi$), normal reflection amplitudes satisfy $r^{e'e'}_{\uparrow\downarrow}=r^{h'h' *}_{\uparrow\downarrow}$, $r^{e'e'}_{\downarrow\uparrow}=r^{h'h' *}_{\downarrow\uparrow}$ and, Andreev reflection amplitudes satisfy $r^{e'h'}_{\uparrow\downarrow}=r^{h'e' *}_{\uparrow\downarrow}$, $r^{e'h'}_{\downarrow\uparrow}=r^{h'e' *}_{\downarrow\uparrow}$ with $\text{Im}[r^{e'h'}_{\uparrow\downarrow}]\gg1$, $\text{Im}[r^{e'h'}_{\downarrow\uparrow}]\gg1$. We find that surface MST correlations are finite for both even-$\nu$ and odd-$\nu$ while bulk MST correlations vanish for both even-$\nu$ and odd-$\nu$.
\begin{align}
\label{even-mixed-topo-pp1-zero-mbs}
\mbox{Surface even-$\nu$ MST:}\,\,f_{3}^{E,S}(x,\bar{x},\nu\rightarrow0)&=
\frac{\eta}{16}\Bigg(\big(r_{\uparrow\downarrow}^{h'e'}+r_{\downarrow\uparrow}^{h'e'}\big)\Bigg[\frac{1}{\text{Im}[q_e\gamma_e]}+\frac{\gamma_e^2}{\text{Im}[q_h\gamma_e]}\Bigg]e^{-i(q_{e}x-q_{h}\bar{x})}-\big(r_{\uparrow\downarrow}^{e'h'}+r_{\downarrow\uparrow}^{e'h'}\big)\nonumber\\&\times\Bigg[\frac{1}{\text{Im}[q_e\gamma_e]}+\frac{(\gamma_e^{*})^2}{\text{Im}[q_h\gamma_e]}\Bigg]e^{i(q_{h}x-q_{e}\bar{x})}\Bigg),\,\, \mbox{for}\,\, x<0,\\
\mbox{Surface odd-$\nu$ MST:}\,\,f_{3}^{O,S}(x,\bar{x},\nu\rightarrow0)&=
\frac{\eta}{16}\Bigg(\big(r_{\uparrow\downarrow}^{e'e'}+r_{\downarrow\uparrow}^{e'e'}\big)\Bigg[\frac{1}{\text{Im}[q_e\gamma_e]}+\frac{|\gamma_e|^2}{\text{Im}[q_h\gamma_e]}\Bigg] e^{-iq_{e}(x+\bar{x})}-\big(r_{\uparrow\downarrow}^{h'h'}+r_{\downarrow\uparrow}^{h'h'}\big)\nonumber\\&\times\Bigg[\frac{1}{\text{Im}[q_e\gamma_e]}+\frac{|\gamma_e|^2}{\text{Im}[q_h\gamma_e]}\Bigg]e^{iq_{h}(x+\bar{x})}-\big(r_{\uparrow\downarrow}^{h'e'}+r_{\downarrow\uparrow}^{h'e'}\big)\Bigg[\frac{1}{\text{Im}[q_e\gamma_e]}-\frac{\gamma_e^2}{\text{Im}[q_h\gamma_e]}\Bigg]\nonumber\\&\times e^{-i(q_{e}x-q_{h}\bar{x})}+\big(r_{\uparrow\downarrow}^{e'h'}+r_{\downarrow\uparrow}^{e'h'}\big)\Bigg[\frac{1}{\text{Im}[q_e\gamma_e]}-\frac{(\gamma_e^{*})^2}{\text{Im}[q_h\gamma_e]}\Bigg]e^{i(q_{h}x-q_{e}\bar{x})}\Bigg),\,\,\mbox{for}\, x<0.
\label{odd-mixed-topo-pp-zero-mbs}
\end{align}
Surface and bulk components of even-$\nu$ MST and odd-$\nu$ MST correlations, at $\nu\rightarrow0$ in the superconducting region, are plotted as a function of position $x$ in Fig.~4. Similar to Fig.~3, we consider three cases: (a) trivial regime, (b) topological regime when MBS are absent, and (c) topological regime when MBS are present. From Fig.~4(a), we see that, in the trivial regime, surface even-$\nu$ MST correlations are finite and show a decay without any oscillation, but odd-$\nu$ MST correlations vanish. In Fig.~4(b), in the topological regimes when MBS are absent ($\varphi=\pi/5$), both even-$\nu$ MST and odd-$\nu$ MST correlations are zero.
\begin{figure}[ht]
\centering{\includegraphics[width=0.99\textwidth]{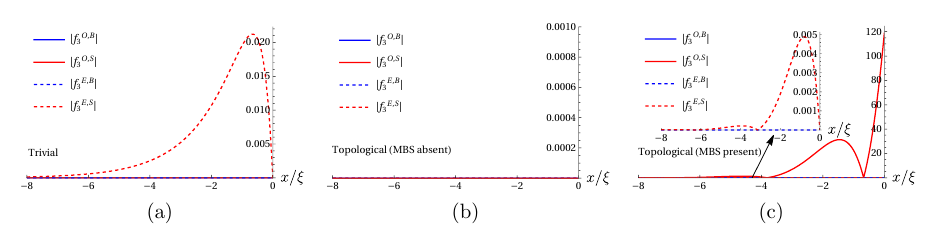}}
\vskip -0.2 in \caption{\small \sl The absolute values of the bulk and surface contributions to even-$\nu$ MST and odd-$\nu$ MST correlations within the left superconductor vs. position $x$ for (a) trivial regime, (b) topological regime when MBS are absent ($\varphi=\pi/5$) and, (c) topological regime when MBS are present ($\varphi=\pi$). Parameters: $\mathcal{S}=1/2$, $J=3$, $\Delta_{p_{x}}=\sqrt{2}$, $\mu_{p_{x}}=-1$ (for (a)), $\mu_{p_{x}}=1$ (for (b) and (c)), $\bar{x}=0$, $\varphi=\pi/5$ (for (a) and (b)), $\varphi=\pi$ (for (c)), $\nu\rightarrow0$. In Figs.~4(a-c), $x$ has been scaled by the superconducting coherence length $\xi$ for dimensionless representation.}
\end{figure}
Finally, in Fig.~4(c), we see that when MBS are present, surface odd-$\nu$ MST correlations are finite with very large magnitudes while surface even-$\nu$ MST correlations are very small. Further, surface even-$\nu$ MST correlations are finite in the trivial regime when MBS are absent, as well as in the topological regime when MBS are present and, thus, do not distinguish MBS. Surface odd-$\nu$ MST correlations are finite in the topological regime when MBS are present but vanish in the absence of MBS. \textbf{Therefore, induced surface odd-$\nu$ MST correlations imply presence of MBS. This is the main result of our work.}
{\section{Effect of disorder on Majorana induced odd-frequency correlations}
In this section we discuss the effects of disorder on our results. To study the impact of disorder on our results, we consider a finite length JJ where two normal metals with a spin flipper (SF) are sandwiched between two nodal $p_{x}$ superconductors, as shown in Fig.~5. Normal metal-$p_{x}$ superconductor interfaces are modeled with $\delta$-like potential barriers of strength $V_{1}$ and $V_{2}$. The scattering of an incident electron-like quasiparticle with spin-up is presented in Fig.~5. In this double-barrier 1D system, multiple scattering of electrons and holes occurs due to the interface barriers. In Ref.~\cite{mel}, the authors demonstrate that conductance
is enhanced in a 1D Normal metal-Insulator-Normal metal-Insulator-Superconductor (N-I$_1$-N-I$_2$-S) double-barrier junction due to multiple scattering\cite{duh}. This phenomenon is similar to what happens in a disordered Normal metal-Insulator-Superconductor (N-I-S) junction, where conductance can be
enhanced due to repeated scattering of electrons by disorder, as shown in Ref.~\cite{hek}. This enhancement in conductance suggests
that multiple scattering in a double-barrier junction introduces disorder effects.}

{The BdG Hamiltonian for nodal $p$-wave superconductor ($p_{x}$)-Insulator ($I_1$)-Normal metal ($N_{1}$)-Spin flipper (SF)-Normal metal ($N_{2}$)-Insulator ($I_2$)-nodal $p$-wave superconductor ($p_{x}$) junction, as shown in Fig.~5, is expressed as follows,
\begin{equation}
H_{BdG}^{\mbox{$p_{x}$-$I_1$-$N_1$-SF-$N_2$-$I_2$-$p_{x}$}}(x)=
\begin{pmatrix}
H_{Q}\hat{I} & \Delta_{L}p  \hat{\sigma}_{x} \\
\Delta_{L}^{*}p \hat{\sigma}_{x} & -H_{Q}\hat{I}
\end{pmatrix},
\label{hamp}
\end{equation}
with $H_{Q}=-\frac{\hbar^2}{2m^{*}}\frac{\partial^2}{\partial x^2}+V_{1}\delta(x+L/2)+V_{2}\delta(x-L/2)-J_{0}\delta(x)\vec s\cdot\vec{\mathcal{S}}-\mu_{r}'$.
\begin{figure}[ht]
\centering{\includegraphics[width=0.78\textwidth]{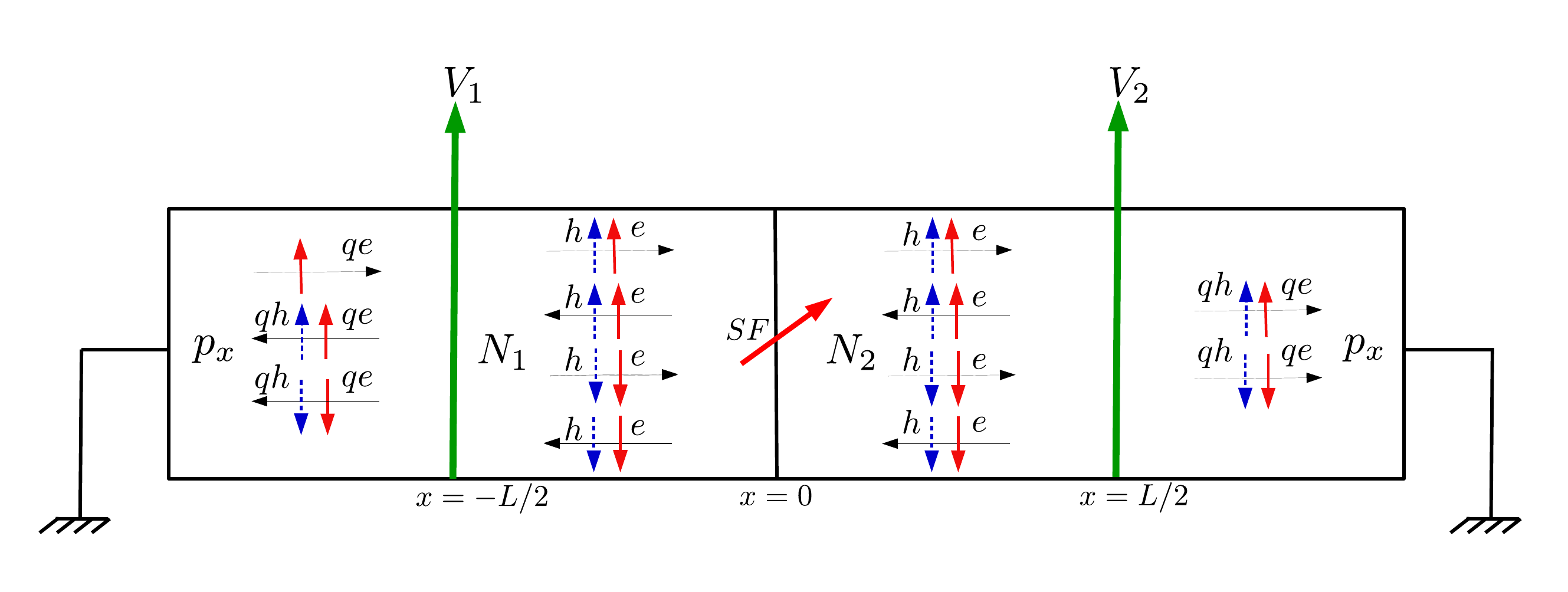}}
\vskip -0.2 in \caption{\small \sl {Josephson junction composed of two normal metals and a SF at $x=0$ embedded between two nodal $p_{x}$ superconductors. Normal metal-$p_x$ superconductor interfaces at $x=\pm L/2$ are modeled with $\delta$-like potential barriers of strength $V_1$ and $V_2$. Two $p_{x}$ superconductors are grounded. The scattering process of an incident spin-up electron-like quasiparticle (qe) is depicted and SF's spin being oriented along any arbitrary direction.}}
\end{figure}
$V_1$ represents the strength of $\delta$-like potential barrier at $p_x$-$N_1$ interface, while $V_2$ denotes the strength of $\delta$-like potential barrier at $N_2$-$p_x$ interface. We have used dimensionless parameters $Z_1=\frac{2m^{*}V_{1}}{\hbar^2 k_{\mu_{p_x}}}$ and $Z_2=\frac{2m^{*}V_{2}}{\hbar^2 k_{\mu_{p_x}}}$ to quantify interface transparencies\cite{BTK}. $\mu_{r}'$ corresponds to the chemical potential for normal metal ($r=N$) or $p_x$ superconductor ($r=p_x$) and $\mu_{r}=\frac{\hbar^2\mu_{r}^{\prime}}{m^{*}\Delta_{p_x}^2}$ is the normalized chemical potential. Further, the gap parameter $\Delta_L$ has the following form $\Delta_L=\Delta_{p_x}'[e^{i\varphi_{L}}\theta(-x-L/2)+e^{i\varphi_{R}}\theta(x-L/2)]$.}

{By diagonalizing the Hamiltonian \eqref{hamp}, we obtain the wavefunctions in distinct regions of the $p_x$-$I_1$-$N_1$-SF-$N_2$-$I_2$-$p_x$ JJ corresponding to different scattering processes. The detailed wavefunctions are mentioned in Appendix A. In section IV we discuss in detail the method to calculate the retarded Green's function and induced even-/odd-$\nu$ ST pairing amplitudes in different regions of the $p_x$-$I_1$-$N_1$-SF-$N_2$-$I_2$-$p_x$ JJ. In Fig.~6, we present the surface contributions of even-$\nu$ EST and odd-$\nu$ EST correlations in the left superconductor as a function of position $x$ for different values of $Z_1$ and $Z_2$ in the $\nu\rightarrow0$ limit. The three phases: (a) the trivial phase, (b) the topological phase without MBS, and (c) the topological phase with MBS are then analyzed in Fig.~6. In the trivial phase, the magnitude of surface even-$\nu$ EST correlations
changes with disorder as seen for different values of $Z_1$ and $Z_2$.
\begin{figure}[ht]
\centering{\includegraphics[width=0.99\textwidth]{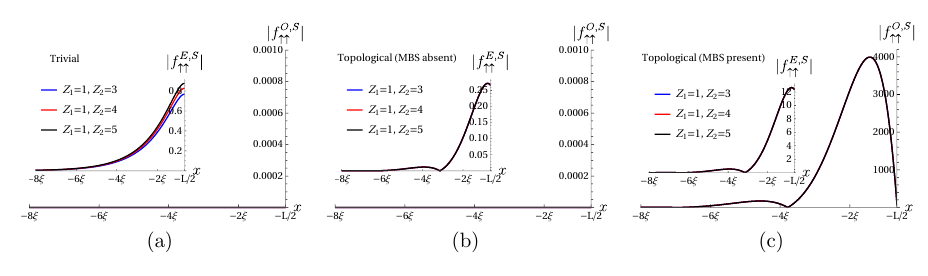}}
\vskip -0.2 in \caption{\small \sl {The absolute values of the surface contribution to even-$\nu$ EST and odd-$\nu$ EST correlations within the left superconductor vs. position $x$ for different values of interface transparencies $Z_1$ and $Z_2$. We consider three cases: (a) trivial regime, (b) topological regime when MBS are absent ($\varphi=\pi/5$), and, (c) topological regime when MBS are present ($\varphi=\pi$). Parameters: $\mathcal{S}=1/2$, $J=3$, $\Delta_{p_{x}}=\sqrt{2}$, $\mu_{p_{x}}=-1$ (for (a)), $\mu_{p_{x}}=1$ (for (b) and (c)), $\mu_{N}=10$, $L=1.3\xi$, $\bar{x}=0$, $\varphi=\pi/5$ (for (a) and (b)), $\varphi=\pi$ (for (c)), $\nu\rightarrow0$.}}
\end{figure}
In contrast, in the topological phase when MBS are absent, their magnitude remains independent of disorder. The magnitude of surface odd-$\nu$ EST correlations is zero regardless of disorder in the absence of MBS. In the topological phase when MBS are present, both surface even-$\nu$ EST and surface odd-$\nu$ EST correlations are finite and their magnitudes are very large as well as independent of disorder as seen for different $Z_1$ and $Z_2$ values.
Surface even-$\nu$ EST correlations are finite for various values of $Z_1$ and $Z_2$ in both the trivial and topological phases, regardless of the presence of MBS, but in topological phase their magnitudes are independent of disorder. Surface odd-$\nu$ EST correlations are non-zero only in the topological phase when MBS are present and vanish in their absence. Therefore, surface-induced odd-$\nu$ EST correlations serve as an effective detector of MBS, even in the presence of disorder.}

{
\begin{figure}[ht]
\centering{\includegraphics[width=0.99\textwidth]{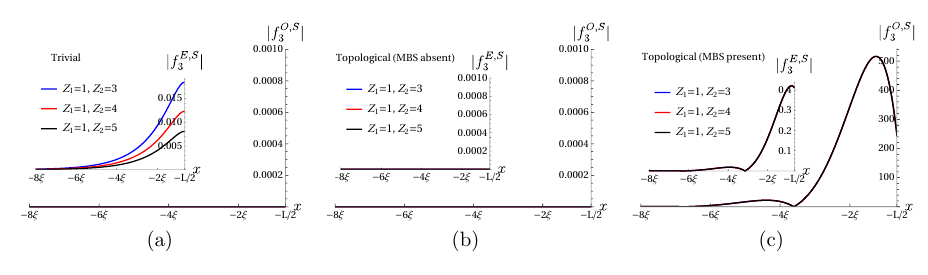}}
\vskip -0.2 in \caption{\small \sl {The absolute values of the surface contribution to even-$\nu$ MST and odd-$\nu$ MST correlations within the left superconductor vs. position $x$ for different values of interface transparencies $Z_1$ and $Z_2$. We consider three cases: (a) trivial regime, (b) topological regime when MBS are absent ($\varphi=\pi/5$), and, (c) topological regime when MBS are present ($\varphi=\pi$). Parameters: $\mathcal{S}=1/2$, $J=3$, $\Delta_{p_{x}}=\sqrt{2}$, $\mu_{p_{x}}=-1$ (for (a)), $\mu_{p_{x}}=1$ (for (b) and (c)), $\mu_{N}=10$, $L=1.3\xi$, $\bar{x}=0$, $\varphi=\pi/5$ (for (a) and (b)), $\varphi=\pi$ (for (c)), $\nu\rightarrow0$.}}
\end{figure}
Next, in Fig.~7, we plot the surface even-$\nu$ MST and surface odd-$\nu$ MST correlations in the left superconductor as a function of position $x$ for
different values of $Z_1$ and $Z_2$ in the $\nu\rightarrow0$ limit.
Similar to Fig.~6, we consider three phases: (a) the trivial phase, (b)
the topological phase without MBS, and (c) the topological phase with MBS. In
the trivial phase, the magnitude of surface even-$\nu$ MST correlations changes with disorder as seen for different values of $Z_1$ and $Z_2$, while the magnitude of surface odd-$\nu$ MST correlations is zero
regardless of disorder. In the topological phase when MBS are absent, both surface even-$\nu$ MST and surface odd-$\nu$ MST correlations are zero regardless of disorder. Finally, in the topological phase
when MBS are present, both surface even-$\nu$ MST and surface odd-$\nu$ MST correlations are finite and their magnitudes are huge as well as independent of disorder as seen for different $Z_1$ and $Z_2$ values. Surface
even-$\nu$ MST correlations are finite for different values of $Z_1$ and $Z_2$ in both the trivial and topological
phases, irrespective of the presence of MBS, but in topological phase they are independent of disorder, and their magnitudes are enhanced multifold in topological phase with MBS. Surface odd-$\nu$ MST correlations are non-zero only in the topological phase when MBS are present and they vanish when MBS are absent. Thus, surface-induced odd-$\nu$ MST correlations can serve as a true detector of MBS even in the presence of disorder. This indicates the robustness of surface induced odd-$\nu$ spin-triplet pairing to the deleterious effects of disorder.
\section{Analysis}}
We compare our results in Table I both when SF is present and when it is absent. Even-$\nu$ SS and odd-$\nu$ SS correlations always vanish in both the trivial and topological regimes irrespective of spin-flip scattering and therefore we do not put them in Table I. We notice that only even-$\nu$ EST correlations are finite in bulk in the presence as well as in the absence of MBS irrespective of spin-flip scattering. Thus, the bulk component of EST correlations does not contribute to the identification of MBS. Surface contributions of even-$\nu$ EST correlations are finite both in trivial and topological regimes in the presence of spin-flip scattering. Surface components of even-$\nu$ MST correlations are finite in the topological regimes when MBS are present and also in the trivial regimes in the presence of spin-flip scattering. Thus, even-$\nu$ EST and even-$\nu$ MST correlations can not help in detecting MBS. Surface contributions of odd-$\nu$ EST correlations are nonzero in the presence of MBS; however, they vanish when MBS are absent regardless of spin-flip scattering. Surface components of odd-$\nu$ MST correlations are present when MBS occur, but they are absent when MBS are absent in the presence of spin-flip scattering. Thus, surface-induced odd-$\nu$ EST and odd-$\nu$ MST correlations can distinguish MBS in the presence of spin-flip scattering. In our work, when SF is absent, surface odd-$\nu$ MST correlations will also be absent. Thus, spin-flip scattering helps in detecting MBS via inducing surface odd-$\nu$ MST correlation in the presence of MBS.
\begin{table}[h]
\caption{Comparing odd- and even-$\nu$ correlations at $\nu\rightarrow0$ in presence and absence of SF}
\begin{tabular}{|p{2cm}|p{2cm}|p{2cm}|p{2.5cm}|p{2cm}|p{2cm}|p{2cm}|p{2cm}|}
\hline
& & \multicolumn{2}{|c|}{Topological (MBS present)} & \multicolumn{2}{|c|}{Topological (MBS absent)} & \multicolumn{2}{|c|}{Trivial}\\
\hline
& & SF present & SF absent & SF present & SF absent & SF present & SF absent\\
\hline
\multirow{2}{*}{Odd-$\nu$ EST}&Bulk & Absent & Absent & Absent & Absent & Absent & Absent
\\\cline{2-8}& {Surface} & \textcolor{red}{Present} & \textcolor{red}{Present} & \textcolor{red}{Absent} & \textcolor{red}{Absent} & \textcolor{red}{Absent} & \textcolor{red}{Absent} \\
\hline
\multirow{2}{*}{Even-$\nu$ EST}&Bulk & Present & Present & Present & Present & Present & Present
\\\cline{2-8}&Surface & Present & Present  & Present & Present & Present & Absent \\
\hline
\multirow{2}{*}{Odd-$\nu$ MST} &Bulk & Absent & Absent & Absent & Absent & Absent & Absent
\\\cline{2-8} &  {Surface} & \textcolor{red}{\textbf{Present}} & \textcolor{red}{Absent} & \textcolor{red}{Absent} & \textcolor{red}{Absent} & \textcolor{red}{Absent} & \textcolor{red}{Absent} \\
\hline
\multirow{2}{*}{Even-$\nu$ MST} &Bulk & Absent & Absent & Absent & Absent & Absent & Absent
\\\cline{2-8}&Surface & Present & Absent & Absent  & Absent & Present & Absent \\
\hline
\end{tabular}
\end{table}

Next, in Table II we explain the reasons behind our results. In the trivial regime, for $\nu\rightarrow0$, and for MBS absent, normal reflection amplitudes without spin-flip are zero, i.e., $r_{\uparrow\uparrow}^{e'e'}= r_{\downarrow\downarrow}^{e'e'}= r_{\uparrow\uparrow}^{h'h'}= r_{\downarrow\downarrow}^{h'h'}=0$ while Andreev reflection amplitudes without spin-flip satisfy $r_{\uparrow\uparrow}^{e'h'}=r_{\uparrow\uparrow}^{h'e'}\neq0$,  $r_{\downarrow\downarrow}^{e'h'}=r_{\downarrow\downarrow}^{h'e'}\neq0$ which lead to finite surface even-$\nu$ EST correlations with vanishing surface odd-$\nu$ EST correlations, see Eq.~\eqref{even-equal-pp1-zero}. Further, in the trivial regime, normal reflection amplitudes with spin-flip are zero, i.e., $r_{\uparrow\downarrow}^{e'e'}= r_{\downarrow\uparrow}^{e'e'}= r_{\uparrow\downarrow}^{h'h'}= r_{\downarrow\uparrow}^{h'h'}=0$ while Andreev reflection amplitudes with spin-flip satisfy $r_{\uparrow\downarrow}^{e'h'}=r_{\uparrow\downarrow}^{h'e'}\neq0$, $r_{\downarrow\uparrow}^{e'h'}=r_{\downarrow\uparrow}^{h'e'}\neq0$ which lead to finite surface even-$\nu$ MST correlations and vanishing surface odd-$\nu$ MST correlations, as evident from Eq.~\eqref{even-mixed-pp1-zero}.

In the topological regime for $\nu\rightarrow0$ and when MBS are absent ($\varphi\neq\pi$), normal reflection amplitudes without spin-flip satisfy $r_{\uparrow\uparrow}^{e'e'}= r_{\downarrow\downarrow}^{e'e'}= r_{\uparrow\uparrow}^{h'h'}= r_{\downarrow\downarrow}^{h'h'}=0$ while Andreev reflection amplitudes without spin-flip are equal, i.e., $r_{\uparrow\uparrow}^{e'h'}= r_{\downarrow\downarrow}^{e'h'}= r_{\uparrow\uparrow}^{h'e'}= r_{\downarrow\downarrow}^{h'e'}\neq0$ which lead to finite surface even-$\nu$ EST correlations and vanishing surface odd-$\nu$ EST correlations, see Eq.~\eqref{even-equal-topo-pp1-zero}. In the topological regime, when MBS are absent, normal reflection amplitudes with spin-flip satisfy $r_{\uparrow\downarrow}^{e'e'}= r_{\downarrow\uparrow}^{e'e'}= r_{\uparrow\downarrow}^{h'h'}= r_{\downarrow\uparrow}^{h'h'}=0$ while spin flip Andreev reflection amplitudes are again vanishing, i.e., $r_{\uparrow\downarrow}^{e'h'}= r_{\downarrow\uparrow}^{e'h'}= r_{\uparrow\downarrow}^{h'e'}= r_{\downarrow\uparrow}^{h'e'}=0$ which lead to vanishing surface even-$\nu$ MST and surface odd-$\nu$ MST correlations.

Finally, in the topological regime for $\nu\rightarrow0$ and when MBS are present ($\varphi=\pi$), normal reflection amplitudes without spin-flip satisfy $r_{\uparrow\uparrow}^{e'e'}=r_{\uparrow\uparrow}^{h'h' *}$, $r_{\downarrow\downarrow}^{e'e'}=r_{\downarrow\downarrow}^{h'h' *}$ while Andreev reflection amplitudes without spin-flip satisfy $r_{\uparrow\uparrow}^{e'h'}=r_{\uparrow\uparrow}^{h'e' *}$, $r_{\downarrow\downarrow}^{e'h'}=r_{\downarrow\downarrow}^{h'e' *}$ which lead to finite surface even-$\nu$ EST and surface odd-$\nu$ EST correlations, see Eqs.~\eqref{even-equal-topo-pp1-zero-mbs}, \eqref{odd-equal-topo-pp-zero-mbs}. In this regime, normal reflection amplitudes with spin-flip satisfy $r_{\uparrow\downarrow}^{e'e'}=r_{\uparrow\downarrow}^{h'h' *}$, $r_{\downarrow\uparrow}^{e'e'}=r_{\downarrow\uparrow}^{h'h' *}$ while Andreev reflection amplitudes with spin-flip satisfy $r_{\uparrow\downarrow}^{e'h'}=r_{\uparrow\downarrow}^{h'e' *}$, $r_{\downarrow\uparrow}^{e'h'}=r_{\downarrow\uparrow}^{h'e' *}$ which lead to finite surface even-$\nu$ MST and finite surface odd-$\nu$ MST correlations, as evident from Eqs.~\eqref{even-mixed-topo-pp1-zero-mbs}, \eqref{odd-mixed-topo-pp-zero-mbs}. It is important to note that both normal and Andreev reflection amplitudes follow distinct conditions in the presence and absence of MBS. These differing conditions lead to distinguishable results under these two scenarios.
\begin{table}[ht]
\caption{Normal and Andreev reflection amplitudes at $\nu\rightarrow0$ in trivial and topological regimes (MBS absent and MBS present}
\begin{tabular}
{|p{1.6cm}|p{4cm}|p{4cm}|p{4cm}|p{4cm}|}
\hline
& Normal reflection amplitudes without spin-flip & Normal reflection amplitudes with spin-flip & Andreev reflection amplitudes without spin-flip & Andreev reflection amplitudes with spin-flip\\
\hline
Trivial & $r_{\uparrow\uparrow}^{e'e'}= r_{\downarrow\downarrow}^{e'e'}= r_{\uparrow\uparrow}^{h'h'}= r_{\downarrow\downarrow}^{h'h'}=0$, i.e., they are vanishing & $r_{\uparrow\downarrow}^{e'e'}= r_{\downarrow\uparrow}^{e'e'}= r_{\uparrow\downarrow}^{h'h'}= r_{\downarrow\uparrow}^{h'h'}=0$, i.e., they are vanishing & $r_{\uparrow\uparrow}^{e'h'}=r_{\uparrow\uparrow}^{h'e'}$,  $r_{\downarrow\downarrow}^{e'h'}=r_{\downarrow\downarrow}^{h'e'}$, they are finite (real) & $r_{\uparrow\downarrow}^{e'h'}=r_{\uparrow\downarrow}^{h'e'}$, $r_{\downarrow\uparrow}^{e'h'}=r_{\downarrow\uparrow}^{h'e'}$, they are finite (real)\\
\hline
Topological (MBS absent) & $r_{\uparrow\uparrow}^{e'e'}= r_{\downarrow\downarrow}^{e'e'}= r_{\uparrow\uparrow}^{h'h'}= r_{\downarrow\downarrow}^{h'h'}=0$, i.e., they are vanishing &  $r_{\uparrow\downarrow}^{e'e'}= r_{\downarrow\uparrow}^{e'e'}= r_{\uparrow\downarrow}^{h'h'}= r_{\downarrow\uparrow}^{h'h'}=0$, i.e., they are vanishing & $r_{\uparrow\uparrow}^{e'h'}= r_{\downarrow\downarrow}^{e'h'}= r_{\uparrow\uparrow}^{h'e'}= r_{\downarrow\downarrow}^{h'e'}$, they are finite (imaginary) & $r_{\uparrow\downarrow}^{e'h'}= r_{\downarrow\uparrow}^{e'h'}= r_{\uparrow\downarrow}^{h'e'}= r_{\downarrow\uparrow}^{h'e'}=0$, i.e., they are vanishing\\
\hline
Topological (MBS present) & $r_{\uparrow\uparrow}^{e'e'}=r_{\uparrow\uparrow}^{h'h' *}$, $r_{\downarrow\downarrow}^{e'e'}=r_{\downarrow\downarrow}^{h'h' *}$, they are finite (complex) & $r_{\uparrow\downarrow}^{e'e'}=r_{\uparrow\downarrow}^{h'h' *}$, $r_{\downarrow\uparrow}^{e'e'}=r_{\downarrow\uparrow}^{h'h' *}$, they are finite (complex) & $r_{\uparrow\uparrow}^{e'h'}=r_{\uparrow\uparrow}^{h'e' *}$, $r_{\downarrow\downarrow}^{e'h'}=r_{\downarrow\downarrow}^{h'e' *}$, they are finite (complex) & $r_{\uparrow\downarrow}^{e'h'}=r_{\uparrow\downarrow}^{h'e' *}$, $r_{\downarrow\uparrow}^{e'h'}=r_{\downarrow\uparrow}^{h'e' *}$, they are finite (complex)\\
\hline
\end{tabular}
\end{table}

{In Table III, we analyze the impact of disorder on surface induced odd-$\nu$ spin-triplet pairing. The effects of disorder have been dealt with in detail in Sec.~V and Figs.~6, 7.
 In the trivial regime, surface odd-$\nu$ EST and surface odd-$\nu$ MST correlations vanish regardless of disorder, whereas surface even-$\nu$ EST and surface even-$\nu$ MST correlations are finite and vary in magnitude with disorder. In the topological regime when MBS are absent, surface odd-$\nu$ EST and surface even-/odd-$\nu$ MST correlations vanish regardless of disorder, while surface even-$\nu$ EST correlations are finite and their magnitude is unaffected by disorder. However, in the topological regime when MBS are present, all four types of correlations (surface odd-$\nu$ EST, surface even-$\nu$ EST, surface odd-$\nu$ MST, and surface even-$\nu$ MST) are finite and their magnitudes are independent of disorder. Therefore, even in the presence of disorder, surface odd-$\nu$ EST and surface odd-$\nu$ MST correlations are reliable indicators for MBS detection, as they are finite only in the presence of MBS and vanish in their absence. This implies that our results are robust against disorder.}
\begin{table}[ht]{
\caption{{Effects of disorder on surface odd-/even-$\nu$ superconducting pairing at $\nu\rightarrow0$ in trivial and topological regimes (MBS absent and MBS present), see Sec. V and Figs. 6. 7 for more details.}}
\begin{tabular}
{|p{3cm}|p{4.8cm}|p{4.8cm}|p{4.8cm}|}
\hline
& Trivial & Topological (MBS absent) & Topological (MBS present)\\
\hline
Surface odd-$\nu$ EST & They are vanishing regardless of disorder & They are vanishing regardless of disorder & They are finite and their magnitude is independent of disorder\\
\hline
Suface even-$\nu$ EST & They are finite and their magnitude changes with disorder & They are finite and their magnitude is independent of disorder & They are finite and their magnitude is independent of disorder\\
\hline
Surface odd-$\nu$ MST & They are vanishing regardless of disorder & They are vanishing regardless of disorder & They are finite and their magnitude is independent of disorder\\
\hline
Surface even-$\nu$ MST & They are finite and their magnitude changes with disorder & They are vanishing regardless of disorder & They are finite and their magnitude is independent of disorder\\
\hline
\end{tabular}}
\end{table}

In Ref.~\cite{taka}, it is found that within the topological phase, the spatial variation of the odd-$\nu$ EST pairing amplitude coincides with that of the local density of states at zero energy in a semi-infinite Kitaev chain. This suggests a direct link between the wave function of Majorana fermions and the odd-$\nu$ EST pairing amplitude at low frequencies. Further, in Ref.~\cite{Lee}, the authors propose that the interaction between Majorana zero modes and a spin-polarized nanowire leads to the emergence of odd-$\nu$ EST pairing in the nanowire. In addition, in Ref.~\cite{Kuzm}, the authors examine the stability of odd-$\nu$ EST pairing induced in a nanowire coupled with Majorana zero modes, particularly when the coupling between them exhibits complexity. Nevertheless, in Refs.~\cite{taka}-\cite{Kuzm}, odd-$\nu$ MST pairing does not emerge, and they do not offer any methodology for MBS detection through odd-$\nu$ pairing. This distinction sets our work apart from theirs.

\section{EXPERIMENTAL DETECTION \& CONCLUSION}
Our system as depicted in Fig.~1 can be implemented in a laboratory setting. $p_{x}$-wave pairing is experimentally found in the quasi 1D organic superconductors (TMTSF)$_{2}$PF$_{6}$\cite{leb,lee}. Replacing the spin flipper at the $p_{x}$-$p_{x}$ superconductor interface must be technically feasible. From an experimental perspective, magnetic molecules such as the Mn$_{4}$O$_{3}$ complex, featuring a spin quantum number of $S=9/2$\cite{wern}, offer a partial analog for the spin flipper. It is worth noting that the interior dynamics of such a high-spin molecule may exhibit notable distinctions from those of our spin flipper. Nonetheless, the spin flipper can effectively emulate half-integer spin states, along with capturing the related spin magnetic moment of the molecule. This enables a substantial approximation of electron interactions with such entities.

{Experimentally, the signature of surface odd-$\nu$ ST correlations can be probed by
observing the zero-frequency peak in the LDOS and LMDOS. In the topological regime, energy-bound states exhibit a crossing at $\nu\rightarrow0$ limit and phase difference $\varphi=\pm\pi$, indicative of the presence of Majorana zero modes within the junction. These energy-bound states naturally manifest as a zero-frequency peak in the LDOS and LMDOS. This is depicted in Figs.~8 and 9, where we plot LDOS and LMDOS as a function of frequency for the $p_x$-$I_1$-$N_1$-SF-$N_2$-$I_2$-$p_x$ JJ (see Fig.~5) at the $p_x$-$N_1$ interface, considering different values of $Z_1$ and $Z_2$. The formula for calculating LDOS and LMDOS is mentioned in Appendix D. We choose three phases: (a) the trivial phase, (b) the topological phase without MBS, and (c) the topological phase
with MBS. We notice a zero-frequency ($\nu=0$) peak in both LDOS and LMDOS in the topological phase when MBS are present. However, no such peak is present in the topological phase without MBS or in the trivial phase.
\begin{figure}[ht]
\centering{\includegraphics[width=0.99\textwidth]{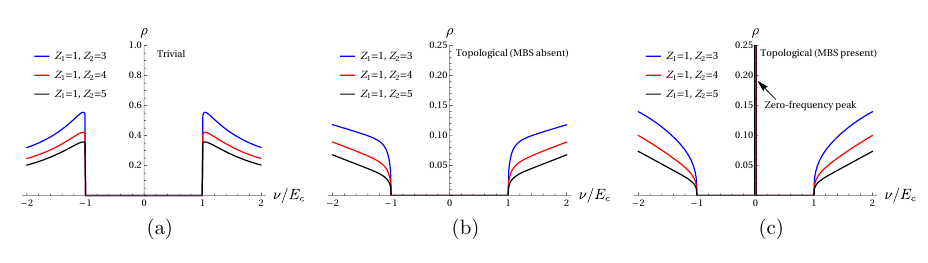}}
\vskip -0.2 in \caption{\small \sl
{Frequency dependence of LDOS at $p_x$-$N_1$ interface for different values of interface transparencies $Z_1$ and $Z_2$. We consider three cases: (a) trivial regime, (b) topological regime when MBS are absent ($\varphi=\pi/5$), and, (c) topological regime when MBS are present ($\varphi=\pi$). Parameters: $\mathcal{S}=1/2$, $J=3$, $\Delta_{p_{x}}=\sqrt{2}$, $\mu_{p_{x}}=-1$ (for (a)), $\mu_{p_{x}}=1$ (for (b) and (c)), $\mu_{N}=10$, $E_c=1$, $L=1.3\xi$, $x=\bar{x}=-L/2$, $\varphi=\pi/5$ (for (a) and (b)), $\varphi=\pi$ (for (c)).}}
\end{figure}
\begin{figure}[ht]
\centering{\includegraphics[width=0.99\textwidth]{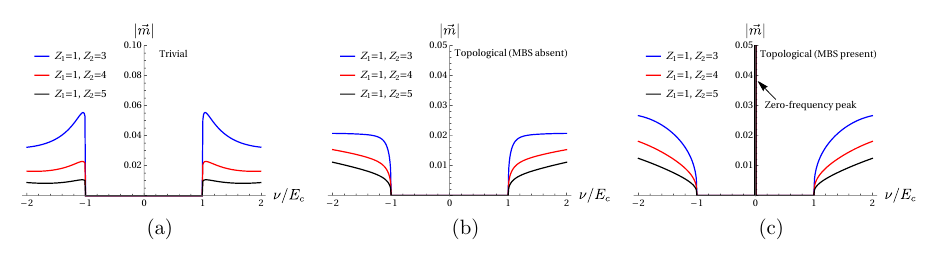}}
\vskip -0.2 in \caption{\small \sl
{Frequency dependence of LMDOS at $p_x$-$N_1$ interface for different values of interface transparencies $Z_1$ and $Z_2$. We consider three cases: (a) trivial regime, (b) topological regime when MBS are absent ($\varphi=\pi/5$), and, (c) topological regime when MBS are present ($\varphi=\pi$). Parameters: $\mathcal{S}=1/2$, $J=3$, $\Delta_{p_{x}}=\sqrt{2}$, $\mu_{p_{x}}=-1$ (for (a)), $\mu_{p_{x}}=1$ (for (b) and (c)), $\mu_{N}=10$, $E_c=1$, $L=1.3\xi$, $x=\bar{x}=-L/2$, $\varphi=\pi/5$ (for (a) and (b)), $\varphi=\pi$ (for (c)).}}
\end{figure}
Remarkably, surface odd-$\nu$ ST pairing is finite and exhibits significantly large magnitudes in the topological regime when MBS are present, as illustrated in Figs.~3(c), 4(c), 6(c) and 7(c). Therefore,
by observing the zero-frequency peak in LDOS and LMDOS, one can detect surface odd-$\nu$ ST pairing in our setup. }

{Further, the signature of surface odd-$\nu$ ST pairing can be observed through the measurement of total DC Josephson current\cite{Parhi}. This is shown in Fig.~10, where total DC Josephson current is plotted as a function of junction length $L$ in case of $p_x$-$I_1$-$N_1$-SF-$N_2$-$I_2$-$p_x$ JJ for different values of $Z_1$ and $Z_2$. The formula for calculating the DC Josephson current using the Furusaki-Tsukuda formalism\cite{furu,costa,sci} is provided in Appendix E. Similarly, as before, we consider three phases: (a) the trivial phase, (b) the topological phase without MBS, and (c) the topological phase with MBS in Fig.~10. We see that the magnitude of the total DC Josephson current is enhanced almost five-fold in the presence of MBS, wherein surface odd-$\nu$ ST pairing also exhibits multifold enhancement. Therefore, the enhancement of the total DC Josephson current serves as an indirect signature of surface odd-$\nu$ ST pairing in our setup.
\begin{figure}[ht]
\centering{\includegraphics[width=0.99\textwidth]{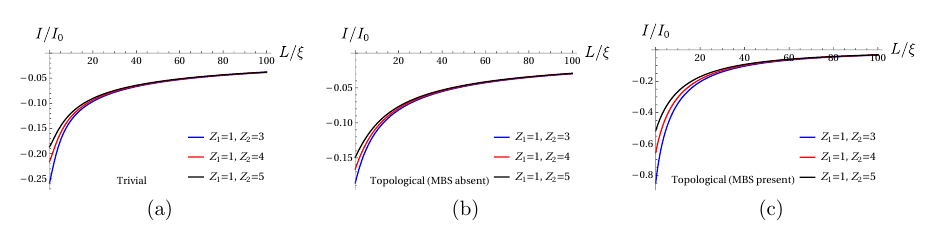}}
\vskip -0.2 in \caption{\small \sl {DC Josephson current as a function of junction length $L$ for different values of interface transparencies $Z_1$ and $Z_2$. We consider three cases: (a) trivial regime, (b) topological regime when MBS are absent ($\varphi=\pi/5$), and, (c) topological regime when MBS are present ($\varphi=\pi$). Parameters: $\mathcal{S}=1/2$, $J=3$, $\Delta_{p_{x}}=\sqrt{2}$, $\mu_{p_{x}}=-4$ (for (a)), $\mu_{p_{x}}=8.5$ (for (b) and (c)), $\mu_{N}=1000$, $k_{\mu_N}L=\pi$, $\varphi=\pi/5$ (for (a) and (b)), $\varphi=\pi$ (for (c)), $E_c=4$, $I_0=eE_{c}/\hbar$, $T\rightarrow0$.}}
\end{figure}}

In summary, our investigation reveals that surface odd-$\nu$ ST correlations provide a distinct signature of the presence of MBS in $p_{x}$ superconductor-SF-$p_{x}$ superconductor Josephson junction. Surface even-$\nu$ ST correlations, on the other hand, remain finite irrespective of the existence of MBS, rendering them unsuitable for MBS detection. In contrast, surface odd-$\nu$ ST correlations exhibit finite values exclusively in the presence of MBS, and they vanish in the absence of MBS. Spin flip scattering plays a crucial role in detecting MBS via inducing surface odd-$\nu$ MST correlations in their presence.
While the relationship between MBS and odd-$\nu$ pairing is explored in Refs.~\cite{tsi}-\cite{Kuzm}, odd-$\nu$ MST correlations do not emerge, and they do not offer a method to detect MBS through odd-$\nu$ pairing. This sets our work apart from them.  {Finally, the surface induced odd-$\nu$ spin-triplet pairing in presence of MBS (topological regime) are immune to disorder and can be effectively detected via LDOS/LMDOS zero frequency peak and multifold enhancement of the Josephson supercurrent in this regime.} Thus, our findings underscore the utility of surface odd-$\nu$ ST correlations as an effective tool for discerning the presence of MBS.

\section*{ACKNOWLEDGMENTS}
This research received funding from the following grants: 1. SERB Grant No. CRG/2019/006258 for the study of Josephson junctions with strained Dirac materials and their applications in quantum information processing, and 2. SERB MATRICS Grant No. MTR/2018/000070 for investigating Nash equilibrium versus Pareto optimality in N-Player games.

\appendix
{\section*{APPENDIX A: Wavefunctions \& Boundary conditions}}
In this Appendix, we present the detailed wavefunctions and boundary conditions  {for the $p_x$-SF-$p_x$ junction and the $p_x$-$I_1$-$N_1$-SF-$N_2$-$I_2$-$p_x$ junction, as shown in Figs.~1 and 5, respectively.}
\subsection{$p_x$-SF-$p_x$ junctions}
We diagonalize Hamiltonian \eqref{hamm} and obtain the wavefunctions in distinct regions {for $p_x$-SF-$p_x$ junction.} These wavefunctions are:
\begin{equation}
\small
\begin{split}
\Phi_{1}(x)&=
\begin{cases}
    \psi_{e\uparrow}^{S_{L}}{ e}^{iq_{e}x}\phi_{m'}^{\mathcal{S}}+r_{\uparrow\uparrow}^{e'e'}\psi_{e\uparrow}^{\prime S_{L}}{ e}^{-iq_{e}x}\phi_{m'}^{\mathcal{S}}+r_{\uparrow\downarrow}^{e'e'}\psi_{e\downarrow}^{\prime S_{L}}{ e}^{-iq_{e}x}\phi_{m'+1}^{\mathcal{S}}+r_{\uparrow\uparrow}^{e'h'}\psi_{h\uparrow}^{S_{L}}{ e}^{i{q}_{h}x}\phi_{m'}^{\mathcal{S}}+r_{\uparrow\downarrow}^{e'h'}\psi_{h\downarrow}^{S_{L}}{ e}^{i{q}_{h}x}\phi_{m'+1}^{\mathcal{S}}\,,  & x<0 \\
     \tilde{t}_{\uparrow\uparrow}^{e'e'}\psi_{e\uparrow}^{S_{R}}{ e}^{iq_{e}x}\phi_{m'}^{\mathcal{S}}+\tilde{t}_{\uparrow\downarrow}^{e'e'}\psi_{e\downarrow}^{S_{R}}{ e}^{iq_{e}x}\phi_{m'+1}^{\mathcal{S}}+\tilde{t}_{\uparrow\uparrow}^{e'h'}\psi_{h\uparrow}^{\prime S_{R}}{ e}^{-iq_{h}x}\phi_{m'}^{\mathcal{S}}+\tilde{t}_{\uparrow\downarrow}^{e'h'}\psi_{h\downarrow}^{\prime S_{R}}{ e}^{-iq_{h}x}\phi_{m'+1}^{\mathcal{S}}
 \,,& x>0
\end{cases}
\\
\Phi_{2}(x)&=
\begin{cases}
    \psi_{e\downarrow}^{S_{L}}{ e}^{iq_{e}x}\phi_{m'}^{\mathcal{S}}+r_{\downarrow\uparrow}^{e'e'}\psi_{e\uparrow}^{\prime S_{L}}{ e}^{-iq_{e}x}\phi_{m'-1}^{\mathcal{S}}+r_{\downarrow\downarrow}^{e'e'}\psi_{e\downarrow}^{\prime S_{L}}{ e}^{-iq_{e}x}\phi_{m'}^{\mathcal{S}}+r_{\downarrow\uparrow}^{e'h'}\psi_{h\uparrow}^{S_{L}}{ e}^{i{q}_{h}x}\phi_{m'-1}^{\mathcal{S}}+r_{\downarrow\downarrow}^{e'h'}\psi_{h\downarrow}^{S_{L}}{ e}^{i{q}_{h}x}\phi_{m'}^{\mathcal{S}}\,,  & x<0 \\
    \tilde{t}_{\downarrow\uparrow}^{e'e'}\psi_{e\uparrow}^{S_{R}}{ e}^{iq_{e}x}\phi_{m'-1}^{\mathcal{S}}+\tilde{t}_{\downarrow\downarrow}^{e'e'}\psi_{e\downarrow}^{S_{R}}{ e}^{iq_{e}x}\phi_{m'}^{\mathcal{S}}+\tilde{t}_{\downarrow\uparrow}^{e'h'}\psi_{h\uparrow}^{\prime S_{R}}{ e}^{-iq_{h}x}\phi_{m'-1}^{\mathcal{S}}+\tilde{t}_{\downarrow\downarrow}^{e'h'}\psi_{h\downarrow}^{\prime S_{R}}{ e}^{-iq_{h}x}\phi_{m'}^{\mathcal{S}}\,,& x>0
\end{cases}\\
\Phi_{3}(x)&=
\begin{cases}
    \psi_{h\uparrow}^{\prime S_{L}}{ e}^{-i{q}_{h}x}\phi_{m'}^{\mathcal{S}}+r_{\uparrow\uparrow}^{h'e'}\psi_{e\uparrow}^{\prime S_{L}}{ e}^{-i{q}_{e}x}\phi_{m'}^{\mathcal{S}}+r_{\uparrow\downarrow}^{h'e'}\psi_{e\downarrow}^{\prime S_{L}}{ e}^{-i{q}_{e}x}\phi_{m'+1}^{\mathcal{S}}+r_{\uparrow\uparrow}^{h'h'}\psi_{h\uparrow}^{S_{L}}{ e}^{iq_{h}x}\phi_{m'}^{\mathcal{S}}+r_{\uparrow\downarrow}^{h'h'}\psi_{h\downarrow}^{S_{L}}{ e}^{iq_{h}x}\phi_{m'+1}^{\mathcal{S}}\,,  & x<0 \\
    \tilde{t}_{\uparrow\uparrow}^{h'e'}\psi_{e\uparrow}^{S_{R}}{ e}^{iq_{e}x}\phi_{m'}^{\mathcal{S}}+\tilde{t}_{\uparrow\downarrow}^{h'e'}\psi_{e\downarrow}^{S_{R}}{ e}^{iq_{e}x}\phi_{m'+1}^{\mathcal{S}}+\tilde{t}_{\uparrow\uparrow}^{h'h'}\psi_{h\uparrow}^{\prime S_{R}}{ e}^{-iq_{h}x}\phi_{m'}^{\mathcal{S}}+\tilde{t}_{\uparrow\downarrow}^{h'h'}\psi_{h\downarrow}^{\prime S_{R}}{ e}^{-iq_{h}x}\phi_{m'+1}^{\mathcal{S}}\,,& x>0
\end{cases}\\
\Phi_{4}(x)&=
\begin{cases}
    \psi_{h\downarrow}^{\prime S_{L}}{ e}^{-i{q}_{h}x}\phi_{m'}^{\mathcal{S}}+r_{\downarrow\uparrow}^{h'e'}\psi_{e\uparrow}^{\prime S_{L}}{ e}^{-i{q}_{e}x}\phi_{m'-1}^{\mathcal{S}}+r_{\downarrow\downarrow}^{h'e'}\psi_{e\downarrow}^{\prime S_{L}}{ e}^{-i{q}_{e}x}\phi_{m'}^{\mathcal{S}}+r_{\downarrow\uparrow}^{h'h'}\psi_{h\uparrow}^{S_{L}}{ e}^{iq_{h}x}\phi_{m'-1}^{\mathcal{S}}+r_{\downarrow\downarrow}^{h'h'}\psi_{h\downarrow}^{S_{L}}{ e}^{iq_{h}x}\phi_{m'}^{\mathcal{S}}\,,  & x<0 \\
    \tilde{t}_{\downarrow\uparrow}^{h'e'}\psi_{e\uparrow}^{S_{R}}{ e}^{iq_{e}x}\phi_{m'-1}^{\mathcal{S}}+\tilde{t}_{\downarrow\downarrow}^{h'e'}\psi_{e\downarrow}^{S_{R}}{ e}^{iq_{e}x}\phi_{m'}^{\mathcal{S}}+\tilde{t}_{\downarrow\uparrow}^{h'h'}\psi_{h\uparrow}^{\prime S_{R}}{ e}^{-iq_{h}x}\phi_{m'-1}^{\mathcal{S}}+\tilde{t}_{\downarrow\downarrow}^{h'h'}\psi_{h\downarrow}^{\prime S_{R}}{ e}^{-iq_{h}x}\phi_{m'}^{\mathcal{S}}\,,& x>0
\end{cases}
\\
\Phi_{5}(x)&=
\begin{cases}
  t_{\uparrow\uparrow}^{e'e'}\psi_{e\uparrow}^{\prime S_{L}}{e}^{-iq_{e}x}\phi_{m'}^{\mathcal{S}}+t_{\uparrow\downarrow}^{e'e'}\psi_{e\downarrow}^{\prime S_{L}}{ e}^{-iq_{e}x}\phi_{m'+1}^{\mathcal{S}}+t_{\uparrow\uparrow}^{e'h'}\psi_{h\uparrow}^{S_{L}}{ e}^{i{q}_{h}x}\phi_{m'}^{\mathcal{S}}+t_{\uparrow\downarrow}^{e'h'}\psi_{h\downarrow}^{S_{L}}{ e}^{i{q}_{h}x}\phi_{m'+1}^{\mathcal{S}}\,,& x<0 \\
   \psi_{e\uparrow}^{\prime S_{R}}{ e}^{-iq_{e}x}\phi_{m'}^{\mathcal{S}}+\tilde{r}_{\uparrow\uparrow}^{e'e'}\psi_{e\uparrow}^{S_{R}}{ e}^{iq_{e}x}\phi_{m'}^{\mathcal{S}}+\tilde{r}_{\uparrow\downarrow}^{e'e'}\psi_{e\downarrow}^{S_{R}}{ e}^{iq_{e}x}\phi_{m'+1}^{\mathcal{S}}+\tilde{r}_{\uparrow\uparrow}^{e'h'}\psi_{h\uparrow}^{\prime S_{R}}{ e}^{-iq_{h}x}\phi_{m'}^{\mathcal{S}}+\tilde{r}_{\uparrow\downarrow}^{e'h'}\psi_{h\downarrow}^{\prime S_{R}}{ e}^{-iq_{h}x}\phi_{m'+1}^{\mathcal{S}}\,,& x>0
\end{cases}
\\
\Phi_{6}(x)&=
\begin{cases}
  t_{\downarrow\uparrow}^{e'e'}\psi_{e\uparrow}^{\prime S_{L}}{ e}^{-iq_{e}x}\phi_{m'-1}^{\mathcal{S}}+t_{\downarrow\downarrow}^{e'e'}\psi_{e\downarrow}^{\prime S_{L}}{ e}^{-iq_{e}x}\phi_{m'}^{\mathcal{S}}+t_{\downarrow\uparrow}^{e'h'}\psi_{h\uparrow}^{S_{L}}{ e}^{i{q}_{h}x}\phi_{m'-1}^{\mathcal{S}}+t_{\downarrow\downarrow}^{e'h'}\psi_{h\downarrow}^{S_{L}}{ e}^{i{q}_{h}x}\phi_{m'}^{\mathcal{S}}\,, & x<0 \\
   \psi_{e\downarrow}^{\prime S_{R}}{ e}^{-iq_{e}x}\phi_{m'}^{\mathcal{S}}+\tilde{r}_{\downarrow\uparrow}^{e'e'}\psi_{e\uparrow}^{S_{R}}{ e}^{iq_{e}x}\phi_{m'-1}^{\mathcal{S}}+\tilde{r}_{\downarrow\downarrow}^{e'e'}\psi_{e\downarrow}^{S_{R}}{ e}^{iq_{e}x}\phi_{m'}^{\mathcal{S}}+\tilde{r}_{\downarrow\uparrow}^{e'h'}\psi_{h\uparrow}^{\prime S_{R}}{ e}^{-iq_{h}x}\phi_{m'-1}^{\mathcal{S}}+\tilde{r}_{\downarrow\downarrow}^{e'h'}\psi_{h\downarrow}^{\prime S_{R}}{ e}^{-iq_{h}x}\phi_{m'}^{\mathcal{S}}\,,& x>0
\end{cases}\\
\Phi_{7}(x)&=
\begin{cases}
  t_{\uparrow\uparrow}^{h'e'}\psi_{e\uparrow}^{\prime S_{L}}{ e}^{-i{q}_{e}x}\phi_{m'}^{\mathcal{S}}+t_{\uparrow\downarrow}^{h'e'}\psi_{e\downarrow}^{\prime S_{L}}{ e}^{-i{q}_{e}x}\phi_{m'+1}^{\mathcal{S}}+t_{\uparrow\uparrow}^{h'h'}\psi_{h\uparrow}^{S_{L}}{ e}^{iq_{h}x}\phi_{m'}^{\mathcal{S}}+t_{\uparrow\downarrow}^{h'h'}\psi_{h\downarrow}^{S_{L}}{ e}^{iq_{h}x}\phi_{m'+1}^{\mathcal{S}}\,,  & x<0 \\
   \psi_{h\uparrow}^{S_{R}}{ e}^{iq_{h}x}\phi_{m'}^{\mathcal{S}}+\tilde{r}_{\uparrow\uparrow}^{h'e'}\psi_{e\uparrow}^{S_{R}}{ e}^{iq_{e}x}\phi_{m'}^{\mathcal{S}}+\tilde{r}_{\uparrow\downarrow}^{h'e'}\psi_{e\downarrow}^{S_{R}}{ e}^{iq_{e}x}\phi_{m'+1}^{\mathcal{S}}+\tilde{r}_{\uparrow\uparrow}^{h'h'}\psi_{h\uparrow}^{\prime S_{R}}{ e}^{-iq_{h}x}\phi_{m'}^{\mathcal{S}}+\tilde{r}_{\uparrow\downarrow}^{h'h'}\psi_{h\downarrow}^{\prime S_{R}}{ e}^{-iq_{h}x}\phi_{m'+1}^{\mathcal{S}}\,,& x>0
\end{cases}
\\
\Phi_{8}(x)&=
\begin{cases}
  t_{\downarrow\uparrow}^{h'e'}\psi_{e\uparrow}^{\prime S_{L}}{ e}^{-i{q}_{e}x}\phi_{m'-1}^{\mathcal{S}}+t_{\downarrow\downarrow}^{h'e'}\psi_{e\downarrow}^{\prime S_{L}}{ e}^{-i{q}_{e}x}\phi_{m'}^{\mathcal{S}}+t_{\downarrow\uparrow}^{h'h'}\psi_{h\uparrow}^{S_{L}}{ e}^{iq_{h}x}\phi_{m'-1}^{\mathcal{S}}+t_{\downarrow\downarrow}^{h'h'}\psi_{h\downarrow}^{S_{L}}{ e}^{iq_{h}x}\phi_{m'}^{\mathcal{S}}\,,  & x<0 \\
   \psi_{h\downarrow}^{S_{R}}{ e}^{iq_{h}x}\phi_{m'}^{\mathcal{S}}+\tilde{r}_{\downarrow\uparrow}^{h'e'}\psi_{e\uparrow}^{S_{R}}{ e}^{iq_{e}x}\phi_{m'-1}^{\mathcal{S}}+\tilde{r}_{\downarrow\downarrow}^{h'e'}\psi_{e\downarrow}^{S_{R}}{ e}^{iq_{e}x}\phi_{m'}^{\mathcal{S}}+\tilde{r}_{\downarrow\uparrow}^{h'h'}\psi_{h\uparrow}^{\prime S_{R}}{ e}^{-iq_{h}x}\phi_{m'-1}^{\mathcal{S}}+\tilde{r}_{\downarrow\downarrow}^{h'h'}\psi_{h\downarrow}^{\prime S_{R}}{ e}^{-iq_{h}x}\phi_{m'}^{\mathcal{S}}\,,& x>0
\end{cases}
\end{split}
\label{waveJ}
\end{equation}
where $\psi_{e\uparrow}^{S_{L}}=\frac{1}{\sqrt{|\gamma_e|^2+1}}\begin{pmatrix}
                        \gamma_{e}\\
                        0\\
                        0\\
                        1
                       \end{pmatrix}$, $\psi_{e\downarrow}^{S_{L}}=\frac{1}{\sqrt{|\gamma_e|^2+1}}\begin{pmatrix}
                        0\\
                        \gamma_{e}\\
                        1\\
                        0
                       \end{pmatrix}$, $\psi_{h\uparrow}^{S_{L}}=\frac{1}{\sqrt{|\gamma_h|^2+1}}\begin{pmatrix}
                        \gamma_{h}\\
                        0\\
                        0\\
                        1
                       \end{pmatrix}$, $\psi_{h\downarrow}^{S_{L}}=\frac{1}{\sqrt{|\gamma_h|^2+1}}\begin{pmatrix}
                        0\\
                        \gamma_{h}\\
                        1\\
                        0
                       \end{pmatrix}$, $\psi_{e\uparrow}^{\prime S_{L}}=\frac{1}{\sqrt{|\gamma_e|^2+1}}\begin{pmatrix}
                        -\gamma_{e}\\
                        0\\
                        0\\
                        1
                       \end{pmatrix}$, $\psi_{e\downarrow}^{\prime S_{L}}=\frac{1}{\sqrt{|\gamma_e|^2+1}}\begin{pmatrix}
                        0\\
                        -\gamma_{e}\\
                        1\\
                        0
                       \end{pmatrix}$, $\psi_{h\uparrow}^{\prime S_{L}}=\frac{1}{\sqrt{|\gamma_h|^2+1}}\begin{pmatrix}
                        -\gamma_{h}\\
                        0\\
                        0\\
                        1
                       \end{pmatrix}$, $\psi_{h\downarrow}^{\prime S_{L}}=\frac{1}{\sqrt{|\gamma_h|^2+1}}\begin{pmatrix}
                        0\\
                        -\gamma_{h}\\
                        1\\
                        0
                       \end{pmatrix}$, $\psi_{e\uparrow}^{S_{R}}=\begin{pmatrix}
                        \gamma_{e}e^{i\varphi}\\
                        0\\
                        0\\
                        1
                       \end{pmatrix}$, $\psi_{e\downarrow}^{S_{R}}=\frac{1}{\sqrt{|\gamma_e|^2+1}}\begin{pmatrix}
                        0\\
                        \gamma_{e}e^{i\varphi}\\
                        1\\
                        0
                       \end{pmatrix}$, $\psi_{h\uparrow}^{S_{R}}=\frac{1}{\sqrt{|\gamma_h|^2+1}}\begin{pmatrix}
                        \gamma_{h}e^{i\varphi}\\
                        0\\
                        0\\
                        1
                       \end{pmatrix}$, $\psi_{h\downarrow}^{S_{R}}=\frac{1}{\sqrt{|\gamma_h|^2+1}}\begin{pmatrix}
                        0\\
                        \gamma_{h}e^{i\varphi}\\
                        1\\
                        0
                       \end{pmatrix}$, $\psi_{e\uparrow}^{\prime S_{R}}=\frac{1}{\sqrt{|\gamma_e|^2+1}}\begin{pmatrix}
                        -\gamma_{e}e^{i\varphi}\\
                        0\\
                        0\\
                        1
                       \end{pmatrix}$, $\psi_{e\downarrow}^{\prime S_{R}}=\frac{1}{\sqrt{|\gamma_e|^2+1}}\begin{pmatrix}
                        0\\
                        -\gamma_{e}e^{i\varphi}\\
                        1\\
                        0
                       \end{pmatrix}$, $\psi_{h\uparrow}^{\prime S_{R}}=\frac{1}{\sqrt{|\gamma_h|^2+1}}\begin{pmatrix}
                        -\gamma_{h}e^{i\varphi}\\
                        0\\
                        0\\
                        1
                       \end{pmatrix}$, $\psi_{h\downarrow}^{\prime S_{R}}=\frac{1}{\sqrt{|\gamma_h|^2+1}}\begin{pmatrix}
                        0\\
                        -\gamma_{h}e^{i\varphi}\\
                        1\\
                        0
                       \end{pmatrix}$, and $\gamma_{e,h}=(\nu+q_{e,h}^{2}-\mu_{p_{x}}')/(\Delta_{p_{x}}' q_{e,h})$ with $\hbar=2m^{*}=1$. In Eq.~\eqref{waveJ}, $\Phi_{1}$, $\Phi_{2}$,  $\Phi_{3}$, $\Phi_{4}$ denote the wavefunctions when electron-like quasiparticle with spin-up, electron-like quasiparticle with spin-down, hole-like quasiparticle with spin-up and hole-like quasiparticle with spin-down are injected from left $p_{x}$ superconductor, respectively, while $\Phi_{5}$, $\Phi_{6}$,  $\Phi_{7}$, $\Phi_{8}$ denote the wavefunctions when the corresponding quasiparticles are injected
                       from right $p_{x}$ superconductor, respectively.
                       $r_{ij}^{nn}$ and $\tilde{r}_{ij}^{nn}$ are the reflection amplitudes in left $p_{x}$ superconductor and right $p_{x}$ superconductor respectively, while $t_{ij}^{nn}$ and $\tilde{t}_{ij}^{nn}$ are the transmission amplitudes in left $p_{x}$ superconductor and right $p_{x}$ superconductor respectively with $i,j\in{\{\uparrow,\downarrow\}}$ and $n\in{\{e',h'\}}$.
Further, in Eq.~\eqref{waveJ}, $\phi_{m'}^{\mathcal{S}}$ denotes the eigenfunction of the SF, with $\mathcal{S}$ representing its spin and $m'$ corresponding to its magnetic moment. The action of $\mathcal{S}^{z}$ is given as- $\mathcal{S}^{z}\phi_{m'}^{\mathcal{S}}=\hbar m'\phi_{m'}^{\mathcal{S}}$. After diagonalizing the Hamiltonian $(H_{BdG}^{\mbox{$p_{x}$-SF-$p_{x}$}})^{*}(-k)$ instead of $H_{BdG}^{\mbox{$p_{x}$-SF-$p_{x}$}}(k)$, we will obtain the conjugate process $\tilde{\Phi}_{i}$ necessary to construct the retarded Green's function in next section. For our model we notice that $\tilde{\psi}_{e\uparrow}^{S_{L}}=\frac{1}{\sqrt{|\gamma_e|^2+1}}\begin{pmatrix}
                        -\gamma_{e}\\
                        0\\
                        0\\
                        1
                       \end{pmatrix}$, $\tilde{\psi}_{e\downarrow}^{S_{L}}=\frac{1}{\sqrt{|\gamma_e|^2+1}}\begin{pmatrix}
                        0\\
                        -\gamma_{e}\\
                        1\\
                        0
                       \end{pmatrix}$, $\tilde{\psi}_{h\uparrow}^{S_{L}}=\frac{1}{\sqrt{|\gamma_h|^2+1}}\begin{pmatrix}
                        -\gamma_{h}\\
                        0\\
                        0\\
                        1
                       \end{pmatrix}$, $\tilde{\psi}_{h\downarrow}^{S_{L}}=\frac{1}{\sqrt{|\gamma_h|^2+1}}\begin{pmatrix}
                        0\\
                        -\gamma_{h}\\
                        1\\
                        0
                       \end{pmatrix}$, $\tilde{\psi}_{e\uparrow}^{\prime S_{L}}=\frac{1}{\sqrt{|\gamma_e|^2+1}}\begin{pmatrix}
                        \gamma_{e}\\
                        0\\
                        0\\
                        1
                       \end{pmatrix}$, $\tilde{\psi}_{e\downarrow}^{\prime S_{L}}=\frac{1}{\sqrt{|\gamma_e|^2+1}}\begin{pmatrix}
                        0\\
                        \gamma_{e}\\
                        1\\
                        0
                       \end{pmatrix}$, $\tilde{\psi}_{h\uparrow}^{\prime S_{L}}=\frac{1}{\sqrt{|\gamma_h|^2+1}}\begin{pmatrix}
                        \gamma_{h}\\
                        0\\
                        0\\
                        1
                       \end{pmatrix}$, $\tilde{\psi}_{h\downarrow}^{\prime S_{L}}=\frac{1}{\sqrt{|\gamma_h|^2+1}}\begin{pmatrix}
                        0\\
                        \gamma_{h}\\
                        1\\
                        0
                       \end{pmatrix}$, $\tilde{\psi}_{e\uparrow}^{S_{R}}=\frac{1}{\sqrt{|\gamma_e|^2+1}}\begin{pmatrix}
                        -\gamma_{e}e^{-i\varphi}\\
                        0\\
                        0\\
                        1
                       \end{pmatrix}$, $\tilde{\psi}_{e\downarrow}^{S_{R}}=\frac{1}{\sqrt{|\gamma_e|^2+1}}\begin{pmatrix}
                        0\\
                        -\gamma_{e}e^{-i\varphi}\\
                        1\\
                        0
                       \end{pmatrix}$, $\tilde{\psi}_{h\uparrow}^{S_{R}}=\frac{1}{\sqrt{|\gamma_h|^2+1}}\begin{pmatrix}
                        -\gamma_{h}e^{-i\varphi}\\
                        0\\
                        0\\
                        1
                       \end{pmatrix}$, $\tilde{\psi}_{h\downarrow}^{S_{R}}=\frac{1}{\sqrt{|\gamma_h|^2+1}}\begin{pmatrix}
                        0\\
                        -\gamma_{h}e^{-i\varphi}\\
                        1\\
                        0
                       \end{pmatrix}$, $\tilde{\psi}_{e\uparrow}^{\prime S_{R}}=\frac{1}{\sqrt{|\gamma_e|^2+1}}\begin{pmatrix}
                        \gamma_{e}e^{-i\varphi}\\
                        0\\
                        0\\
                        1
                       \end{pmatrix}$, $\tilde{\psi}_{e\downarrow}^{\prime S_{R}}=\frac{1}{\sqrt{|\gamma_e|^2+1}}\begin{pmatrix}
                        0\\
                        \gamma_{e}e^{-i\varphi}\\
                        1\\
                        0
                       \end{pmatrix}$, $\tilde{\psi}_{h\uparrow}^{\prime S_{R}}=\frac{1}{\sqrt{|\gamma_h|^2+1}}\begin{pmatrix}
                        \gamma_{h}e^{-i\varphi}\\
                        0\\
                        0\\
                        1
                       \end{pmatrix}$, $\tilde{\psi}_{h\downarrow}^{\prime S_{R}}=\frac{1}{\sqrt{|\gamma_h|^2+1}}\begin{pmatrix}
                        0\\
                        \gamma_{h}e^{-i\varphi}\\
                        1\\
                        0
                       \end{pmatrix}$. Further, $\xi=\hbar^2/(m^{*}\Delta_{p_{x}}')$ represents the superconducting coherence length\cite{tsi}.
We have verified that for each kind of incident quasi-particle (electron-like/hole-like) the probability conservation $|r_{ij}^{nn}|^2+|\tilde{t}_{ij}^{nn}|^2=1$ and $|\tilde{r}_{ij}^{nn}|^2+|t_{ij}^{nn}|^2=1$ both below and above the gap.

At the $p_{x}$-$p_{x}$ interface ($x=0$), the boundary conditions are\cite{tsi}:
\begin{equation}
\label{bc1}
\Phi_{l}|_{x<0}=\Phi_{l}|_{x>0}\,\,\, \mbox{and,}\,\,\, \frac{d\Phi_{l}|_{x>0}}{dx}-\frac{d\Phi_{l}|_{x<0}}{dx}=\frac{im^{*}\Delta_{p_{x}}'}{\hbar^2}\begin{bmatrix}
0 & (1-e^{i\varphi})\sigma_{x}\\
(-1+e^{-i\varphi})\sigma_{x} & 0                                                                                                                              \end{bmatrix}\Phi_{l}|_{x=0}-\frac{2m^{*}J_{0}}{\hbar^2}\vec{s}.\vec{\mathcal{S}}\Phi_{l}|_{x=0},(l=1,2,...,8),
\end{equation}
where $\vec{s}.\vec{\mathcal{S}}=s^{z}\mathcal{S}^{z}+(s^{+}\mathcal{S}^{-}+s^{-}\mathcal{S}^{+})/2$ is the exchange operator of the SF's Hamiltonian\cite{AJP} with $s^{\pm}=s_{x}\pm is_{y}$ for electron-like or hole-like quasiparticle and $\mathcal{S}^{\pm}=\mathcal{S}_{x}\pm i\mathcal{S}_{y}$ for the SF. For wave functions involving electron-like quasiparticles with spin up and spin down, actions of $\vec{s}.\vec{\mathcal{S}}$ are
\begin{equation}
\vec{s}.\vec{\mathcal{S}}\psi_{e\uparrow}^{S_{L}}\phi_{m'}^{\mathcal{S}}=\frac{\hbar^2m'}{2}\psi_{e\uparrow}^{S_{L}}\phi_{m'}^{\mathcal{S}}+\frac{\hbar^2f}{2}\psi_{e\downarrow}^{S_{L}}\phi_{m'+1}^{\mathcal{S}},\,\,\,\mbox{and,}\,\,\,\,
\label{euu}
\vec{s}.\vec{\mathcal{S}} \psi_{e\downarrow}^{S_{L}}\phi_{m'}^{\mathcal{S}}=-\frac{\hbar^2m'}{2}\psi_{e\downarrow}^{S_{L}}\phi_{m'}^{\mathcal{S}}+\frac{\hbar^2f'}{2}\psi_{e\uparrow}^{S_{L}}\phi_{m'-1}^{\mathcal{S}}.
\end{equation}
Similarly, for wave functions involving hole-like quasiparticles with spin up and spin down, actions of $\vec{s}.\vec{\mathcal{S}}$ are
\begin{equation}
\vec{s}.\vec{\mathcal{S}}\psi_{h\uparrow}^{\prime S_{L}}\phi_{m'}^{\mathcal{S}}=\frac{\hbar^2m'}{2}\psi_{h\uparrow}^{\prime S_{L}}\phi_{m'}^{\mathcal{S}}+\frac{\hbar^2f}{2}\psi_{h\downarrow}^{\prime S_{L}}\phi_{m'+1}^{\mathcal{S}},\,\,\,\mbox{and,}\,\,\,\,
\label{huu}
\vec{s}.\vec{\mathcal{S}} \psi_{h\downarrow}^{\prime S_{L}}\phi_{m'}^{\mathcal{S}}=-\frac{\hbar^2m'}{2}\psi_{h\downarrow}^{\prime S_{L}}\phi_{m'}^{\mathcal{S}}+\frac{\hbar^2f'}{2}\psi_{h\uparrow}^{\prime S_{L}}\phi_{m'-1}^{\mathcal{S}}.
\end{equation}
In Eqs.~\eqref{euu} and \eqref{huu},  $f=\sqrt{(\mathcal{S}-m')(\mathcal{S}+m'+1)}$ represents the probability of spin-flip for the incident process of a spin-up electron-like or hole-like quasiparticle, where $f'=\sqrt{(\mathcal{S}+m')(\mathcal{S}-m'+1)}$ represents the probability of spin-flip for the incident process of a spin-down electron-like or hole-like quasiparticle. By employing the equations above and solving the boundary condition (Eq.~\eqref{bc1}), we obtain a set of eight equations for scattering processes, as shown in Eq.~\eqref{waveJ}. Various scattering amplitudes $r_{ij}^{nn}$, $\tilde{r}_{ij}^{nn}$, $\tilde{t}_{ij}^{nn}$, $t_{ij}^{nn}$ for each kind of incident quasiparticle (electron-like/hole-like) are determined from these $8$ equations. We have verified the detailed balance condition\cite{furu} for Andreev reflection, i.e., $\frac{r_{ij}^{e'h'}(-\varphi,\nu)}{q_{e}}=\frac{r_{ij}^{h'e'}(\varphi,\nu)}{q_{h}}$ and, $\frac{\tilde{r}_{ij}^{e'h'}(-\varphi,\nu)}{q_{e}}=\frac{\tilde{r}_{ij}^{h'e'}(\varphi,\nu)}{q_{h}}$, which implies the accuracy of our calculation.
{\subsection{$p_x$-$I_1$-$N_1$-SF-$N_2$-$I_2$-$p_x$ junctions}
We diagonalize Hamiltonian \eqref{hamp} and obtain the wavefunctions in distinct regions for $p_x$-$I_1$-$N_1$-SF-$N_2$-$I_2$-$p_x$ junction. These
wavefunctions are:
\begin{equation}
\footnotesize
\begin{split}
\Phi_{1}(x)&=
\begin{cases}
    \psi_{e\uparrow}^{S_{L}}{ e}^{iq_{e}(x+L/2)}\phi_{m'}^{\mathcal{S}}+r_{\uparrow\uparrow}^{\prime e'e'}\psi_{e\uparrow}^{\prime S_{L}}{ e}^{-iq_{e}(x+L/2)}\phi_{m'}^{\mathcal{S}}+r_{\uparrow\downarrow}^{\prime e'e'}\psi_{e\downarrow}^{\prime S_{L}}{ e}^{-iq_{e}(x+L/2)}\phi_{m'+1}^{\mathcal{S}}+r_{\uparrow\uparrow}^{\prime e'h'}\psi_{h\uparrow}^{S_{L}}{ e}^{i{q}_{h}(x+L/2)}\phi_{m'}^{\mathcal{S}}\\+r_{\uparrow\downarrow}^{\prime e'h'}\psi_{h\downarrow}^{S_{L}}{ e}^{i{q}_{h}(x+L/2)}\phi_{m'+1}^{\mathcal{S}}\,,   & x<-L/2 \\
    \psi_{e\uparrow}^{N}a_{\uparrow\uparrow}^{\prime e'e'}e^{ik_e(x+L/2)}\phi_{m'}^{\mathcal{S}}+\psi_{e\downarrow}^{N}a_{\uparrow\downarrow}^{\prime e'e'}e^{ik_e(x+L/2)}\phi_{m'+1}^{\mathcal{S}}+\psi_{e\uparrow}^{N}b_{\uparrow\uparrow}^{\prime e'e'}e^{-ik_ex}\phi_{m'}^{\mathcal{S}}+\psi_{e\downarrow}^{N}b_{\uparrow\downarrow}^{\prime e'e'}e^{-ik_ex}\phi_{m'+1}^{\mathcal{S}}\\+\psi_{h\uparrow}^{N}c_{\uparrow\uparrow}^{\prime e'h'}e^{-ik_h(x+L/2)}\phi_{m'}^{\mathcal{S}}+\psi_{h\downarrow}^{N}c_{\uparrow\downarrow}^{\prime e'h'}e^{-ik_h(x+L/2)}\phi_{m'+1}^{\mathcal{S}}+\psi_{h\uparrow}^{N}d_{\uparrow\uparrow}^{\prime e'h'}e^{ik_hx}\phi_{m'}^{\mathcal{S}}+\psi_{h\downarrow}^{N}d_{\uparrow\downarrow}^{\prime e'h'}e^{ik_hx}\phi_{m'+1}^{\mathcal{S}}\,, & -L/2<x<0 \\
    \psi_{e\uparrow}^{N}e_{\uparrow\uparrow}^{\prime e'e'}e^{ik_ex}\phi_{m'}^{\mathcal{S}}+\psi_{e\downarrow}^{N}e_{\uparrow\downarrow}^{\prime e'e'}e^{ik_ex}\phi_{m'+1}^{\mathcal{S}}+\psi_{e\uparrow}^{N}f_{\uparrow\uparrow}^{\prime e'e'}e^{-ik_e(x-L/2)}\phi_{m'}^{\mathcal{S}}+\psi_{e\downarrow}^{N}f_{\uparrow\downarrow}^{\prime e'e'}e^{-ik_e(x-L/2)}\phi_{m'+1}^{\mathcal{S}}\\
    +\psi_{h\uparrow}^{N}g_{\uparrow\uparrow}^{\prime e'h'}e^{-ik_hx}\phi_{m'}^{\mathcal{S}}+\psi_{h\downarrow}^{N}g_{\uparrow\downarrow}^{\prime e'h'}e^{-ik_hx}\phi_{m'+1}^{\mathcal{S}}+\psi_{h\uparrow}^{N}h_{\uparrow\uparrow}^{\prime e'h'}e^{ik_h(x-L/2)}\phi_{m'}^{\mathcal{S}}+\psi_{h\downarrow}^{N}h_{\uparrow\downarrow}^{\prime e'h'}e^{ik_h(x-L/2)}\phi_{m'+1}^{\mathcal{S}}\,, & 0<x<L/2 \\
     \tilde{t}_{\uparrow\uparrow}^{\prime e'e'}\psi_{e\uparrow}^{S_{R}}{ e}^{iq_{e}(x-L/2)}\phi_{m'}^{\mathcal{S}}+\tilde{t}_{\uparrow\downarrow}^{\prime e'e'}\psi_{e\downarrow}^{S_{R}}{ e}^{iq_{e}(x-L/2)}\phi_{m'+1}^{\mathcal{S}}+\tilde{t}_{\uparrow\uparrow}^{\prime e'h'}\psi_{h\uparrow}^{\prime S_{R}}{ e}^{-iq_{h}(x-L/2)}\phi_{m'}^{\mathcal{S}}+\tilde{t}_{\uparrow\downarrow}^{\prime e'h'}\psi_{h\downarrow}^{\prime S_{R}}{ e}^{-iq_{h}(x-L/2)}\phi_{m'+1}^{\mathcal{S}}
 \,,& x>L/2
\end{cases}\\
\end{split}\nonumber
\end{equation}
\begin{equation}
\footnotesize
\begin{split}
\Phi_{2}(x)&=
\begin{cases}
    \psi_{e\downarrow}^{S_{L}}{ e}^{iq_{e}(x+L/2)}\phi_{m'}^{\mathcal{S}}+r_{\downarrow\uparrow}^{\prime e'e'}\psi_{e\uparrow}^{\prime S_{L}}{ e}^{-iq_{e}(x+L/2)}\phi_{m'-1}^{\mathcal{S}}+r_{\downarrow\downarrow}^{\prime e'e'}\psi_{e\downarrow}^{\prime S_{L}}{ e}^{-iq_{e}(x+L/2)}\phi_{m'}^{\mathcal{S}}+r_{\downarrow\uparrow}^{\prime e'h'}\psi_{h\uparrow}^{S_{L}}{ e}^{i{q}_{h}(x+L/2)}\phi_{m'-1}^{\mathcal{S}}\\+r_{\downarrow\downarrow}^{\prime e'h'}\psi_{h\downarrow}^{S_{L}}{ e}^{i{q}_{h}(x+L/2)}\phi_{m'}^{\mathcal{S}}\,,  & x<-L/2 \\
    \psi_{e\uparrow}^{N}a_{\downarrow\uparrow}^{\prime e'e'}e^{ik_e(x+L/2)}\phi_{m'}^{\mathcal{S}}+\psi_{e\downarrow}^{N}a_{\downarrow\downarrow}^{\prime e'e'}e^{ik_e(x+L/2)}\phi_{m'+1}^{\mathcal{S}}+\psi_{e\uparrow}^{N}b_{\downarrow\uparrow}^{\prime e'e'}e^{-ik_ex}\phi_{m'}^{\mathcal{S}}+\psi_{e\downarrow}^{N}b_{\downarrow\downarrow}^{\prime e'e'}e^{-ik_ex}\phi_{m'+1}^{\mathcal{S}}\\+\psi_{h\uparrow}^{N}c_{\downarrow\uparrow}^{\prime e'h'}e^{-ik_h(x+L/2)}\phi_{m'}^{\mathcal{S}}+\psi_{h\downarrow}^{N}c_{\downarrow\downarrow}^{\prime e'h'}e^{-ik_h(x+L/2)}\phi_{m'+1}^{\mathcal{S}}+\psi_{h\uparrow}^{N}d_{\downarrow\uparrow}^{\prime e'h'}e^{ik_hx}\phi_{m'}^{\mathcal{S}}+\psi_{h\downarrow}^{N}d_{\downarrow\downarrow}^{\prime e'h'}e^{ik_hx}\phi_{m'+1}^{\mathcal{S}}\,, & -L/2<x<0 \\
    \psi_{e\uparrow}^{N}e_{\downarrow\uparrow}^{\prime e'e'}e^{ik_ex}\phi_{m'}^{\mathcal{S}}+\psi_{e\downarrow}^{N}e_{\downarrow\downarrow}^{\prime e'e'}e^{ik_ex}\phi_{m'+1}^{\mathcal{S}}+\psi_{e\uparrow}^{N}f_{\downarrow\uparrow}^{\prime e'e'}e^{-ik_e(x-L/2)}\phi_{m'}^{\mathcal{S}}+\psi_{e\downarrow}^{N}f_{\downarrow\downarrow}^{\prime e'e'}e^{-ik_e(x-L/2)}\phi_{m'+1}^{\mathcal{S}}\\
    +\psi_{h\uparrow}^{N}g_{\downarrow\uparrow}^{\prime e'h'}e^{-ik_hx}\phi_{m'}^{\mathcal{S}}+\psi_{h\downarrow}^{N}g_{\downarrow\downarrow}^{\prime e'h'}e^{-ik_hx}\phi_{m'+1}^{\mathcal{S}}+\psi_{h\uparrow}^{N}h_{\downarrow\uparrow}^{\prime e'h'}e^{ik_h(x-L/2)}\phi_{m'}^{\mathcal{S}}+\psi_{h\downarrow}^{N}h_{\downarrow\downarrow}^{\prime e'h'}e^{ik_h(x-L/2)}\phi_{m'+1}^{\mathcal{S}}\,, & 0<x<L/2 \\
    \tilde{t}_{\downarrow\uparrow}^{\prime e'e'}\psi_{e\uparrow}^{S_{R}}{ e}^{iq_{e}(x-L/2)}\phi_{m'-1}^{\mathcal{S}}+\tilde{t}_{\downarrow\downarrow}^{\prime e'e'}\psi_{e\downarrow}^{S_{R}}{ e}^{iq_{e}(x-L/2)}\phi_{m'}^{\mathcal{S}}+\tilde{t}_{\downarrow\uparrow}^{\prime e'h'}\psi_{h\uparrow}^{\prime S_{R}}{\rm e}^{-iq_{h}(x-L/2)}\phi_{m'-1}^{\mathcal{S}}+\tilde{t}_{\downarrow\downarrow}^{\prime e'h'}\psi_{h\downarrow}^{\prime S_{R}}{\rm e}^{-iq_{h}(x-L/2)}\phi_{m'}^{\mathcal{S}}\,,& x>L/2
\end{cases}\\
\Phi_{3}(x)&=
\begin{cases}
    \psi_{h\uparrow}^{\prime S_{L}}{e}^{-i{q}_{h}(x+L/2)}\phi_{m'}^{\mathcal{S}}+r_{\uparrow\uparrow}^{\prime h'e'}\psi_{e\uparrow}^{\prime S_{L}}{e}^{-i{q}_{e}(x+L/2)}\phi_{m'}^{\mathcal{S}}+r_{\uparrow\downarrow}^{\prime h'e'}\psi_{e\downarrow}^{\prime S_{L}}{e}^{-i{q}_{e}(x+L/2)}\phi_{m'+1}^{\mathcal{S}}+r_{\uparrow\uparrow}^{\prime h'h'}\psi_{h\uparrow}^{S_{L}}{e}^{iq_{h}(x+L/2)}\phi_{m'}^{\mathcal{S}}\\+r_{\uparrow\downarrow}^{\prime h'h'}\psi_{h\downarrow}^{S_{L}}{e}^{iq_{h}(x+L/2)}\phi_{m'+1}^{\mathcal{S}}\,,  & x<-L/2 \\
    \psi_{e\uparrow}^{N}a_{\uparrow\uparrow}^{\prime h'e'}e^{ik_e(x+L/2)}\phi_{m'}^{\mathcal{S}}+\psi_{e\downarrow}^{N}a_{\uparrow\downarrow}^{\prime h'e'}e^{ik_e(x+L/2)}\phi_{m'+1}^{\mathcal{S}}+\psi_{e\uparrow}^{N}b_{\uparrow\uparrow}^{\prime h'e'}e^{-ik_ex}\phi_{m'}^{\mathcal{S}}+\psi_{e\downarrow}^{N}b_{\uparrow\downarrow}^{\prime h'e'}e^{-ik_ex}\phi_{m'+1}^{\mathcal{S}}\\+\psi_{h\uparrow}^{N}c_{\uparrow\uparrow}^{\prime h'h'}e^{-ik_h(x+L/2)}\phi_{m'}^{\mathcal{S}}+\psi_{h\downarrow}^{N}c_{\uparrow\downarrow}^{\prime h'h'}e^{-ik_h(x+L/2)}\phi_{m'+1}^{\mathcal{S}}+\psi_{h\uparrow}^{N}d_{\uparrow\uparrow}^{\prime h'h'}e^{ik_hx}\phi_{m'}^{\mathcal{S}}+\psi_{h\downarrow}^{N}d_{\uparrow\downarrow}^{\prime h'h'}e^{ik_hx}\phi_{m'+1}^{\mathcal{S}}\,, & -L/2<x<0 \\
    \psi_{e\uparrow}^{N}e_{\uparrow\uparrow}^{\prime h'e'}e^{ik_ex}\phi_{m'}^{\mathcal{S}}+\psi_{e\downarrow}^{N}e_{\uparrow\downarrow}^{\prime h'e'}e^{ik_ex}\phi_{m'+1}^{\mathcal{S}}+\psi_{e\uparrow}^{N}f_{\uparrow\uparrow}^{\prime h'e'}e^{-ik_e(x-L/2)}\phi_{m'}^{\mathcal{S}}+\psi_{e\downarrow}^{N}f_{\uparrow\downarrow}^{\prime h'e'}e^{-ik_e(x-L/2)}\phi_{m'+1}^{\mathcal{S}}\\
    +\psi_{h\uparrow}^{N}g_{\uparrow\uparrow}^{\prime h'h'}e^{-ik_hx}\phi_{m'}^{\mathcal{S}}+\psi_{h\downarrow}^{N}g_{\uparrow\downarrow}^{\prime h'h'}e^{-ik_hx}\phi_{m'+1}^{\mathcal{S}}+\psi_{h\uparrow}^{N}h_{\uparrow\uparrow}^{\prime h'h'}e^{ik_h(x-L/2)}\phi_{m'}^{\mathcal{S}}+\psi_{h\downarrow}^{N}h_{\uparrow\downarrow}^{\prime h'h'}e^{ik_h(x-L/2)}\phi_{m'+1}^{\mathcal{S}}\,, & 0<x<L/2 \\
    \tilde{t}_{\uparrow\uparrow}^{\prime h'e'}\psi_{e\uparrow}^{S_{R}}{e}^{iq_{e}(x-L/2)}\phi_{m'}^{\mathcal{S}}+\tilde{t}_{\uparrow\downarrow}^{\prime h'e'}\psi_{e\downarrow}^{S_{R}}{e}^{iq_{e}(x-L/2)}\phi_{m'+1}^{\mathcal{S}}+\tilde{t}_{\uparrow\uparrow}^{\prime h'h'}\psi_{h\uparrow}^{\prime S_{R}}{ e}^{-iq_{h}(x-L/2)}\phi_{m'}^{\mathcal{S}}+\tilde{t}_{\uparrow\downarrow}^{\prime h'h'}\psi_{h\downarrow}^{\prime S_{R}}{e}^{-iq_{h}(x-L/2)}\phi_{m'+1}^{\mathcal{S}}\,,& x>L/2
\end{cases}\\
\Phi_{4}(x)&=
\begin{cases}
    \psi_{h\downarrow}^{\prime S_{L}}{ e}^{-i{q}_{h}(x+L/2)}\phi_{m'}^{\mathcal{S}}+r_{\downarrow\uparrow}^{\prime h'e'}\psi_{e\uparrow}^{\prime S_{L}}{ e}^{-i{q}_{e}(x+L/2)}\phi_{m'-1}^{\mathcal{S}}+r_{\downarrow\downarrow}^{\prime h'e'}\psi_{e\downarrow}^{\prime S_{L}}{ e}^{-i{q}_{e}(x+L/2)}\phi_{m'}^{\mathcal{S}}+r_{\downarrow\uparrow}^{\prime h'h'}\psi_{h\uparrow}^{S_{L}}{ e}^{iq_{h}(x+L/2)}\phi_{m'-1}^{\mathcal{S}}\\+r_{\downarrow\downarrow}^{\prime h'h'}\psi_{h\downarrow}^{S_{L}}{ e}^{iq_{h}(x+L/2)}\phi_{m'}^{\mathcal{S}}\,,  & x<-L/2 \\
    \psi_{e\uparrow}^{N}a_{\downarrow\uparrow}^{\prime h'e'}e^{ik_e(x+L/2)}\phi_{m'}^{\mathcal{S}}+\psi_{e\downarrow}^{N}a_{\downarrow\downarrow}^{\prime h'e'}e^{ik_e(x+L/2)}\phi_{m'+1}^{\mathcal{S}}+\psi_{e\uparrow}^{N}b_{\downarrow\uparrow}^{\prime h'e'}e^{-ik_ex}\phi_{m'}^{\mathcal{S}}+\psi_{e\downarrow}^{N}b_{\downarrow\downarrow}^{\prime h'e'}e^{-ik_ex}\phi_{m'+1}^{\mathcal{S}}\\+\psi_{h\uparrow}^{N}c_{\downarrow\uparrow}^{\prime h'h'}e^{-ik_h(x+L/2)}\phi_{m'}^{\mathcal{S}}+\psi_{h\downarrow}^{N}c_{\downarrow\downarrow}^{\prime h'h'}e^{-ik_h(x+L/2)}\phi_{m'+1}^{\mathcal{S}}+\psi_{h\uparrow}^{N}d_{\downarrow\uparrow}^{\prime h'h'}e^{ik_hx}\phi_{m'}^{\mathcal{S}}+\psi_{h\downarrow}^{N}d_{\downarrow\downarrow}^{\prime h'h'}e^{ik_hx}\phi_{m'+1}^{\mathcal{S}}\,, & -L/2<x<0 \\
    \psi_{e\uparrow}^{N}e_{\downarrow\uparrow}^{\prime h'e'}e^{ik_ex}\phi_{m'}^{\mathcal{S}}+\psi_{e\downarrow}^{N}e_{\downarrow\downarrow}^{\prime h'e'}e^{ik_ex}\phi_{m'+1}^{\mathcal{S}}+\psi_{e\uparrow}^{N}f_{\downarrow\uparrow}^{\prime h'e'}e^{-ik_e(x-L/2)}\phi_{m'}^{\mathcal{S}}+\psi_{e\downarrow}^{N}f_{\downarrow\downarrow}^{\prime h'e'}e^{-ik_e(x-L/2)}\phi_{m'+1}^{\mathcal{S}}\\
    +\psi_{h\uparrow}^{N}g_{\downarrow\uparrow}^{\prime h'h'}e^{-ik_hx}\phi_{m'}^{\mathcal{S}}+\psi_{h\downarrow}^{N}g_{\downarrow\downarrow}^{\prime h'h'}e^{-ik_hx}\phi_{m'+1}^{\mathcal{S}}+\psi_{h\uparrow}^{N}h_{\downarrow\uparrow}^{\prime h'h'}e^{ik_h(x-L/2)}\phi_{m'}^{\mathcal{S}}+\psi_{h\downarrow}^{N}h_{\downarrow\downarrow}^{\prime h'h'}e^{ik_h(x-L/2)}\phi_{m'+1}^{\mathcal{S}}\,, & 0<x<L/2 \\
    \tilde{t}_{\downarrow\uparrow}^{\prime h'e'}\psi_{e\uparrow}^{S_{R}}{ e}^{iq_{e}(x-L/2)}\phi_{m'-1}^{\mathcal{S}}+\tilde{t}_{\downarrow\downarrow}^{\prime h'e'}\psi_{e\downarrow}^{S_{R}}{ e}^{iq_{e}(x-L/2)}\phi_{m'}^{\mathcal{S}}+\tilde{t}_{\downarrow\uparrow}^{\prime h'h'}\psi_{h\uparrow}^{\prime S_{R}}{ e}^{-iq_{h}(x-L/2)}\phi_{m'-1}^{\mathcal{S}}+\tilde{t}_{\downarrow\downarrow}^{\prime h'h'}\psi_{h\downarrow}^{\prime S_{R}}{ e}^{-iq_{h}(x-L/2)}\phi_{m'}^{\mathcal{S}}\,,& x>L/2
\end{cases}
\\
\Phi_{5}(x)&=
\begin{cases}
  t_{\uparrow\uparrow}^{\prime e'e'}\psi_{e\uparrow}^{\prime S_{L}}{ e}^{-iq_{e}(x+L/2)}\phi_{m'}^{\mathcal{S}}+t_{\uparrow\downarrow}^{\prime e'e'}\psi_{e\downarrow}^{\prime S_{L}}{ e}^{-iq_{e}(x+L/2)}\phi_{m'+1}^{\mathcal{S}}+t_{\uparrow\uparrow}^{\prime e'h'}\psi_{h\uparrow}^{S_{L}}{ e}^{i{q}_{h}(x+L/2)}\phi_{m'}^{\mathcal{S}}+t_{\uparrow\downarrow}^{\prime e'h'}\psi_{h\downarrow}^{S_{L}}{ e}^{i{q}_{h}(x+L/2)}\phi_{m'+1}^{\mathcal{S}}\,,& x<-L/2 \\
  \psi_{e\uparrow}^{N}\tilde{a}_{\uparrow\uparrow}^{\prime e'e'}e^{ik_e(x+L/2)}\phi_{m'}^{\mathcal{S}}+\psi_{e\downarrow}^{N}\tilde{a}_{\uparrow\downarrow}^{\prime e'e'}e^{ik_e(x+L/2)}\phi_{m'+1}^{\mathcal{S}}+\psi_{e\uparrow}^{N}\tilde{b}_{\uparrow\uparrow}^{\prime e'e'}e^{-ik_ex}\phi_{m'}^{\mathcal{S}}+\psi_{e\downarrow}^{N}\tilde{b}_{\uparrow\downarrow}^{\prime e'e'}e^{-ik_ex}\phi_{m'+1}^{\mathcal{S}}\\+\psi_{h\uparrow}^{N}\tilde{c}_{\uparrow\uparrow}^{\prime e'h'}e^{-ik_h(x+L/2)}\phi_{m'}^{\mathcal{S}}+\psi_{h\downarrow}^{N}\tilde{c}_{\uparrow\downarrow}^{\prime e'h'}e^{-ik_h(x+L/2)}\phi_{m'+1}^{\mathcal{S}}+\psi_{h\uparrow}^{N}\tilde{d}_{\uparrow\uparrow}^{\prime e'h'}e^{ik_hx}\phi_{m'}^{\mathcal{S}}+\psi_{h\downarrow}^{N}\tilde{d}_{\uparrow\downarrow}^{\prime e'h'}e^{ik_hx}\phi_{m'+1}^{\mathcal{S}}\,, & -L/2<x<0 \\
  \psi_{e\uparrow}^{N}\tilde{e}_{\uparrow\uparrow}^{\prime e'e'}e^{ik_ex}\phi_{m'}^{\mathcal{S}}+\psi_{e\downarrow}^{N}\tilde{e}_{\uparrow\downarrow}^{\prime e'e'}e^{ik_ex}\phi_{m'+1}^{\mathcal{S}}+\psi_{e\uparrow}^{N}\tilde{f}_{\uparrow\uparrow}^{\prime e'e'}e^{-ik_e(x-L/2)}\phi_{m'}^{\mathcal{S}}+\psi_{e\downarrow}^{N}\tilde{f}_{\uparrow\downarrow}^{\prime e'e'}e^{-ik_e(x-L/2)}\phi_{m'+1}^{\mathcal{S}}\\
    +\psi_{h\uparrow}^{N}\tilde{g}_{\uparrow\uparrow}^{\prime e'h'}e^{-ik_hx}\phi_{m'}^{\mathcal{S}}+\psi_{h\downarrow}^{N}\tilde{g}_{\uparrow\downarrow}^{\prime e'h'}e^{-ik_hx}\phi_{m'+1}^{\mathcal{S}}+\psi_{h\uparrow}^{N}\tilde{h}_{\uparrow\uparrow}^{\prime e'h'}e^{ik_h(x-L/2)}\phi_{m'}^{\mathcal{S}}+\psi_{h\downarrow}^{N}\tilde{h}_{\uparrow\downarrow}^{\prime e'h'}e^{ik_h(x-L/2)}\phi_{m'+1}^{\mathcal{S}}\,, & 0<x<L/2 \\
   \psi_{e\uparrow}^{\prime S_{R}}{ e}^{-iq_{e}(x-L/2)}\phi_{m'}^{\mathcal{S}}+\tilde{r}_{\uparrow\uparrow}^{\prime e'e'}\psi_{e\uparrow}^{S_{R}}{ e}^{iq_{e}(x-L/2)}\phi_{m'}^{\mathcal{S}}+\tilde{r}_{\uparrow\downarrow}^{\prime e'e'}\psi_{e\downarrow}^{S_{R}}{ e}^{iq_{e}(x-L/2)}\phi_{m'+1}^{\mathcal{S}}+\tilde{r}_{\uparrow\uparrow}^{\prime e'h'}\psi_{h\uparrow}^{\prime S_{R}}{ e}^{-iq_{h}(x-L/2)}\phi_{m'}^{\mathcal{S}}\\+\tilde{r}_{\uparrow\downarrow}^{\prime e'h'}\psi_{h\downarrow}^{\prime S_{R}}{ e}^{-iq_{h}(x-L/2)}\phi_{m'+1}^{\mathcal{S}}\,,& x>L/2
\end{cases}
\\
\Phi_{6}(x)&=
\begin{cases}
  t_{\downarrow\uparrow}^{\prime e'e'}\psi_{e\uparrow}^{\prime S_{L}}{ e}^{-iq_{e}(x+L/2)}\phi_{m'-1}^{\mathcal{S}}+t_{\downarrow\downarrow}^{\prime e'e'}\psi_{e\downarrow}^{\prime S_{L}}{ e}^{-iq_{e}(x+L/2)}\phi_{m'}^{\mathcal{S}}+t_{\downarrow\uparrow}^{\prime e'h'}\psi_{h\uparrow}^{S_{L}}{ e}^{i{q}_{h}(x+L/2)}\phi_{m'-1}^{\mathcal{S}}+t_{\downarrow\downarrow}^{\prime e'h'}\psi_{h\downarrow}^{S_{L}}{ e}^{i{q}_{h}(x+L/2)}\phi_{m'}^{\mathcal{S}}\,, & x<-L/2 \\
  \psi_{e\uparrow}^{N}\tilde{a}_{\downarrow\uparrow}^{\prime e'e'}e^{ik_e(x+L/2)}\phi_{m'}^{\mathcal{S}}+\psi_{e\downarrow}^{N}\tilde{a}_{\downarrow\downarrow}^{\prime e'e'}e^{ik_e(x+L/2)}\phi_{m'+1}^{\mathcal{S}}+\psi_{e\uparrow}^{N}\tilde{b}_{\downarrow\uparrow}^{\prime e'e'}e^{-ik_ex}\phi_{m'}^{\mathcal{S}}+\psi_{e\downarrow}^{N}\tilde{b}_{\downarrow\downarrow}^{\prime e'e'}e^{-ik_ex}\phi_{m'+1}^{\mathcal{S}}\\+\psi_{h\uparrow}^{N}\tilde{c}_{\downarrow\uparrow}^{\prime e'h'}e^{-ik_h(x+L/2)}\phi_{m'}^{\mathcal{S}}+\psi_{h\downarrow}^{N}\tilde{c}_{\downarrow\downarrow}^{\prime e'h'}e^{-ik_h(x+L/2)}\phi_{m'+1}^{\mathcal{S}}+\psi_{h\uparrow}^{N}\tilde{d}_{\downarrow\uparrow}^{\prime e'h'}e^{ik_hx}\phi_{m'}^{\mathcal{S}}+\psi_{h\downarrow}^{N}\tilde{d}_{\downarrow\downarrow}^{\prime e'h'}e^{ik_hx}\phi_{m'+1}^{\mathcal{S}}\,, & -L/2<x<0 \\
  \psi_{e\uparrow}^{N}\tilde{e}_{\downarrow\uparrow}^{\prime e'e'}e^{ik_ex}\phi_{m'}^{\mathcal{S}}+\psi_{e\downarrow}^{N}\tilde{e}_{\downarrow\downarrow}^{\prime e'e'}e^{ik_ex}\phi_{m'+1}^{\mathcal{S}}+\psi_{e\uparrow}^{N}\tilde{f}_{\downarrow\uparrow}^{\prime e'e'}e^{-ik_e(x-L/2)}\phi_{m'}^{\mathcal{S}}+\psi_{e\downarrow}^{N}\tilde{f}_{\downarrow\downarrow}^{\prime e'e'}e^{-ik_e(x-L/2)}\phi_{m'+1}^{\mathcal{S}}\\
    +\psi_{h\uparrow}^{N}\tilde{g}_{\downarrow\uparrow}^{\prime e'h'}e^{-ik_hx}\phi_{m'}^{\mathcal{S}}+\psi_{h\downarrow}^{N}\tilde{g}_{\downarrow\downarrow}^{\prime e'h'}e^{-ik_hx}\phi_{m'+1}^{\mathcal{S}}+\psi_{h\uparrow}^{N}\tilde{h}_{\downarrow\uparrow}^{\prime e'h'}e^{ik_h(x-L/2)}\phi_{m'}^{\mathcal{S}}+\psi_{h\downarrow}^{N}\tilde{h}_{\downarrow\downarrow}^{\prime e'h'}e^{ik_h(x-L/2)}\phi_{m'+1}^{\mathcal{S}}\,, & 0<x<L/2 \\
   \psi_{e\downarrow}^{\prime S_{R}}{ e}^{-iq_{e}(x-L/2)}\phi_{m'}^{\mathcal{S}}+\tilde{r}_{\downarrow\uparrow}^{\prime e'e'}\psi_{e\uparrow}^{S_{R}}{ e}^{iq_{e}(x-L/2)}\phi_{m'-1}^{\mathcal{S}}+\tilde{r}_{\downarrow\downarrow}^{\prime e'e'}\psi_{e\downarrow}^{S_{R}}{ e}^{iq_{e}(x-L/2)}\phi_{m'}^{\mathcal{S}}+\tilde{r}_{\downarrow\uparrow}^{\prime e'h'}\psi_{h\uparrow}^{\prime S_{R}}{ e}^{-iq_{h}(x-L/2)}\phi_{m'-1}^{\mathcal{S}}\\+\tilde{r}_{\downarrow\downarrow}^{\prime e'h'}\psi_{h\downarrow}^{\prime S_{R}}{ e}^{-iq_{h}(x-L/2)}\phi_{m'}^{\mathcal{S}}\,,& x>L/2
\end{cases}\\
\Phi_{7}(x)&=
\begin{cases}
  t_{\uparrow\uparrow}^{\prime h'e'}\psi_{e\uparrow}^{\prime S_{L}}{ e}^{-i{q}_{e}(x+L/2)}\phi_{m'}^{\mathcal{S}}+t_{\uparrow\downarrow}^{\prime h'e'}\psi_{e\downarrow}^{\prime S_{L}}{ e}^{-i{q}_{e}(x+L/2)}\phi_{m'+1}^{\mathcal{S}}+t_{\uparrow\uparrow}^{\prime h'h'}\psi_{h\uparrow}^{S_{L}}{ e}^{iq_{h}(x+L/2)}\phi_{m'}^{\mathcal{S}}+t_{\uparrow\downarrow}^{\prime h'h'}\psi_{h\downarrow}^{S_{L}}{ e}^{iq_{h}(x+L/2)}\phi_{m'+1}^{\mathcal{S}}\,,  & x<-L/2 \\
  \psi_{e\uparrow}^{N}\tilde{a}_{\uparrow\uparrow}^{\prime h'e'}e^{ik_e(x+L/2)}\phi_{m'}^{\mathcal{S}}+\psi_{e\downarrow}^{N}\tilde{a}_{\uparrow\downarrow}^{\prime h'e'}e^{ik_e(x+L/2)}\phi_{m'+1}^{\mathcal{S}}+\psi_{e\uparrow}^{N}\tilde{b}_{\uparrow\uparrow}^{\prime h'e'}e^{-ik_ex}\phi_{m'}^{\mathcal{S}}+\psi_{e\downarrow}^{N}\tilde{b}_{\uparrow\downarrow}^{\prime h'e'}e^{-ik_ex}\phi_{m'+1}^{\mathcal{S}}\\+\psi_{h\uparrow}^{N}\tilde{c}_{\uparrow\uparrow}^{\prime h'h'}e^{-ik_h(x+L/2)}\phi_{m'}^{\mathcal{S}}+\psi_{h\downarrow}^{N}\tilde{c}_{\uparrow\downarrow}^{\prime h'h'}e^{-ik_h(x+L/2)}\phi_{m'+1}^{\mathcal{S}}+\psi_{h\uparrow}^{N}\tilde{d}_{\uparrow\uparrow}^{\prime h'h'}e^{ik_hx}\phi_{m'}^{\mathcal{S}}+\psi_{h\downarrow}^{N}\tilde{d}_{\uparrow\downarrow}^{\prime h'h'}e^{ik_hx}\phi_{m'+1}^{\mathcal{S}}\,, & -L/2<x<0 \\
  \psi_{e\uparrow}^{N}\tilde{e}_{\uparrow\uparrow}^{\prime h'e'}e^{ik_ex}\phi_{m'}^{\mathcal{S}}+\psi_{e\downarrow}^{N}\tilde{e}_{\uparrow\downarrow}^{\prime h'e'}e^{ik_ex}\phi_{m'+1}^{\mathcal{S}}+\psi_{e\uparrow}^{N}\tilde{f}_{\uparrow\uparrow}^{\prime h'e'}e^{-ik_e(x-L/2)}\phi_{m'}^{\mathcal{S}}+\psi_{e\downarrow}^{N}\tilde{f}_{\uparrow\downarrow}^{\prime h'e'}e^{-ik_e(x-L/2)}\phi_{m'+1}^{\mathcal{S}}\\
    +\psi_{h\uparrow}^{N}\tilde{g}_{\uparrow\uparrow}^{\prime h'h'}e^{-ik_hx}\phi_{m'}^{\mathcal{S}}+\psi_{h\downarrow}^{N}\tilde{g}_{\uparrow\downarrow}^{\prime h'h'}e^{-ik_hx}\phi_{m'+1}^{\mathcal{S}}+\psi_{h\uparrow}^{N}\tilde{h}_{\uparrow\uparrow}^{\prime h'h'}e^{ik_h(x-L/2)}\phi_{m'}^{\mathcal{S}}+\psi_{h\downarrow}^{N}\tilde{h}_{\uparrow\downarrow}^{\prime h'h'}e^{ik_h(x-L/2)}\phi_{m'+1}^{\mathcal{S}}\,, & 0<x<L/2 \\
   \psi_{h\uparrow}^{S_{R}}{ e}^{iq_{h}(x-L/2)}\phi_{m'}^{\mathcal{S}}+\tilde{r}_{\uparrow\uparrow}^{\prime h'e'}\psi_{e\uparrow}^{S_{R}}{ e}^{iq_{e}(x-L/2)}\phi_{m'}^{\mathcal{S}}+\tilde{r}_{\uparrow\downarrow}^{\prime h'e'}\psi_{e\downarrow}^{S_{R}}{ e}^{iq_{e}(x-L/2)}\phi_{m'+1}^{\mathcal{S}}+\tilde{r}_{\uparrow\uparrow}^{\prime h'h'}\psi_{h\uparrow}^{\prime S_{R}}{ e}^{-iq_{h}(x-L/2)}\phi_{m'}^{\mathcal{S}}\\+\tilde{r}_{\uparrow\downarrow}^{\prime h'h'}\psi_{h\downarrow}^{\prime S_{R}}{ e}^{-iq_{h}(x-L/2)}\phi_{m'+1}^{\mathcal{S}}\,,& x>L/2
\end{cases}
\\
\Phi_{8}(x)&=
\begin{cases}
  t_{\downarrow\uparrow}^{\prime h'e'}\psi_{e\uparrow}^{\prime S_{L}}{ e}^{-i{q}_{e}(x+L/2)}\phi_{m'-1}^{\mathcal{S}}+t_{\downarrow\downarrow}^{\prime h'e'}\psi_{e\downarrow}^{\prime S_{L}}{ e}^{-i{q}_{e}(x+L/2)}\phi_{m'}^{\mathcal{S}}+t_{\downarrow\uparrow}^{\prime h'h'}\psi_{h\uparrow}^{S_{L}}{ e}^{iq_{h}(x+L/2)}\phi_{m'-1}^{\mathcal{S}}+t_{\downarrow\downarrow}^{\prime h'h'}\psi_{h\downarrow}^{S_{L}}{ e}^{iq_{h}(x+L/2)}\phi_{m'}^{\mathcal{S}}\,,  & x<-L/2 \\
  \psi_{e\uparrow}^{N}\tilde{a}_{\downarrow\uparrow}^{\prime h'e'}e^{ik_e(x+L/2)}\phi_{m'}^{\mathcal{S}}+\psi_{e\downarrow}^{N}\tilde{a}_{\downarrow\downarrow}^{\prime h'e'}e^{ik_e(x+L/2)}\phi_{m'+1}^{\mathcal{S}}+\psi_{e\uparrow}^{N}\tilde{b}_{\downarrow\uparrow}^{\prime h'e'}e^{-ik_ex}\phi_{m'}^{\mathcal{S}}+\psi_{e\downarrow}^{N}\tilde{b}_{\downarrow\downarrow}^{\prime h'e'}e^{-ik_ex}\phi_{m'+1}^{\mathcal{S}}\\+\psi_{h\uparrow}^{N}\tilde{c}_{\downarrow\uparrow}^{\prime h'h'}e^{-ik_h(x+L/2)}\phi_{m'}^{\mathcal{S}}+\psi_{h\downarrow}^{N}\tilde{c}_{\downarrow\downarrow}^{\prime h'h'}e^{-ik_h(x+L/2)}\phi_{m'+1}^{\mathcal{S}}+\psi_{h\uparrow}^{N}\tilde{d}_{\downarrow\uparrow}^{\prime h'h'}e^{ik_hx}\phi_{m'}^{\mathcal{S}}+\psi_{h\downarrow}^{N}\tilde{d}_{\downarrow\downarrow}^{\prime h'h'}e^{ik_hx}\phi_{m'+1}^{\mathcal{S}}\,, & -L/2<x<0 \\
  \psi_{e\uparrow}^{N}\tilde{e}_{\downarrow\uparrow}^{\prime h'e'}e^{ik_ex}\phi_{m'}^{\mathcal{S}}+\psi_{e\downarrow}^{N}\tilde{e}_{\downarrow\downarrow}^{\prime h'e'}e^{ik_ex}\phi_{m'+1}^{\mathcal{S}}+\psi_{e\uparrow}^{N}\tilde{f}_{\downarrow\uparrow}^{\prime h'e'}e^{-ik_e(x-L/2)}\phi_{m'}^{\mathcal{S}}+\psi_{e\downarrow}^{N}\tilde{f}_{\downarrow\downarrow}^{\prime h'e'}e^{-ik_e(x-L/2)}\phi_{m'+1}^{\mathcal{S}}\\
    +\psi_{h\uparrow}^{N}\tilde{g}_{\downarrow\uparrow}^{\prime h'h'}e^{-ik_hx}\phi_{m'}^{\mathcal{S}}+\psi_{h\downarrow}^{N}\tilde{g}_{\downarrow\downarrow}^{\prime h'h'}e^{-ik_hx}\phi_{m'+1}^{\mathcal{S}}+\psi_{h\uparrow}^{N}\tilde{h}_{\downarrow\uparrow}^{\prime h'h'}e^{ik_h(x-L/2)}\phi_{m'}^{\mathcal{S}}+\psi_{h\downarrow}^{N}\tilde{h}_{\downarrow\downarrow}^{\prime h'h'}e^{ik_h(x-L/2)}\phi_{m'+1}^{\mathcal{S}}\,, & 0<x<L/2 \\
   \psi_{h\downarrow}^{S_{R}}{ e}^{iq_{h}(x-L/2)}\phi_{m'}^{\mathcal{S}}+\tilde{r}_{\downarrow\uparrow}^{\prime h'e'}\psi_{e\uparrow}^{S_{R}}{ e}^{iq_{e}(x-L/2)}\phi_{m'-1}^{\mathcal{S}}+\tilde{r}_{\downarrow\downarrow}^{\prime h'e'}\psi_{e\downarrow}^{S_{R}}{ e}^{iq_{e}(x-L/2)}\phi_{m'}^{\mathcal{S}}+\tilde{r}_{\downarrow\uparrow}^{\prime h'h'}\psi_{h\uparrow}^{\prime S_{R}}{ e}^{-iq_{h}(x-L/2)}\phi_{m'-1}^{\mathcal{S}}\\+\tilde{r}_{\downarrow\downarrow}^{\prime h'h'}\psi_{h\downarrow}^{\prime S_{R}}{ e}^{-iq_{h}(x-L/2)}\phi_{m'}^{\mathcal{S}}\,,& x>L/2
\end{cases}
\end{split}
\label{waveJF}
\end{equation}
where the expressions for $\psi_{e\uparrow}^{S_{L(R)}}$, $\psi_{e\downarrow}^{S_{L(R)}}$, $\psi_{h\uparrow}^{S_{L(R)}}$, $\psi_{h\downarrow}^{S_{L(R)}}$, $\psi_{e\uparrow}^{\prime S_{L(R)}}$, $\psi_{e\downarrow}^{\prime S_{L(R)}}$, $\psi_{h\uparrow}^{\prime S_{L(R)}}$, $\psi_{h\downarrow}^{\prime S_{L(R)}}$ are mentioned below Eq.~\eqref{waveJ} and, $\psi_{e\uparrow}^{N}=\begin{pmatrix}
                        1\\
                        0\\
                        0\\
                        0
                       \end{pmatrix}$, $\psi_{e\downarrow}^{N}=\begin{pmatrix}
                        0\\
                        1\\
                        0\\
                        0
                       \end{pmatrix}$, $\psi_{h\uparrow}^{N}=\begin{pmatrix}
                        0\\
                        0\\
                        0\\
                        1
                       \end{pmatrix}$, $\psi_{h\downarrow}^{N}=\begin{pmatrix}
                        0\\
                        0\\
                        1\\
                        0
                       \end{pmatrix}$.
$r_{ij}^{\prime nn}$ and $\tilde{r}_{ij}^{\prime nn}$ represent the reflection amplitudes in left $p_{x}$ superconductor and right $p_{x}$ superconductor respectively, while $t_{ij}^{\prime nn}$ and $\tilde{t}_{ij}^{\prime nn}$ represent the transmission amplitudes in left $p_{x}$ superconductor and right $p_{x}$ superconductor respectively with $i,j\in{\{\uparrow,\downarrow\}}$ and $n\in{\{e',h'\}}$. $k_{e,h}=\sqrt{\frac{2m^{*}}{\hbar^2}(\mu_N^{\prime}\pm\nu)}$ represent the wavevectors in normal metal. For $|\nu|\ll\mu_N^{\prime}$, $k_{e,h}$ can be written as $k_{e,h}=k_{\mu_{N}}\pm\frac{\nu}{\xi k_{\mu_{N}}\Delta_{p_x}^{\prime}}$, where $k_{\mu_N}=\sqrt{\frac{2m^{*}\mu_{N}^{\prime}}{\hbar^2}}$.
After diagonalizing the Hamiltonian $(H_{BdG}^{\mbox{$p_{x}$-$I_1$-$N_1$-SF-$N_2$-$I_2$-$p_{x}$}})^{*}(-k)$ instead of $H_{BdG}^{\mbox{$p_{x}$-$I_1$-$N_1$-SF-$N_2$-$I_2$-$p_{x}$}}(k)$, we will obtain the conjugate process $\tilde{\Phi}_{i}$ necessary to construct the retarded Green's function in next section. We find that $\tilde{\psi}_{e\uparrow}^{N}=\psi_{e\uparrow}^{N}$, $\tilde{\psi}_{e\downarrow}^{N}=\psi_{e\downarrow}^{N}$, $\tilde{\psi}_{h\uparrow}^{N}=\psi_{h\uparrow}^{N}$, $\tilde{\psi}_{h\downarrow}^{N}=\psi_{h\downarrow}^{N}$ and the expressions for $\tilde{\psi}_{e\uparrow}^{S_{L(R)}}$, $\tilde{\psi}_{e\downarrow}^{S_{L(R)}}$, $\tilde{\psi}_{h\uparrow}^{S_{L(R)}}$, $\tilde{\psi}_{h\downarrow}^{S_{L(R)}}$, $\tilde{\psi}_{e\uparrow}^{\prime S_{L(R)}}$, $\tilde{\psi}_{e\downarrow}^{\prime S_{L(R)}}$, $\tilde{\psi}_{h\uparrow}^{\prime S_{L(R)}}$, $\tilde{\psi}_{h\downarrow}^{\prime S_{L(R)}}$ are the same as those for the $p_x$-SF-$p_x$ junction.}

{At the $p_x$-$N_1$ interface ($x=-L/2$), the boundary conditions are:
\begin{equation}
\label{bc11}
\Phi_{l}|_{x<-L/2}=\Phi_{l}|_{-L/2<x<0}\,\,\, \mbox{and,}\,\,\, \frac{d\Phi_{l}|_{-L/2<x<0}}{dx}-\frac{d\Phi_{l}|_{x<-L/2}}{dx}=\frac{im^{*}\Delta_{p_{x}}'}{\hbar^2}\begin{bmatrix}
0 & \sigma_{x}\\
-\sigma_{x} & 0                                                                                                                              \end{bmatrix}\Phi_{l}|_{x=-L/2}+\frac{2m^{*}V_{1}}{\hbar^2}\Phi_{l}|_{x=-L/2},(l=1,2,...,8).
\end{equation}
Similarly, at the $N_1$-$N_2$ interface ($x=0$), the boundary conditions are:
\begin{equation}
\label{bc12}
\Phi_{l}|_{-L/2<x<0}=\Phi_{l}|_{0<x<L/2}\,\,\, \mbox{and,}\,\,\, \frac{d\Phi_{l}|_{0<x<L/2}}{dx}-\frac{d\Phi_{l}|_{-L/2<x<0}}{dx}=-\frac{2m^{*}J_{0}}{\hbar^2}\vec{s}.\vec{\mathcal{S}}\Phi_{l}|_{x=0},(l=1,2,...,8).
\end{equation}
Finally, at the $N_2$-$p_x$ interface ($x=L/2$), the boundary conditions are:
\begin{equation}
\label{bc13}
\Phi_{l}|_{0<x<L/2}=\Phi_{l}|_{x>L/2}\,\,\, \mbox{and,}\,\,\, \frac{d\Phi_{l}|_{x>L/2}}{dx}-\frac{d\Phi_{l}|_{0<x<L/2}}{dx}=\frac{im^{*}\Delta_{p_{x}}'}{\hbar^2}\begin{bmatrix}
0 & -e^{i\varphi}\sigma_{x}\\
e^{-i\varphi}\sigma_{x} & 0                                                                                                                              \end{bmatrix}\Phi_{l}|_{x=L/2}+\frac{2m^{*}V_{2}}{\hbar^2}\Phi_{l}|_{x=L/2},(l=1,2,...,8).
\end{equation}
For wave functions involving electron with spin up and spin down, actions of $\vec{s}.\vec{\mathcal{S}}$ are
\begin{equation}
\vec{s}.\vec{\mathcal{S}}\psi_{e\uparrow}^{N}\varphi_{m'}^{\mathcal{S}}=\frac{\hbar^2m'}{2}\psi_{e\uparrow}^{N}\varphi_{m'}^{\mathcal{S}}+\frac{\hbar^2f}{2}\psi_{e\downarrow}^{N}\varphi_{m'+1}^{\mathcal{S}},\,\,\,\mbox{and,}\,\,\,\,
\label{eu}
\vec{s}.\vec{\mathcal{S}} \psi_{e\downarrow}^{N}\varphi_{m'}^{\mathcal{S}}=-\frac{\hbar^2m'}{2}\psi_{e\downarrow}^{N}\varphi_{m'}^{\mathcal{S}}+\frac{\hbar^2f'}{2}\psi_{e\uparrow}^{N}\varphi_{m'-1}^{\mathcal{S}}.
\end{equation}
Similarly, for wave functions involving hole with spin up and spin down, actions of $\vec{s}.\vec{\mathcal{S}}$ are
\begin{equation}
\vec{s}.\vec{\mathcal{S}}\psi_{h\uparrow}^{N}\varphi_{m'}^{\mathcal{S}}=\frac{\hbar^2m'}{2}\psi_{h\uparrow}^{N}\varphi_{m'}^{\mathcal{S}}+\frac{\hbar^2f}{2}\psi_{h\downarrow}^{N}\varphi_{m'+1}^{\mathcal{S}},\,\,\,\mbox{and,}\,\,\,\,
\vec{s}.\vec{\mathcal{S}}\psi_{h\downarrow}^{N}\varphi_{m'}^{\mathcal{S}}=-\frac{\hbar^2m'}{2}\psi_{h\downarrow}^{N}\varphi_{m'}^{\mathcal{S}}+\frac{\hbar^2f'}{2}\psi_{h\uparrow}^{N}\varphi_{m'-1}^{\mathcal{S}}.
\end{equation}
By employing the equations above and solving the boundary conditions (Eqs.~\eqref{bc11}-\eqref{bc13}), we obtain a set of $24$ equations for scattering processes, as shown in Eq.~\eqref{waveJF}. Different scattering amplitudes $r_{ij}^{\prime nn}$, $\tilde{r}_{ij}^{\prime nn}$, $\tilde{t}_{ij}^{\prime nn}$, $t_{ij}^{\prime nn}$ for each kind of incident quasiparticle (electron-like/hole-like) are obtained from these $24$ equations.}
\section*{Appendix B: Green's functions}
In this Appendix, we present the detailed calculations of Green's functions. Following Refs.~\cite{cay1} and \cite{cay2}, the retarded Green's functions $\mathcal{G}^{r}(x,\bar{x},\nu)$ can be expressed as-
\begin{equation}
\label{RGF}
\begin{split}
\mathcal{G}^{r}(x,\bar{x},\nu)=
\begin{cases}
\Phi_{1}(x)[\alpha_{11}\tilde{\Phi}_{5}^{T}(\bar{x})+\alpha_{12}\tilde{\Phi}_{6}^{T}(\bar{x})+\alpha_{13}\tilde{\Phi}_{7}^{T}(\bar{x})+\alpha_{14}\tilde{\Phi}_{8}^{T}(\bar{x})]\\
+
\Phi_{2}(x)[\alpha_{21}\tilde{\Phi}_{5}^{T}(\bar{x})+\alpha_{22}\tilde{\Phi}_{6}^{T}(\bar{x})+\alpha_{23}\tilde{\Phi}_{7}^{T}(\bar{x})+\alpha_{24}\tilde{\Phi}_{8}^{T}(\bar{x})]\\
+
\Phi_{3}(x)[\alpha_{31}\tilde{\Phi}_{5}^{T}(\bar{x})+\alpha_{32}\tilde{\Phi}_{6}^{T}(\bar{x})+\alpha_{33}\tilde{\Phi}_{7}^{T}(\bar{x})+\alpha_{34}\tilde{\Phi}_{8}^{T}(\bar{x})]\\
+
\Phi_{4}(x)[\alpha_{41}\tilde{\Phi}_{5}^{T}(\bar{x})+\alpha_{42}\tilde{\Phi}_{6}^{T}(\bar{x})+\alpha_{43}\tilde{\Phi}_{7}^{T}(\bar{x})+\alpha_{44}\tilde{\Phi}_{8}^{T}(\bar{x})]
\,,\quad x>\bar{x}&\\
\Phi_{5}(x)[\beta_{11}\tilde{\Phi}_{1}^{T}(\bar{x})+\beta_{12}\tilde{\Phi}_{2}^{T}(\bar{x})+\beta_{13}\tilde{\Phi}_{3}^{T}(\bar{x})+\beta_{14}\tilde{\Phi}_{4}^{T}(\bar{x})]\\
+\Phi_{6}(x)[\beta_{21}\tilde{\Phi}_{1}^{T}(\bar{x})+\beta_{22}\tilde{\Phi}_{2}^{T}(\bar{x})+\beta_{23}\tilde{\Phi}_{3}^{T}(\bar{x})+\beta_{24}\tilde{\Phi}_{4}^{T}(\bar{x})]\\
+\Phi_{7}(x)[\beta_{31}\tilde{\Phi}_{1}^{T}(\bar{x})+\beta_{32}\tilde{\Phi}_{2}^{T}(\bar{x})+\beta_{33}\tilde{\Phi}_{3}^{T}(\bar{x})+\beta_{34}\tilde{\Phi}_{4}^{T}(\bar{x})]\\
+\Phi_{8}(x)[\beta_{41}\tilde{\Phi}_{1}^{T}(\bar{x})+\beta_{42}\tilde{\Phi}_{2}^{T}(\bar{x})+\beta_{43}\tilde{\Phi}_{3}^{T}(\bar{x})+\beta_{44}\tilde{\Phi}_{4}^{T}(\bar{x})]\,, \quad x<\bar{x}.&
\end{cases}
\end{split}
\end{equation}
In Eq.~\eqref{RGF}, $\alpha_{ij}$ and $\beta_{mn}$ ($i,j\in1,...4$ and $m,n\in1,...4$) are determined from the equation of motion of the Green's function,
\begin{equation}
[\nu-H_{BdG}^{\mbox{$p_{x}$-SF-$p_{x}$}}(x)]\mathcal{G}^{r}(x,\bar{x},\nu)=\delta(x-\bar{x}).
\label{rgf1}
\end{equation}
If we integrate Eq.~\eqref{rgf1} with respect to $x$ over the vicinity of $x=\bar{x}$, we get,
\begin{equation}
\label{conditionGRSO}
\begin{split}
[\mathcal{G}^{r}(x>\bar{x})]_{x=\bar{x}}=[\mathcal{G}^{r}(x<\bar{x})]_{x=\bar{x}},\,\,\,
[\frac{d}{dx}\mathcal{G}^{r}(x>\bar{x})]_{x=\bar{x}}-[\frac{d}{dx}\mathcal{G}^{r}(x<\bar{x})]_{x=\bar{x}}=\eta\tau_{z}\sigma_{0},
\end{split}
\end{equation}
where $\eta=\frac{2m^{*}}{\hbar^2}$. $\mathcal{G}^{r}$ is represented as,
\begin{equation}
\label{GF}
\mathcal{G}^{r}(x,\bar{x},\nu)=
\begin{bmatrix}
\mathcal{G}^{r}_{ee}&\mathcal{G}^{r}_{eh}\\
\mathcal{G}^{r}_{he}&\mathcal{G}^{r}_{hh}
\end{bmatrix},
\end{equation}
where $\mathcal{G}^{r}_{ee}$, $\mathcal{G}^{r}_{eh}$, $\mathcal{G}^{r}_{he}$, $\mathcal{G}^{r}_{hh}$ are matrices. The {normal and anomalous components} of the $\mathcal{G}^{r}$, necessary for calculating {LDOS/LMDOS and} pairing amplitudes when spin-flip scattering is considered, are given by
\begin{equation}
\label{greh}
{\mathcal{G}^{r}_{ee}=\begin{bmatrix}
                      [\mathcal{G}^{r}_{ee}]_{\uparrow\uparrow} &[\mathcal{G}^{r}_{ee}]_{\uparrow\downarrow}\\
                      [\mathcal{G}^{r}_{ee}]_{\downarrow\uparrow}&[\mathcal{G}^{r}_{ee}]_{\downarrow\downarrow}
                      \end{bmatrix}\,\,\,\, \mbox{and,}}\,\,\,\,
\mathcal{G}^{r}_{eh}=\begin{bmatrix}
                      [\mathcal{G}^{r}_{eh}]_{\uparrow\downarrow} &[\mathcal{G}^{r}_{eh}]_{\uparrow\uparrow}\\
                      [\mathcal{G}^{r}_{eh}]_{\downarrow\downarrow}&[\mathcal{G}^{r}_{eh}]_{\downarrow\uparrow}
                     \end{bmatrix}\,\,\,\mbox{{respectively}}.
\end{equation}
We calculate EST and MST correlations using $\mathcal{G}_{eh}^{r}$. Retarded Green's functions $\mathcal{G}^{r}$ are derived by substituting the wavefunctions from Eq.~\eqref{waveJ} {(Eq.~\eqref{waveJF})} into Eq.~\eqref{RGF} in case of {$p_x$-SF-$p_x$ ($p_x$-$I_1$-$N_1$-SF-$N_2$-$I_2$-$p_x$) junction,} with $r_{ij}^{nn}$ {($r_{ij}^{\prime nn}$)} determined using Eq.~\eqref{bc1} {(Eqs.~\eqref{bc11}-\eqref{bc13}).} In {case of $p_x$-SF-$p_x$} junction, in the trivial regime for {$\mathcal{G}_{ee}^{r}$ and} $\mathcal{G}_{eh}^{r}$ we get,
\begin{equation}
\begin{split}
{[\mathcal{G}^{r}_{ee}]_{\uparrow\uparrow}}&{=\frac{i\eta}{2(q_{h}\gamma_h-q_{e}\gamma_e)}\Big(\gamma_ee^{iq_{e}|x-\bar{x}|}+\gamma_he^{-iq_{h}|x-\bar{x}|}-r_{\uparrow\uparrow}^{e'e'}\gamma_ee^{-iq_e(x+\bar{x})}-r_{\uparrow\uparrow}^{h'h'}\gamma_he^{iq_h(x+\bar{x})}+r_{\uparrow\uparrow}^{e'h'}\gamma_h\sqrt{\frac{1-\gamma_e^2}{1-\gamma_h^2}}e^{i(q_{h}x-q_{e}\bar{x})}}\\&{+r_{\uparrow\uparrow}^{h'e'}\gamma_e\sqrt{\frac{1-\gamma_h^2}{1-\gamma_e^2}}e^{-i(q_{e}x-q_{h}\bar{x})}\Big),}\\
{[\mathcal{G}^{r}_{ee}]_{\downarrow\downarrow}}&{=\frac{i\eta}{2(q_{h}\gamma_h-q_{e}\gamma_e)}\Big(\gamma_ee^{iq_{e}|x-\bar{x}|}+\gamma_he^{-iq_{h}|x-\bar{x}|}-r_{\downarrow\downarrow}^{e'e'}\gamma_ee^{-iq_e(x+\bar{x})}-r_{\downarrow\downarrow}^{h'h'}\gamma_he^{iq_h(x+\bar{x})}+r_{\downarrow\downarrow}^{e'h'}\gamma_h\sqrt{\frac{1-\gamma_e^2}{1-\gamma_h^2}}e^{i(q_{h}x-q_{e}\bar{x})}}\\&{+r_{\downarrow\downarrow}^{h'e'}\gamma_e\sqrt{\frac{1-\gamma_h^2}{1-\gamma_e^2}}e^{-i(q_{e}x-q_{h}\bar{x})}\Big),}\\
{[\mathcal{G}^{r}_{ee}]_{\uparrow\downarrow}}&{=\frac{i\eta}{2(q_{h}\gamma_h-q_{e}\gamma_e)}\Big(-r_{\downarrow\uparrow}^{e'e'}\gamma_ee^{-iq_e(x+\bar{x})}-r_{\downarrow\uparrow}^{h'h'}\gamma_he^{iq_h(x+\bar{x})}+r_{\downarrow\uparrow}^{e'h'}\gamma_h\sqrt{\frac{1-\gamma_e^2}{1-\gamma_h^2}}e^{i(q_{h}x-q_{e}\bar{x})}+r_{\downarrow\uparrow}^{h'e'}\gamma_e\sqrt{\frac{1-\gamma_h^2}{1-\gamma_e^2}}}\\&{\times e^{-i(q_{e}x-q_{h}\bar{x})}\Big),}\\
{[\mathcal{G}^{r}_{ee}]_{\downarrow\uparrow}}&{=\frac{i\eta}{2(q_{h}\gamma_h-q_{e}\gamma_e)}\Big(-r_{\uparrow\downarrow}^{e'e'}\gamma_ee^{-iq_e(x+\bar{x})}-r_{\uparrow\downarrow}^{h'h'}\gamma_he^{iq_h(x+\bar{x})}+r_{\uparrow\downarrow}^{e'h'}\gamma_h\sqrt{\frac{1-\gamma_e^2}{1-\gamma_h^2}}e^{i(q_{h}x-q_{e}\bar{x})}+r_{\uparrow\downarrow}^{h'e'}\gamma_e\sqrt{\frac{1-\gamma_h^2}{1-\gamma_e^2}}}\\&{\times e^{-i(q_{e}x-q_{h}\bar{x})}\Big),}\\
[\mathcal{G}^{r}_{eh}]_{\uparrow\uparrow}&=\frac{i\eta}{2(q_{h}\gamma_e-q_{e}\gamma_h)}\Big(-\gamma_e\gamma_he^{iq_{e}|x-\bar{x}|}\mbox{sgn}(x-\bar{x})+\gamma_e\gamma_he^{-iq_{h}|x-\bar{x}|}+r_{\uparrow\uparrow}^{e'e'}\gamma_e\gamma_he^{-iq_{e}(x+\bar{x})}-r_{\uparrow\uparrow}^{h'h'}\gamma_e\gamma_h e^{iq_{h}(x+\bar{x})}\\&-r_{\uparrow\uparrow}^{e'h'}\gamma_h^2\sqrt{\frac{1-\gamma_e^2}{1-\gamma_h^2}}e^{i(q_{h}x-q_{e}\bar{x})}+r_{\uparrow\uparrow}^{h'e'}\gamma_e^2\sqrt{\frac{1-\gamma_h^2}{1-\gamma_e^2}}e^{-i(q_{e}x-q_{h}\bar{x})}\Big),
\end{split}\nonumber
\end{equation}
\begin{equation}
\begin{split}
[\mathcal{G}^{r}_{eh}]_{\downarrow\downarrow}&=\frac{i\eta}{2(q_{h}\gamma_e-q_{e}\gamma_h)}\Big(-\gamma_e\gamma_he^{iq_{e}|x-\bar{x}|}\mbox{sgn}(x-\bar{x})+\gamma_e\gamma_he^{-iq_{h}|x-\bar{x}|}+r_{\downarrow\downarrow}^{e'e'}\gamma_e\gamma_he^{-iq_{e}(x+\bar{x})}-r_{\downarrow\downarrow}^{h'h'}\gamma_e\gamma_h e^{iq_{h}(x+\bar{x})}\\&-r_{\downarrow\downarrow}^{e'h'}\gamma_h^2\sqrt{\frac{1-\gamma_e^2}{1-\gamma_h^2}}e^{i(q_{h}x-q_{e}\bar{x})}+r_{\downarrow\downarrow}^{h'e'}\gamma_e^2\sqrt{\frac{1-\gamma_h^2}{1-\gamma_e^2}}e^{-i(q_{e}x-q_{h}\bar{x})}\Big),\\
[\mathcal{G}^{r}_{eh}]_{\uparrow\downarrow}&=\frac{i\eta}{2(q_{h}\gamma_e-q_{e}\gamma_h)}\Big(r_{\downarrow\uparrow}^{e'e'}\gamma_e\gamma_he^{-iq_{e}(x+\bar{x})}-r_{\downarrow\uparrow}^{h'h'}\gamma_e\gamma_he^{iq_{h}(x+\bar{x})}-r_{\downarrow\uparrow}^{e'h'}\gamma_h^2\sqrt{\frac{1-\gamma_e^2}{1-\gamma_h^2}}e^{i(q_{h}x-q_{e}\bar{x})}+r_{\downarrow\uparrow}^{h'e'}\gamma_e^2\sqrt{\frac{1-\gamma_h^2}{1-\gamma_e^2}}\\&\times e^{-i(q_{e}x-q_{h}\bar{x})}\Big),\\
[\mathcal{G}^{r}_{eh}]_{\downarrow\uparrow}&=\frac{i\eta}{2(q_{h}\gamma_e-q_{e}\gamma_h)}\Big(r_{\uparrow\downarrow}^{e'e'}\gamma_e\gamma_he^{-iq_{e}(x+\bar{x})}-r_{\uparrow\downarrow}^{h'h'}\gamma_e\gamma_he^{iq_{h}(x+\bar{x})}-r_{\uparrow\downarrow}^{e'h'}\gamma_h^2\sqrt{\frac{1-\gamma_e^2}{1-\gamma_h^2}}e^{i(q_{h}x-q_{e}\bar{x})}+r_{\uparrow\downarrow}^{h'e'}\gamma_e^2\sqrt{\frac{1-\gamma_h^2}{1-\gamma_e^2}}\\&\times e^{-i(q_{e}x-q_{h}\bar{x})}\Big).
\end{split}
\end{equation}
In the topological regime for {$\mathcal{G}_{ee}^{r}$ and} $\mathcal{G}_{eh}^{r}$ we get,
\begin{equation}
\begin{split}
{[\mathcal{G}^{r}_{ee}]_{\uparrow\uparrow}}&{=\frac{i\eta}{2(q_{h}\gamma_h-q_{e}\gamma_e)}\Big(\gamma_ee^{iq_{e}|x-\bar{x}|}+\gamma_he^{-iq_{h}|x-\bar{x}|}-r_{\uparrow\uparrow}^{e'e'}\gamma_ee^{-iq_e(x+\bar{x})}-r_{\uparrow\uparrow}^{h'h'}\gamma_he^{iq_h(x+\bar{x})}+r_{\uparrow\uparrow}^{e'h'}\gamma_he^{i(q_{h}x-q_{e}\bar{x})}}\\&{+r_{\uparrow\uparrow}^{h'e'}\gamma_ee^{-i(q_{e}x-q_{h}\bar{x})}\Big),}\\
{[\mathcal{G}^{r}_{ee}]_{\downarrow\downarrow}}&{=\frac{i\eta}{2(q_{h}\gamma_h-q_{e}\gamma_e)}\Big(\gamma_ee^{iq_{e}|x-\bar{x}|}+\gamma_he^{-iq_{h}|x-\bar{x}|}-r_{\downarrow\downarrow}^{e'e'}\gamma_ee^{-iq_e(x+\bar{x})}-r_{\downarrow\downarrow}^{h'h'}\gamma_he^{iq_h(x+\bar{x})}+r_{\downarrow\downarrow}^{e'h'}\gamma_he^{i(q_{h}x-q_{e}\bar{x})}}\\&{+r_{\downarrow\downarrow}^{h'e'}\gamma_ee^{-i(q_{e}x-q_{h}\bar{x})}\Big),}\\
{[\mathcal{G}^{r}_{ee}]_{\uparrow\downarrow}}&{=\frac{i\eta}{2(q_{h}\gamma_h-q_{e}\gamma_e)}\Big(-r_{\downarrow\uparrow}^{e'e'}\gamma_ee^{-iq_e(x+\bar{x})}-r_{\downarrow\uparrow}^{h'h'}\gamma_he^{iq_h(x+\bar{x})}+r_{\downarrow\uparrow}^{e'h'}\gamma_he^{i(q_{h}x-q_{e}\bar{x})}+r_{\downarrow\uparrow}^{h'e'}\gamma_ee^{-i(q_{e}x-q_{h}\bar{x})}\Big),}\\
{[\mathcal{G}^{r}_{ee}]_{\downarrow\uparrow}}&{=\frac{i\eta}{2(q_{h}\gamma_h-q_{e}\gamma_e)}\Big(-r_{\uparrow\downarrow}^{e'e'}\gamma_ee^{-iq_e(x+\bar{x})}-r_{\uparrow\downarrow}^{h'h'}\gamma_he^{iq_h(x+\bar{x})}+r_{\uparrow\downarrow}^{e'h'}\gamma_he^{i(q_{h}x-q_{e}\bar{x})}+r_{\uparrow\downarrow}^{h'e'}\gamma_ee^{-i(q_{e}x-q_{h}\bar{x})}\Big),}\\
[\mathcal{G}^{r}_{eh}]_{\uparrow\uparrow}&=\frac{i\eta}{2(q_{h}\gamma_e-q_{e}\gamma_h)}\Big(-\gamma_e\gamma_he^{iq_{e}|x-\bar{x}|}\mbox{sgn}(x-\bar{x})+\gamma_e\gamma_he^{-iq_{h}|x-\bar{x}|}+r_{\uparrow\uparrow}^{e'e'}\gamma_e\gamma_he^{-iq_{e}(x+\bar{x})}-r_{\uparrow\uparrow}^{h'h'}\gamma_e\gamma_he^{iq_{h}(x+\bar{x})}\\&-r_{\uparrow\uparrow}^{e'h'}\gamma_h^2 e^{i(q_{h}x-q_{e}\bar{x})}+r_{\uparrow\uparrow}^{h'e'}\gamma_e^2e^{-i(q_{e}x-q_{h}\bar{x})}\Big),\\
[\mathcal{G}^{r}_{eh}]_{\downarrow\downarrow}&=\frac{i\eta}{2(q_{h}\gamma_e-q_{e}\gamma_h)}\Big(-\gamma_e\gamma_he^{iq_{e}|x-\bar{x}|}\mbox{sgn}(x-\bar{x})+\gamma_e\gamma_he^{-iq_{h}|x-\bar{x}|}+r_{\downarrow\downarrow}^{e'e'}\gamma_e\gamma_he^{-iq_{e}(x+\bar{x})}-r_{\downarrow\downarrow}^{h'h'}\gamma_e\gamma_he^{iq_{h}(x+\bar{x})}\\&-r_{\downarrow\downarrow}^{e'h'}\gamma_h^2 e^{i(q_{h}x-q_{e}\bar{x})}+r_{\downarrow\downarrow}^{h'e'}\gamma_e^2e^{-i(q_{e}x-q_{h}\bar{x})}\Big),\\
[\mathcal{G}^{r}_{eh}]_{\uparrow\downarrow}&=\frac{i\eta}{2(q_{h}\gamma_e-q_{e}\gamma_h)}\Big(r_{\downarrow\uparrow}^{e'e'}\gamma_e\gamma_he^{-iq_{e}(x+\bar{x})}-r_{\downarrow\uparrow}^{h'h'}\gamma_e\gamma_h e^{iq_{h}(x+\bar{x})}-r_{\downarrow\uparrow}^{e'h'}\gamma_h^2 e^{i(q_{h}x-q_{e}\bar{x})}+r_{\downarrow\uparrow}^{h'e'}\gamma_e^2 e^{-i(q_{e}x-q_{h}\bar{x})}\Big),\\
[\mathcal{G}^{r}_{eh}]_{\downarrow\uparrow}&=\frac{i\eta}{2(q_{h}\gamma_e-q_{e}\gamma_h)}\Big(r_{\uparrow\downarrow}^{e'e'}\gamma_e\gamma_he^{-iq_{e}(x+\bar{x})}-r_{\uparrow\downarrow}^{h'h'}\gamma_e\gamma_h e^{iq_{h}(x+\bar{x})}-r_{\uparrow\downarrow}^{e'h'}\gamma_h^2 e^{i(q_{h}x-q_{e}\bar{x})}+r_{\uparrow\downarrow}^{h'e'}\gamma_e^2 e^{-i(q_{e}x-q_{h}\bar{x})}\Big).
\end{split}
\end{equation}
{In case of $p_x$-$I_1$-$N_1$-SF-$N_2$-$I_2$-$p_x$ junction, in the trivial regime for $\mathcal{G}_{ee}^{r}$ and $\mathcal{G}_{eh}^{r}$ we get,
\begin{equation}
\begin{split}
[\mathcal{G}^{r}_{ee}]_{\uparrow\uparrow}&=\frac{i\eta}{2(q_{h}\gamma_h-q_{e}\gamma_e)}\Big(\gamma_ee^{iq_{e}|x-\bar{x}|}+\gamma_he^{-iq_{h}|x-\bar{x}|}-r_{\uparrow\uparrow}^{\prime e'e'}\gamma_ee^{-iq_e(x+\bar{x}+L)}-r_{\uparrow\uparrow}^{\prime h'h'}\gamma_he^{iq_h(x+\bar{x}+L)}+r_{\uparrow\uparrow}^{\prime e'h'}\gamma_h\sqrt{\frac{1-\gamma_e^2}{1-\gamma_h^2}}\\&\times e^{i(q_{h}(x+L/2)-q_{e}(\bar{x}+L/2))}+r_{\uparrow\uparrow}^{\prime h'e'}\gamma_e\sqrt{\frac{1-\gamma_h^2}{1-\gamma_e^2}}e^{-i(q_{e}(x+L/2)-q_{h}(\bar{x}+L/2))}\Big),\\
\end{split}\nonumber
\end{equation}}
{\begin{equation}
\begin{split}
[\mathcal{G}^{r}_{ee}]_{\downarrow\downarrow}&=\frac{i\eta}{2(q_{h}\gamma_h-q_{e}\gamma_e)}\Big(\gamma_ee^{iq_{e}|x-\bar{x}|}+\gamma_he^{-iq_{h}|x-\bar{x}|}-r_{\downarrow\downarrow}^{\prime e'e'}\gamma_ee^{-iq_e(x+\bar{x}+L)}-r_{\downarrow\downarrow}^{\prime h'h'}\gamma_he^{iq_h(x+\bar{x}+L)}+r_{\downarrow\downarrow}^{\prime e'h'}\gamma_h\sqrt{\frac{1-\gamma_e^2}{1-\gamma_h^2}}\\&\times e^{i(q_{h}(x+L/2)-q_{e}(\bar{x}+L/2))}+r_{\downarrow\downarrow}^{\prime h'e'}\gamma_e\sqrt{\frac{1-\gamma_h^2}{1-\gamma_e^2}}e^{-i(q_{e}(x+L/2)-q_{h}(\bar{x}+L/2))}\Big),\\
[\mathcal{G}^{r}_{ee}]_{\uparrow\downarrow}&=\frac{i\eta}{2(q_{h}\gamma_h-q_{e}\gamma_e)}\Big(-r_{\downarrow\uparrow}^{\prime e'e'}\gamma_ee^{-iq_e(x+\bar{x}+L)}-r_{\downarrow\uparrow}^{\prime h'h'}\gamma_he^{iq_h(x+\bar{x}+L)}+r_{\downarrow\uparrow}^{\prime e'h'}\gamma_h\sqrt{\frac{1-\gamma_e^2}{1-\gamma_h^2}}e^{i(q_{h}(x+L/2)-q_{e}(\bar{x}+L/2))}\\&+r_{\downarrow\uparrow}^{\prime h'e'}\gamma_e\sqrt{\frac{1-\gamma_h^2}{1-\gamma_e^2}}e^{-i(q_{e}(x+L/2)-q_{h}(\bar{x}+L/2))}\Big),\\
[\mathcal{G}^{r}_{ee}]_{\downarrow\uparrow}&=\frac{i\eta}{2(q_{h}\gamma_h-q_{e}\gamma_e)}\Big(-r_{\uparrow\downarrow}^{\prime e'e'}\gamma_ee^{-iq_e(x+\bar{x}+L)}-r_{\uparrow\downarrow}^{\prime h'h'}\gamma_he^{iq_h(x+\bar{x}+L)}+r_{\uparrow\downarrow}^{\prime e'h'}\gamma_h\sqrt{\frac{1-\gamma_e^2}{1-\gamma_h^2}}e^{i(q_{h}(x+L/2)-q_{e}(\bar{x}+L/2))}\\&+r_{\uparrow\downarrow}^{\prime h'e'}\gamma_e\sqrt{\frac{1-\gamma_h^2}{1-\gamma_e^2}}e^{-i(q_{e}(x+L/2)-q_{h}(\bar{x}+L/2))}\Big),\\
[\mathcal{G}^{r}_{eh}]_{\uparrow\uparrow}&=\frac{i\eta}{2(q_{h}\gamma_e-q_{e}\gamma_h)}\Big(-\gamma_e\gamma_he^{iq_{e}|x-\bar{x}|}\mbox{sgn}(x-\bar{x})+\gamma_e\gamma_he^{-iq_{h}|x-\bar{x}|}+r_{\uparrow\uparrow}^{\prime e'e'}\gamma_e\gamma_he^{-iq_{e}(x+\bar{x}+L)}-r_{\uparrow\uparrow}^{\prime h'h'}\gamma_e\gamma_h e^{iq_{h}(x+\bar{x}+L)}\\&-r_{\uparrow\uparrow}^{\prime e'h'}\gamma_h^2\sqrt{\frac{1-\gamma_e^2}{1-\gamma_h^2}}e^{i(q_{h}(x+L/2)-q_{e}(\bar{x}+L/2))}+r_{\uparrow\uparrow}^{\prime h'e'}\gamma_e^2\sqrt{\frac{1-\gamma_h^2}{1-\gamma_e^2}}e^{-i(q_{e}(x+L/2)-q_{h}(\bar{x}+L/2))}\Big),\\
[\mathcal{G}^{r}_{eh}]_{\downarrow\downarrow}&=\frac{i\eta}{2(q_{h}\gamma_e-q_{e}\gamma_h)}\Big(-\gamma_e\gamma_he^{iq_{e}|x-\bar{x}|}\mbox{sgn}(x-\bar{x})+\gamma_e\gamma_he^{-iq_{h}|x-\bar{x}|}+r_{\downarrow\downarrow}^{\prime e'e'}\gamma_e\gamma_he^{-iq_{e}(x+\bar{x}+L)}-r_{\downarrow\downarrow}^{\prime h'h'}\gamma_e\gamma_h e^{iq_{h}(x+\bar{x}+L)}\\&-r_{\downarrow\downarrow}^{\prime e'h'}\gamma_h^2\sqrt{\frac{1-\gamma_e^2}{1-\gamma_h^2}}e^{i(q_{h}(x+L/2)-q_{e}(\bar{x}+L/2))}+r_{\downarrow\downarrow}^{\prime h'e'}\gamma_e^2\sqrt{\frac{1-\gamma_h^2}{1-\gamma_e^2}}e^{-i(q_{e}(x+L/2)-q_{h}(\bar{x}+L/2))}\Big),\\
[\mathcal{G}^{r}_{eh}]_{\uparrow\downarrow}&=\frac{i\eta}{2(q_{h}\gamma_e-q_{e}\gamma_h)}\Big(r_{\downarrow\uparrow}^{\prime e'e'}\gamma_e\gamma_he^{-iq_{e}(x+\bar{x}+L)}-r_{\downarrow\uparrow}^{\prime h'h'}\gamma_e\gamma_he^{iq_{h}(x+\bar{x}+L)}-r_{\downarrow\uparrow}^{\prime e'h'}\gamma_h^2\sqrt{\frac{1-\gamma_e^2}{1-\gamma_h^2}}e^{i(q_{h}(x+L/2)-q_{e}(\bar{x}+L/2))}\\&+r_{\downarrow\uparrow}^{\prime h'e'}\gamma_e^2\sqrt{\frac{1-\gamma_h^2}{1-\gamma_e^2}}e^{-i(q_{e}(x+L/2)-q_{h}(\bar{x}+L/2))}\Big),\\
[\mathcal{G}^{r}_{eh}]_{\downarrow\uparrow}&=\frac{i\eta}{2(q_{h}\gamma_e-q_{e}\gamma_h)}\Big(r_{\uparrow\downarrow}^{\prime e'e'}\gamma_e\gamma_he^{-iq_{e}(x+\bar{x}+L)}-r_{\uparrow\downarrow}^{\prime h'h'}\gamma_e\gamma_he^{iq_{h}(x+\bar{x}+L)}-r_{\uparrow\downarrow}^{\prime e'h'}\gamma_h^2\sqrt{\frac{1-\gamma_e^2}{1-\gamma_h^2}}e^{i(q_{h}(x+L/2)-q_{e}(\bar{x}+L/2))}\\&+r_{\uparrow\downarrow}^{\prime h'e'}\gamma_e^2\sqrt{\frac{1-\gamma_h^2}{1-\gamma_e^2}}e^{-i(q_{e}(x+L/2)-q_{h}(\bar{x}+L/2))}\Big).
\end{split}
\label{greh-triv}
\end{equation}
In the topological regime for $\mathcal{G}_{ee}^{r}$ and $\mathcal{G}_{eh}^{r}$ we get,
\begin{equation}
\begin{split}
[\mathcal{G}^{r}_{ee}]_{\uparrow\uparrow}&=\frac{i\eta}{2(q_{h}\gamma_h-q_{e}\gamma_e)}\Big(\gamma_ee^{iq_{e}|x-\bar{x}|}+\gamma_he^{-iq_{h}|x-\bar{x}|}-r_{\uparrow\uparrow}^{\prime e'e'}\gamma_ee^{-iq_e(x+\bar{x}+L)}-r_{\uparrow\uparrow}^{\prime h'h'}\gamma_he^{iq_h(x+\bar{x}+L)}+r_{\uparrow\uparrow}^{\prime e'h'}\gamma_h\\&\times e^{i(q_{h}(x+L/2)-q_{e}(\bar{x}+L/2))}+r_{\uparrow\uparrow}^{\prime h'e'}\gamma_ee^{-i(q_{e}(x+L/2)-q_{h}(\bar{x}+L/2))}\Big),\\
[\mathcal{G}^{r}_{ee}]_{\downarrow\downarrow}&=\frac{i\eta}{2(q_{h}\gamma_h-q_{e}\gamma_e)}\Big(\gamma_ee^{iq_{e}|x-\bar{x}|}+\gamma_he^{-iq_{h}|x-\bar{x}|}-r_{\downarrow\downarrow}^{\prime e'e'}\gamma_ee^{-iq_e(x+\bar{x}+L)}-r_{\downarrow\downarrow}^{\prime h'h'}\gamma_he^{iq_h(x+\bar{x}+L)}+r_{\downarrow\downarrow}^{\prime e'h'}\gamma_h\\&\times e^{i(q_{h}(x+L/2)-q_{e}(\bar{x}+L/2))}+r_{\downarrow\downarrow}^{\prime h'e'}\gamma_ee^{-i(q_{e}(x+L/2)-q_{h}(\bar{x}+L/2))}\Big),\\
[\mathcal{G}^{r}_{ee}]_{\uparrow\downarrow}&=\frac{i\eta}{2(q_{h}\gamma_h-q_{e}\gamma_e)}\Big(-r_{\downarrow\uparrow}^{\prime e'e'}\gamma_ee^{-iq_e(x+\bar{x}+L)}-r_{\downarrow\uparrow}^{\prime h'h'}\gamma_he^{iq_h(x+\bar{x}+L)}+r_{\downarrow\uparrow}^{\prime e'h'}\gamma_he^{i(q_{h}(x+L/2)-q_{e}(\bar{x}+L/2))}+r_{\downarrow\uparrow}^{\prime h'e'}\gamma_e\\&\times e^{-i(q_{e}(x+L/2)-q_{h}(\bar{x}+L/2))}\Big),\\
\end{split}\nonumber
\end{equation}}
{\begin{equation}
\begin{split}
[\mathcal{G}^{r}_{ee}]_{\downarrow\uparrow}&=\frac{i\eta}{2(q_{h}\gamma_h-q_{e}\gamma_e)}\Big(-r_{\uparrow\downarrow}^{\prime e'e'}\gamma_ee^{-iq_e(x+\bar{x}+L)}-r_{\uparrow\downarrow}^{\prime h'h'}\gamma_he^{iq_h(x+\bar{x}+L)}+r_{\uparrow\downarrow}^{\prime e'h'}\gamma_he^{i(q_{h}(x+L/2)-q_{e}(\bar{x}+L/2))}+r_{\uparrow\downarrow}^{\prime h'e'}\gamma_e\\&\times e^{-i(q_{e}(x+L/2)-q_{h}(\bar{x}+L/2))}\Big),\\
[\mathcal{G}^{r}_{eh}]_{\uparrow\uparrow}&=\frac{i\eta}{2(q_{h}\gamma_e-q_{e}\gamma_h)}\Big(-\gamma_e\gamma_he^{iq_{e}|x-\bar{x}|}\mbox{sgn}(x-\bar{x})+\gamma_e\gamma_he^{-iq_{h}|x-\bar{x}|}+r_{\uparrow\uparrow}^{\prime e'e'}\gamma_e\gamma_he^{-iq_{e}(x+\bar{x}+L)}-r_{\uparrow\uparrow}^{\prime h'h'}\gamma_e\gamma_he^{iq_{h}(x+\bar{x}+L)}\\&-r_{\uparrow\uparrow}^{\prime e'h'}\gamma_h^2e^{i(q_{h}(x+L/2)-q_{e}(\bar{x}+L/2))}+r_{\uparrow\uparrow}^{\prime h'e'}\gamma_e^2e^{-i(q_{e}(x+L/2)-q_{h}(\bar{x}+L/2))}\Big),\\
[\mathcal{G}^{r}_{eh}]_{\downarrow\downarrow}&=\frac{i\eta}{2(q_{h}\gamma_e-q_{e}\gamma_h)}\Big(-\gamma_e\gamma_he^{iq_{e}|x-\bar{x}|}\mbox{sgn}(x-\bar{x})+\gamma_e\gamma_he^{-iq_{h}|x-\bar{x}|}+r_{\downarrow\downarrow}^{\prime e'e'}\gamma_e\gamma_he^{-iq_{e}(x+\bar{x}+L)}-r_{\downarrow\downarrow}^{\prime h'h'}\gamma_e\gamma_he^{iq_{h}(x+\bar{x}+L)}\\&-r_{\downarrow\downarrow}^{\prime e'h'}\gamma_h^2e^{i(q_{h}(x+L/2)-q_{e}(\bar{x}+L/2))}+r_{\downarrow\downarrow}^{\prime h'e'}\gamma_e^2e^{-i(q_{e}(x+L/2)-q_{h}(\bar{x}+L/2))}\Big),\\
[\mathcal{G}^{r}_{eh}]_{\uparrow\downarrow}&=\frac{i\eta}{2(q_{h}\gamma_e-q_{e}\gamma_h)}\Big(r_{\downarrow\uparrow}^{\prime e'e'}\gamma_e\gamma_he^{-iq_{e}(x+\bar{x}+L)}-r_{\downarrow\uparrow}^{\prime h'h'}\gamma_e\gamma_h e^{iq_{h}(x+\bar{x}+L)}-r_{\downarrow\uparrow}^{\prime e'h'}\gamma_h^2 e^{i(q_{h}(x+L/2)-q_{e}(\bar{x}+L/2))}+r_{\downarrow\uparrow}^{\prime h'e'}\gamma_e^2\\&\times e^{-i(q_{e}(x+L/2)-q_{h}(\bar{x}+L/2))}\Big),\\
[\mathcal{G}^{r}_{eh}]_{\downarrow\uparrow}&=\frac{i\eta}{2(q_{h}\gamma_e-q_{e}\gamma_h)}\Big(r_{\uparrow\downarrow}^{\prime e'e'}\gamma_e\gamma_he^{-iq_{e}(x+\bar{x}+L)}-r_{\uparrow\downarrow}^{\prime h'h'}\gamma_e\gamma_h e^{iq_{h}(x+\bar{x}+L)}-r_{\uparrow\downarrow}^{\prime e'h'}\gamma_h^2 e^{i(q_{h}(x+L/2)-q_{e}(\bar{x}+L/2))}+r_{\uparrow\downarrow}^{\prime h'e'}\gamma_e^2\\&\times e^{-i(q_{e}(x+L/2)-q_{h}(\bar{x}+L/2))}\Big).
\end{split}
\label{greh-topo}
\end{equation}}
\section*{Appendix C: Analytical formulas for pairing amplitudes}
In this Appendix we present the analytical formulas for even- and odd-$\nu$ EST and MST correlations at finite $\nu$ in both trivial and topological regimes.
In the \textbf{trivial} regime, both bulk and surface even-$\nu$ EST pairing amplitudes are finite within  the left $p_{x}$ superconductor and can be expressed as:
\begin{align}
\label{even-equal-pp}
\mbox{Bulk even-$\nu$ EST:}\,\,f_{\uparrow\uparrow}^{E,B}(x,\bar{x},\nu)&=\frac{i\eta}{8(q_{h}\gamma_e-q_{e}\gamma_h)}\Big(4\gamma_e\gamma_h \big(e^{-iq_{h}|x-\bar{x}|}-e^{iq_{e}|x-\bar{x}|}\big)\mbox{sgn}(x-\bar{x})\Big)\nonumber\\&=f_{\downarrow\downarrow}^{E,B}(x,\bar{x},\nu),\,\mbox{for}\,\, x<0,\\
\mbox{Surface even-$\nu$ EST:}\,\,f_{\uparrow\uparrow}^{E,S}(x,\bar{x},\nu)&=\frac{i\eta}{8(q_{h}\gamma_e-q_{e}\gamma_h)}\Big(\big(\gamma_h^2\sqrt{\frac{1-\gamma_e^2}{1-\gamma_h^2}}(r_{\uparrow\uparrow}^{e'h'}+r_{\downarrow\downarrow}^{e'h'})+\gamma_e^2\sqrt{\frac{1-\gamma_h^2}{1-\gamma_e^2}}(r_{\uparrow\uparrow}^{h'e'}+r_{\downarrow\downarrow}^{h'e'})\big)\nonumber\\&\times\big(e^{-i(q_{e}x-q_{h}\bar{x})}-e^{-i(q_{e}\bar{x}-q_{h}x)}\big)\Big)=f_{\downarrow\downarrow}^{E,S}(x,\bar{x},\nu),\,\mbox{for}\,\, x<0,
\end{align}
while bulk odd-$\nu$ EST pairing amplitude vanishes even though surface odd-$\nu$ EST pairing amplitude is finite, and can be expressed as,
\begin{align}
\mbox{Surface odd-$\nu$ EST:}\,\,f_{\uparrow\uparrow}^{O,S}(x,\bar{x},\nu)&=-\frac{i\eta}{8(q_{h}\gamma_e-q_{e}\gamma_h)}\Big(2\gamma_e\gamma_h\big((r_{\uparrow\uparrow}^{e'e'}+r_{\downarrow\downarrow}^{e'e'})e^{-iq_{e}(x+\bar{x})}-(r_{\uparrow\uparrow}^{h'h'}+r_{\downarrow\downarrow}^{h'h'})e^{iq_{h}(x+\bar{x})}\big)\nonumber\\&+\big(-\gamma_h^2\sqrt{\frac{1-\gamma_e^2}{1-\gamma_h^2}}(r_{\uparrow\uparrow}^{e'h'}+r_{\downarrow\downarrow}^{e'h'})+\gamma_e^2\sqrt{\frac{1-\gamma_h^2}{1-\gamma_{e}^2}}(r_{\uparrow\uparrow}^{h'e'}+r_{\downarrow\downarrow}^{h'e'})\big)\big(e^{i(q_{h}x-q_{e}\bar{x})}+e^{i(q_{h}\bar{x}-q_{e}x)}\big)\Big)\nonumber\\
&=f_{\downarrow\downarrow}^{O,S}(x,\bar{x},\nu),\, \mbox{for}\,\, x<0.
\label{odd-equal-pp}
\end{align}
\normalsize
However, in the \textbf{topological} regime, both bulk and surface even-$\nu$ EST pairing amplitudes are finite within the left $p_{x}$ superconductor and can be expressed as:
\begin{align}
\mbox{Bulk even-$\nu$ EST:}\,\,\label{even-equal-topo-pp}
f_{\uparrow\uparrow}^{E,B}(x,\bar{x},\nu)&=
\frac{i\eta}{8(q_{e}\gamma_e-q_{h}\gamma_h)}\Big(2\big(e^{-iq_{h}|x-\bar{x}|}-e^{iq_{e}|x-\bar{x}|}\big)\mbox{sgn}(x-\bar{x})\Big)
+\frac{i\eta}{8(q_{h}\gamma_e-q_{e}\gamma_h)}\nonumber\\&\times\Big(2\gamma_e\gamma_h\big(e^{-iq_{h}|x-\bar{x}|}-e^{iq_{e}|x-\bar{x}|}\big)\mbox{sgn}(x-\bar{x})\Big)=f_{\downarrow\downarrow}^{E,B}(x,\bar{x},\nu),\,\,\mbox{for}\,\, x<0,\\\nonumber
\end{align}
\begin{align}
\mbox{Surface even-$\nu$ EST:}\,\,f_{\uparrow\uparrow}^{E,S}(x,\bar{x},\nu)&=
\frac{i\eta}{8(q_{e}\gamma_e-q_{h}\gamma_h)}\Big((r_{\uparrow\uparrow}^{e'h' *}+r_{\downarrow\downarrow}^{e'h' *})e^{-i(q_{e}x-q_{h}\bar{x})}-(r_{\uparrow\uparrow}^{h'e' *}+r_{\downarrow\downarrow}^{h'e' *})e^{i(q_{h}x-q_{e}\bar{x})}\Big)\nonumber\\&+\frac{i\eta}{8(q_{h}\gamma_e-q_{e}\gamma_h)}\Big(-\gamma_h^2(r_{\uparrow\uparrow}^{e'h'}+r_{\downarrow\downarrow}^{e'h'})e^{-i(q_{e}\bar{x}-q_{h}x)}+\gamma_e^2(r_{\uparrow\uparrow}^{h'e'}+r_{\downarrow\downarrow}^{h'e'})e^{-i(q_{e}x-q_{h}\bar{x})}\Big)\nonumber\\&=f_{\downarrow\downarrow}^{E,S}(x,\bar{x},\nu),\,\, \mbox{for}\,\, x<0.\end{align}
But in contrast to the trivial regime wherein bulk odd-$\nu$ EST pairing amplitude vanishes, in the topological regime, both bulk odd-$\nu$ EST pairing amplitude and surface odd-$\nu$ EST pairing amplitude are non-zero, and given by
\begin{align}
\mbox{Bulk odd-$\nu$ EST:}\,\,f_{\uparrow\uparrow}^{O,B}(x,\bar{x},\nu)&=\frac{i\eta}{8(q_{e}\gamma_e-q_{h}\gamma_h)}\Big(-2\big(e^{-iq_{h}|x-\bar{x}|}-e^{iq_{e}|x-\bar{x}|}\big)\mbox{sgn}(x-\bar{x})\Big)+\frac{i\eta}{8(q_{h}\gamma_e-q_{e}\gamma_h)}\nonumber\\&\times\Big(2\gamma_e\gamma_h\big(e^{-iq_{h}|x-\bar{x}|}-e^{iq_{e}|x-\bar{x}|}\big)\mbox{sgn}(x-\bar{x})\Big)=f_{\downarrow\downarrow}^{O,B}(x,\bar{x},\nu),\,\, \mbox{for}\,\, x<0,\\
\mbox{Surface odd-$\nu$ EST:}\,\,f_{\uparrow\uparrow}^{O,S}(x,\bar{x},\nu)&=\frac{i\eta}{8(q_{e}\gamma_e-q_{h}\gamma_h)}\Big((r_{\uparrow\uparrow}^{e'e'}+r_{\downarrow\downarrow}^{e'e'})e^{-iq_{e}(x+\bar{x})}-(r_{\uparrow\uparrow}^{h'h'}+r_{\downarrow\downarrow}^{h'h'})e^{iq_{h}(x+\bar{x})}-(r_{\uparrow\uparrow}^{e'h' *}+r_{\downarrow\downarrow}^{e'h' *})\nonumber\\&\times e^{-i(q_{e}x-q_{h}\bar{x})}+(r_{\uparrow\uparrow}^{h'e' *}+r_{\downarrow\downarrow}^{h'e' *})e^{i(q_{h}x-q_{e}\bar{x})}\Big)+\frac{i\eta}{8(q_{h}\gamma_e-q_{e}\gamma_h)}\Big(\gamma_e\gamma_h(r_{\uparrow\uparrow}^{e'e'}+r_{\downarrow\downarrow}^{e'e'})\nonumber\\&\times e^{-iq_{e}(x+\bar{x})}-\gamma_e\gamma_h(r_{\uparrow\uparrow}^{h'h'}+r_{\downarrow\downarrow}^{h'h'})e^{iq_{h}(x+\bar{x})}-\gamma_h^2(r_{\uparrow\uparrow}^{e'h'}+r_{\downarrow\downarrow}^{e'h'})e^{-i(q_{e}\bar{x}-q_{h}x)}\nonumber\\&+\gamma_e^2(r_{\uparrow\uparrow}^{h'e'}+r_{\downarrow\downarrow}^{h'e'})e^{-i(q_{e}x-q_{h}\bar{x})}\Big)=f_{\downarrow\downarrow}^{O,S}(x,\bar{x},\nu),\,\,\mbox{for}\,\, x<0.
\label{odd-equal-topo-pp}
\end{align}
From Eqs.~\eqref{even-equal-pp}-\eqref{odd-equal-pp}, we notice that in the trivial regime, even-$\nu$ EST correlations survive both in bulk and surface, while odd-$\nu$ EST correlations exist only in surface. In the topological regime, both even-$\nu$ EST and odd-$\nu$ EST correlations exist in bulk as well as in surface as seen from Eqs.~\eqref{even-equal-topo-pp}-\eqref{odd-equal-topo-pp}.

For MST correlations, the bulk components vanish; however, surface components are finite within the left superconductor in both trivial and topological regimes.
In the \textbf{trivial} regime, even-$\nu$ and odd-$\nu$ surface MST components within the left superconductor are given by-
\begin{align}
\mbox{Surface even-$\nu$ MST:}\,\,
f_{3}^{E,S}(x,\bar{x},\nu)&=\frac{i\eta}{8(q_{h}\gamma_e-q_{e}\gamma_h)}\Big(\gamma_h^2\sqrt{\frac{1-\gamma_e^2}{1-\gamma_h^2}}(r_{\uparrow\downarrow}^{e'h'}+r_{\downarrow\uparrow}^{e'h'})+\gamma_e^2\sqrt{\frac{1-\gamma_h^2}{1-\gamma_e^2}}(r_{\uparrow\downarrow}^{h'e'}+r_{\downarrow\uparrow}^{h'e'})\Big)\nonumber\\&\times\Big(e^{i(q_{h}\bar{x}-q_{e}x)}-e^{i(q_{h}x-q_{e}\bar{x})}\Big),\,\mbox{for}\,\,x<0,\\
\mbox{Surface odd-$\nu$ MST:}\,\,f_{3}^{O,S}(x,\bar{x},\nu)&=\frac{i\eta}{8(q_{h}\gamma_e-q_{e}\gamma_h)}\Big(2\gamma_e\gamma_h\big((r_{\uparrow\downarrow}^{e'e'}+r_{\downarrow\uparrow}^{e'e'})e^{-iq_{e}(x+\bar{x})}-(r_{\uparrow\downarrow}^{h'h'}+r_{\downarrow\uparrow}^{h'h'})e^{iq_{h}(x+\bar{x})}\big)\nonumber\\&+\big(-\gamma_h^2\sqrt{\frac{1-\gamma_e^2}{1-\gamma_h^2}}(r_{\uparrow\downarrow}^{e'h'}+r_{\downarrow\uparrow}^{e'h'})+\gamma_e^2\sqrt{\frac{1-\gamma_h^2}{1-\gamma_{e}^2}}(r_{\uparrow\downarrow}^{h'e'}+r_{\downarrow\uparrow}^{h'e'})\big)\big(e^{i(q_{h}\bar{x}-q_{e}x)}+e^{i(q_{h}x-q_{e}\bar{x})}\big)\Big),\nonumber\\&\,\mbox{for}\,\, x<0.
\label{odd-mixed-pp}
\end{align}
In the \textbf{topological} regime, surface even-$\nu$ and odd-$\nu$ MST correlations within the left superconductor are given as,
\begin{align}
\mbox{Surface even-$\nu$ MST:}\,\,f_{3}^{E,S}(x,\bar{x},\nu)&=\frac{i\eta}{8(q_{e}\gamma_e-q_{h}\gamma_h)}\Big((r_{\uparrow\downarrow}^{e'h' *}+r_{\downarrow\uparrow}^{e'h' *})e^{-i(q_{e}x-q_{h}\bar{x})}-(r_{\uparrow\downarrow}^{h'e' *}+r_{\downarrow\uparrow}^{\prime h'e' *})e^{i(q_{h}x-q_{e}\bar{x})}\Big)\nonumber\\&+\frac{i\eta}{8(q_{h}\gamma_e-q_{e}\gamma_h)}\Big(-\gamma_h^2(r_{\uparrow\downarrow}^{e'h'}+r_{\downarrow\uparrow}^{e'h'})e^{-i(q_{e}\bar{x}-q_{h}x)}+\gamma_e^2(r_{\uparrow\downarrow}^{h'e'}+r_{\downarrow\uparrow}^{h'e'})e^{-i(q_{e}x-q_{h}\bar{x})}\Big),\nonumber\\&\quad \mbox{for}\,\, x<0,\\
\mbox{Surface odd-$\nu$ MST:}\,\,f_{3}^{O,S}(x,\bar{x},\nu)&=\frac{i\eta}{8(q_{e}\gamma_e-q_{h}\gamma_h)}\Big((r_{\uparrow\downarrow}^{e'e'}+r_{\downarrow\uparrow}^{e'e'})e^{-iq_{e}(x+\bar{x})}-(r_{\uparrow\downarrow}^{h'h'}+r_{\downarrow\uparrow}^{h'h'})e^{iq_{h}(x+\bar{x})}-(r_{\uparrow\downarrow}^{e'h' *}+r_{\downarrow\uparrow}^{e'h' *})\nonumber\\&\times e^{-i(q_{e}x-q_{h}\bar{x})}+(r_{\uparrow\downarrow}^{h'e' *}+r_{\downarrow\uparrow}^{h'e' *})e^{-i(q_{e}\bar{x}-q_{h}x)}\Big)+\frac{i\eta}{8(q_{h}\gamma_e-q_{e}\gamma_h)}\Big(\gamma_e\gamma_h(r_{\uparrow\downarrow}^{e'e'}+r_{\downarrow\uparrow}^{e'e'})\nonumber\\&\times e^{-iq_{e}(x+\bar{x})}-\gamma_e\gamma_h(r_{\uparrow\downarrow}^{h'h'}+r_{\downarrow\uparrow}^{h'h'})e^{iq_{h}(x+\bar{x})}-\gamma_h^2(r_{\uparrow\downarrow}^{e'h'}+r_{\downarrow\uparrow}^{e'h'})e^{-i(q_{e}\bar{x}-q_{h}x)}\nonumber\\&+\gamma_e^2(r_{\uparrow\downarrow}^{h'e'}+r_{\downarrow\uparrow}^{h'e'})e^{-i(q_{e}x-q_{h}\bar{x})}\Big),\quad \mbox{for}\,\, x<0.
\label{odd-mixed-topo-pp}
\end{align}
{\section*{Appendix D: LDOS and LMDOS}
LDOS $\rho(x,\nu)$ and LMDOS $\vec{m}(x,\nu)$ can be computed\cite{kuzz} from $\mathcal{G}^{r}_{ee}$,
\begin{equation}
\label{lod}
\rho(x,\nu)=-\frac{1}{\pi}\lim_{\epsilon\rightarrow0}\text{Im}[\text{Tr}\{\mathcal{G}^{r}_{ee}(x,x,\nu+i\epsilon)\}],\,\, \mbox{ and }\,\,
\vec{m}(x,\nu)=-\frac{1}{\pi}\lim_{\epsilon\rightarrow0}\text{Im}[\text{Tr}\{\vec{\sigma_{l}}.\mathcal{G}^{r}_{ee}(x,x,\nu+i\epsilon)\}],\,\,(l=1,2,3).
\end{equation}
Using Eq.~\eqref{greh}, we can write LDOS and LMDOS as,
\begin{eqnarray}
\rho(x,\nu)&=&-\frac{1}{\pi}\lim_{\epsilon\rightarrow0}\text{Im}[([\mathcal{G}^{r}_{ee}]_{\uparrow\uparrow}+[\mathcal{G}^{r}_{ee}]_{\downarrow\downarrow})],\\
\vec{m}(x,\nu)&=&-\frac{1}{\pi}\lim_{\epsilon\rightarrow0}\text{Im}[([\mathcal{G}^{r}_{ee}]_{\uparrow\downarrow}+[\mathcal{G}^{r}_{ee}]_{\downarrow\uparrow})\hat{x}+i([\mathcal{G}^{r}_{ee}]_{\uparrow\downarrow}-[\mathcal{G}^{r}_{ee}]_{\downarrow\uparrow})\hat{y}+([\mathcal{G}^{r}_{ee}]_{\uparrow\uparrow}-[\mathcal{G}^{r}_{ee}]_{\downarrow\downarrow})\hat{z}],
\end{eqnarray}
where the expressions for $[\mathcal{G}^{r}_{ee}]_{\uparrow\uparrow}$, $[\mathcal{G}^{r}_{ee}]_{\uparrow\downarrow}$, $[\mathcal{G}^{r}_{ee}]_{\downarrow\uparrow}$, and $[\mathcal{G}^{r}_{ee}]_{\downarrow\downarrow}$ are provided in Eq.~\eqref{greh-triv} for the trivial regime and in Eq.~\eqref{greh-topo} for the topological regime in the case of the $p_x$-$I_1$-$N_1$-SF-$N_2$-$I_2$-$p_x$ JJ. LMDOS is calculated for each of the $2\mathcal{S}+1$ possible values of $m'$ for the spin $\mathcal{S}$ of the SF, and finally, an average is taken over all $m'$ values. We find that the $y$ and $z$ components of LMDOS are zero, and LMDOS is aligned along the $x$ direction.
\section*{Appendix E: DC Josephson current}
The DC Josephson current can be calculated using the Furusaki-Tsukuda technique\cite{furu,costa,sci} as follows:
\begin{equation}
I=\frac{eE_ck_{B}T}{2\hbar}\sum_{\nu_{n}}\frac{q_e(i\nu_n)+q_h(i\nu_n)}{\sqrt{\nu_n^2+E_1^2}}\times\Bigg[\frac{r_{\uparrow\uparrow}^{\prime e'h'}(i\nu_n)+r_{\downarrow\downarrow}^{\prime e'h'}(i\nu_n)}{q_{e}(i\nu_n)}-\frac{r_{\uparrow\uparrow}^{\prime h'e'}(i\nu_n)+r_{\downarrow\downarrow}^{\prime h'e'}(i\nu_n)}{q_{h}(i\nu_n)}\Bigg],
\end{equation}
where $\nu_{n}=(2n+1)\pi k_{B}T$ are fermionic Matsubara frequencies with $n=0,\pm1,\pm2,\pm3,...$ and $E_1=\Delta_{p_x}\sqrt{\mu_{p_x}-\Delta_{p_x}^2/4}$. $k_B$ is the Boltzmann constant. $q_{e,h}(i\nu_n)$, $r_{\uparrow\uparrow}^{\prime e'h'}(i\nu_n)$, $r_{\downarrow\downarrow}^{\prime e'h'}(i\nu_n)$, $r_{\uparrow\uparrow}^{\prime h'e'}(i\nu_n)$ and, $r_{\downarrow\downarrow}^{\prime h'e'}(i\nu_n)$ are obtained from $q_{e,h}$, $r_{\uparrow\uparrow}^{\prime e'h'}$, $r_{\downarrow\downarrow}^{\prime e'h'}$, $r_{\uparrow\uparrow}^{\prime h'e'}$ and, $r_{\downarrow\downarrow}^{\prime h'e'}$ by analytically continuing $\nu$ to $i\nu_{n}$. We perform a numerical summation over the Matsubara frequencies. There are various methods to express the total DC Josephson current formula using the Furusaki-Tsukuda approach\cite{enok,costa}. All these methods yield the same total DC Josephson current. These different methods involve different scattering amplitudes. This is because the Furusaki-Tsukuda procedure adheres to both detailed balance and probability conservation, allowing for multiple representations of the same formula.}

\end{document}